\documentclass[fleqn,usenatbib]{mnras}
\usepackage[dvipsnames]{xcolor}
\usepackage{newtxtext,newtxmath}

\usepackage[T1]{fontenc}

\usepackage{bm}       
\usepackage{hyperref} 
\usepackage{orcidlink}
\DeclareRobustCommand{\NoCaseChange}[1]{#1}

\newcommand{\PabEtitle}{\texorpdfstring{$\bm{\mathcal{P}_{\NoCaseChange{\mathrm{\textit{ab}}}}^{(E)}}$}{PabE}}
\newcommand{\PabBtitle}{\texorpdfstring{$\bm{\mathcal{P}_{\NoCaseChange{\mathrm{\textit{ab}}}}^{(B)}}$}{PabB}}

\DeclareRobustCommand{\VAN}[3]{#2}
\let\VANthebibliography\thebibliography
\def\thebibliography{\DeclareRobustCommand{\VAN}[3]{##3}\VANthebibliography}

\usepackage{graphicx}	
\usepackage{amsmath}	
\usepackage[most]{tcolorbox} 

\newtcolorbox{remark}{colback=blue!2!white,colframe=blue!50!black,title=Remark}

\title[]{\textbf{Interpreting map-based \textit{E/B} spectral properties of CMB foregrounds}}

\author[G. Weymann-Despres et al.]{Gilles Weymann-Despres$^\mathrm{\orcidlink{0000-0002-9281-281X}}$\!,$^{1}$\thanks{E-mail: gilles.weymann-despres@physics.ox.ac.uk}
Léo Vacher$^\mathrm{\orcidlink{0000-0001-9551-1417}}$\!,$^{2,3}$
Michael E. Jones$^\mathrm{\orcidlink{0000-0003-3564-6680
}}$\!,$^{1}$
Angela C. Taylor$^\mathrm{\orcidlink{0000-0002-3309-9081}}$\!,$^{1}$
\newauthor
Carlo Baccigalupi$^\mathrm{\orcidlink{0000-0002-8211-1630}}$\!,$^{2,4,5}$
A.J. Banday$^\mathrm{\orcidlink{0000-0002-3358-0289}}$\!,$^{6}$
Richard D.P. Grumitt$^\mathrm{\orcidlink{0000-0001-9578-6111}}$\!,$^{7}$
Nicoletta Krachmalnicoff$^\mathrm{\orcidlink{0000-0002-5501-8449}}$\!.$^{2,4,5}$\\
\\
$^{1}$Department of Physics, University of Oxford, Denys Wilkinson Building, Keble Road, Oxford OX1 3RH, UK\\
$^{2}$The International School for Advanced Studies (SISSA), via Bonomea 265, I-34136 Trieste, Italy\\
$^{3}$Université Paris-Saclay, CNRS/IN2P3, IJCLab, 91405 Orsay, France\\
$^{4}$The National Institute for Nuclear Physics (INFN), via Valerio 2, I-34127, Trieste, Italy\\
$^{5}$The Institute for Fundamental Physics of the Universe (IFPU), Via Beirut 2, I-34151, Trieste, Italy\\
$^{6}$Univ Toulouse, CNES, CNRS, IRAP, Toulouse, France\\
$^{7}$Global College, Shanghai Jiao Tong University, Shanghai 200240, China
}

\date{Accepted XXX. Received YYY; in original form ZZZ}

\pubyear{\the\year{}}

\begin{document}
\label{firstpage}
\pagerange{\pageref{firstpage}--\pageref{lastpage}}
\maketitle

\begin{abstract}

{Map-space \textit{E}/\textit{B} decompositions of linear polarization are attractive for foreground and CMB analyses because they separate parity families: $B$-family patterns directly contaminate primordial tensor searches, while $E$-family patterns trace coherent Galactic structures. However, the \textit{E}/\textit{B} transform is not fully local and can induce apparent spectral complexity even when the underlying sky is spectrally simple in $\underline{P}=Q+iU$. We quantify this effect for synchrotron emission using complex log--Taylor and moment expansions for $\underline{P}$, its spin-preserving projections $\underline{P}_E$ and $\underline{P}_B$, and its more standard scalar projections $E$ and $B$. We relate the coefficients of these expansions to physical mechanisms such as line-of-sight mixing, synchrotron ageing, and Faraday effects. Using simple sky models, we show how $\underline{P}_E$ and $\underline{P}_B$ reorganize the spectral behaviour of $\underline{P}$ into parity families with a clear geometric meaning. They retain interpretable amplitudes and angles and satisfy the closure relation $\underline{P}=\underline{P}_E+\underline{P}_B$, which extends to all moment orders. By contrast, scalar quantities such as $|E|$ and $|B|$ show larger induced variability of effective spectral parameters and enhanced spectral complexity, while $E+iB$ lacks interpretable polarization amplitude and angle. Finally, we present simple CMB-oriented applications: three-frequency diagnostics to test whether the sky is better described by a power law in $P$ or by separate effective laws in $(P_E,P_B)$, and idealized ILC and masking examples showing that the preferred map-space field depends on where foreground simplicity and residual contamination reside. This framework provides practical diagnostics for choosing foreground modelling, cleaning, and masking strategies in Galactic and CMB $B$-mode analyses.}
\end{abstract}


\section{Introduction}
\label{sec:introduction}

Detecting the faint $B$-mode polarization of the Cosmic Microwave Background (CMB), a key signature of primordial gravitational waves predicted by inflation, remains a central goal in cosmology. However, this signal is obscured by polarized Galactic foregrounds, particularly synchrotron and thermal dust emission. Synchrotron radiation dominates at frequencies below a few tens of GHz and originates from cosmic ray electrons spiralling around the Galactic magnetic field (whose energy spectrum is well characterized, see, \textit{e.g.}, \citealt{Fermi-LAT:2012edv}). Its polarization fraction can reach up to 75\% in uniform fields but is significantly reduced by magnetic field tangling, line-of-sight integration, and Faraday depolarization at low frequencies; see, \textit{e.g.}, \cite{Waelkens:2009}. Improved characterization of this polarized synchrotron foreground is essential for component separation in CMB analysis and for informing the next generation of experiments such as the Simons Observatory \citep{SimonsObservatory:2019qwx} and \textit{LiteBIRD} \citep{LiteBIRD:2022cnt}. 

Understanding synchrotron emission is also key for Galactic science. It provides insight into the structure and dynamics of the Galactic magnetic field: its morphology and spectral behaviour encode information on magnetohydrodynamic turbulence, energy injection and dissipation processes in the interstellar medium, which play a key role in regulating star formation across the Galaxy; see, \textit{e.g.}, \cite{Price:2008rb}.

At each point $\mathbf n$ on the sphere, the measured quantities $Q$ and $U$ are the components of the symmetric, traceless spin-2 tensor $\mathcal{P}_{ab}$, expressed in the local orthonormal tangent basis $(\hat{e}_\theta, \hat{e}_\phi)$, that describes the linear polarization state (see \citealt{Hu:1997hv, Kamionkowski:2015yta}). This tensor can also be expressed in terms of a polarization amplitude $P$ and an angle $\psi\in[-\pi/2,\pi/2]$:
\begin{equation}
\label{eq:polar}
    \mathcal{P}_{ab}\ (\mathbf n) = \frac{1}{\sqrt{2}} 
\begin{pmatrix}
Q & U \\
U & -Q
\end{pmatrix}(\mathbf n)=\frac{P(\mathbf n)}{\sqrt{2}}
\begin{pmatrix}
\cos 2\psi & \sin 2\psi \\
\sin 2\psi & -\cos 2\psi
\end{pmatrix}(\mathbf n).
\end{equation}
In order to separate parity-even (radial or tangential) patterns from parity-odd (curl-like) patterns, polarization is commonly decomposed into $E$ and $B$ modes \citep{Zaldarriaga:1996xe, Seljak:1996gy, Kamionkowski:1996ks}. Similarly, the linear polarization tensor at each point on the sphere can be viewed as a sum of two distinct terms that originate from $E$ and $B$ modes:
\begin{align}
\label{eq:decompo_eb}
    \mathcal{P}_{ab}\ (\mathbf n)=\mathcal{P}_{ab}^{(E)}\ (\mathbf n)+\mathcal{P}_{ab}^{(B)}\ (\mathbf n). 
\end{align}
As shown in Appendix~\ref{app:details_spin2_prop_PEPB}, since both $\mathcal{P}_{ab}^{(E)}$ and $\mathcal{P}_{ab}^{(B)}$ can be expressed exclusively in spin-weighted harmonics of weight $\pm 2$, and since the \textit{E}/\textit{B} decomposition is a linear projector that commutes with rotations when acting on the harmonic coefficients of the full polarization field, each component shares the same spin-2 transformation law as $\mathcal{P}_{ab}$. Therefore, both $\mathcal{P}_{ab}^{(E)}$ and $\mathcal{P}_{ab}^{(B)}$ are themselves spin-2 tensor fields on the sphere. 

Hence, analogously to $Q$ and $U$ in Eq.~\ref{eq:polar}, we can introduce the $E$- and $B$-family Stokes parameters, defined as the components of the local-orthonormal-tangent-basis matrix representations of $\mathcal{P}_{ab}^{(E)}$ and $\mathcal{P}_{ab}^{(B)}$, namely $Q_E$, $U_E$, $Q_B$, and $U_B$, as well as the corresponding polarization amplitudes $P_E$, $P_B$ and polarization angles $\psi_E$ and $\psi_B$ (see \citealt{Rotti:2018pzi, Liu:2018oqp, Liu:2022udr}). In practice, this spin-2 $E/B$ decomposition is useful for tracking, in map space, the gradient-like and curl-like contributions to the polarization through $P_E$ and $P_B$, which can in general originate from distinct physical structures (see \citealt{Liu:2018dbb} or \citealt{Martire:2023ytg}). The more standard construction is instead the spin-0 $E/B$ decomposition, in which the harmonic coefficients $a_{\ell m}^E$ and $a_{\ell m}^B$ are projected as scalar maps $E$ and $B$.\footnote{{\underline{\textbf{Implementation} to compute the $E$/$B$-separated fields:}\\[0.15cm]
The implementation of the fields introduced above is straightforward using the \texttt{healpy} \texttt{map2alm}/\texttt{alm2map} functions, see \cite{Gorski:2004by}:
\begin{align}
\{I,Q,U\} &\;\; \xrightarrow{\;\;{\rm map2alm}\;\;} \{a_{\ell m}^I,a_{\ell m}^E, a_{\ell m}^B\}, \\
\{E, Q_E,U_E\} &\;\; \xleftarrow{\;\;{\rm alm2map}\;\;} \{a_{\ell m}^E,a_{\ell m}^E, 0\},\\
\{B, Q_B,U_B\} &\;\; \xleftarrow{\;\;{\rm alm2map}\;\;} \{a_{\ell m}^B,0,a_{\ell m}^B\}.
\end{align}} (see, \textit{e.g.}, \citealt{BICEP2:2016fge})}. 

{We emphasize that these non-fully-local transforms are strictly well defined only on the full sky. In this conceptual paper we therefore work within this simplified framework and do not address the complications arising from partial-sky analyses. These issues will be discussed in a future paper of the series.}

{Throughout the paper, we use the expression ``\textit{E}/\textit{B}-separated fields'' as a generic term for fields in which the polarization information has been separated into its two parity families. When this terminology could be ambiguous, we explicitly specify the spin of the field. Thus, ``spin-2 \textit{E}/\textit{B}-separated fields'' refers to $\mathcal{P}_{ab}^{(E)}$ and $\mathcal{P}_{ab}^{(B)}$, or equivalently to their Stokes components $(Q_E,U_E)$ and $(Q_B,U_B)$ and associated amplitudes $(P_E,P_B)$. By contrast, ``spin-0 \textit{E}/\textit{B} maps'' refers to the standard scalar maps $E$ and $B$.}

Map-space \textit{E}/\textit{B} separation, in both its spin-preserving and scalar forms, is appealing not only as a mathematical decomposition, but also as a practical way to organize foreground information by parity. Coherent Galactic structures, such as large loops and filamentary features, often appear predominantly in the \textit{E} family, whereas their residual \textit{B}-family contribution is the part directly relevant for primordial tensor searches. This motivates spectral analyses in fields such as $\mathcal{P}_{ab}^{(E)}$ and $\mathcal{P}_{ab}^{(B)}$, rather than only in $(Q,U)$ separately, which are basis dependent, or in the total polarization tensor $\mathcal{P}_{ab}$, which combines the two parity families.

A second motivation is spectral. If different emitting structures overlap on the sky while having both different spectral properties and different \textit{E}/\textit{B} parity content, then the total field $\mathcal{P}_{ab}$ mixes their SEDs. We will discuss that, in such regions, the separated fields $\mathcal{P}_{ab}^{(E)}$ and $\mathcal{P}_{ab}^{(B)}$ can be spectrally simpler than $\mathcal{P}_{ab}$ itself. This situation is not unusual for synchrotron emission: distinct magnetic-field geometries or cosmic-ray electron populations can coexist in projection while contributing differently to the two parity families. One of the goals of this paper is therefore to determine when the spectral modelling should be performed in $\mathcal{P}_{ab}$ and when it is more economical to work directly in $(\mathcal{P}_{ab}^{(E)}, \mathcal{P}_{ab}^{(B)})$.

The central complication is that map-space \textit{E}/\textit{B} separation is non-fully local. The operators that define the spin-2 and spin-0 \textit{E}/\textit{B} transforms act as finite-width convolutions on the sphere, so the value of an \textit{E}/\textit{B}-separated field at a given direction receives contributions from neighbouring structures. The corresponding real-space kernels, and their typical shapes, are shown in Appendix~\ref{app:convolution_kernel}. This lack of full locality implies that \textit{E}/\textit{B} filtering can induce apparent spectral deformations in projected fields, even when the underlying sky is spectrally simple in the total polarization field.

{This lack of full locality implies that \textit{E}/\textit{B} filtering can induce apparent \emph{spectral deformations} in the projected fields, \textit{i.e.}\ departures from the spectral behaviour of the underlying polarization field (we adopt this terminology throughout the paper to avoid confusion with the distinct notion of CMB spectral distortions). Such deformations may manifest as effective curvature, higher-order spectral moments, or frequency-dependent angle evolution, even when the original sky follows a comparatively simple spectral law (for example, a rigid-angle power law in $\mathcal{P}_{ab}$).}

For any spectral analysis, it is advantageous to work with fields of low spectral complexity. Such fields can be described with fewer spectral parameters, and therefore allow more robust fits when only a limited number of frequency channels are available. This is particularly important for foreground-template extrapolation and for parametric component-separation methods, which must assume an explicit spectral model. Low spectral complexity is also desirable for Galactic science: if the effective spectral behaviour varies strongly across the sky and requires many degrees of freedom, then the morphology of a map is not extrapolated faithfully with frequency, and structures identified at one frequency may not have a clear counterpart at another. For these reasons, identifying field representations whose spectra remain as simple as possible is a central practical goal of this work.

In this work we develop a map-space spectral framework designed to quantify and interpret these effects. We introduce complex log--Taylor and complex-moment expansions for generic complex fields, and establish the key properties that make them useful for \textit{E}/\textit{B}-separated analyses. In particular, we show that scalar quantities such as $|E|$ and $|B|$ display the largest induced spatial variability of spectral parameters, while the complex scalar $\underline{S}=E+iB$ lacks a directly interpretable polarization amplitude and angle. By contrast, the spin-2 complex fields $\mathcal{P}_{ab}^{(E)}$ and $\mathcal{P}_{ab}^{(B)}$ display moderate, though non-negligible, induced deformations while retaining a clear geometric meaning: they satisfy the closure relation associated with Eq.~\ref{eq:decompo_eb}, their amplitudes and angles remain interpretable in map space, and the moment formalism promotes this closure relation to all spectral orders. We illustrate these points on a controlled toy model that gathers the main physical mechanisms capable of generating spectral complexity, and on a more realistic \texttt{PySM} synchrotron model \citep{Thorne_2017, Zonca_2021, Pan-ExperimentGalacticScienceGroup:2025vcd}.

{We also introduce simple CMB-oriented applications of the same framework. With three frequency channels, model-comparison diagnostics can test whether a sky, or a given sky region, is better described by a single power law in $P$ or by separate effective laws in $(P_E,P_B)$. Although the present paper only demonstrates this principle on noiseless mock data, such tests could be used on real data to identify where non-standard synchrotron physics, line-of-sight superposition, Faraday effects, or ageing make one representation more economical than another. This information is relevant for building more realistic foreground models, choosing how to extrapolate foreground templates, and selecting the spectral parametrization used in parametric component-separation pipelines. We further show, through idealized ILC and masking examples, that the choice of map-space field affects not only the level of residual $BB$ foreground power, but also where the residuals live on the sky and how they are distributed between the two parity families.}

{We emphasize that the spin-preserving $E/B$ decomposition does not create new information beyond that contained in the full Stokes maps $(Q,U)$. Rather, it reorganizes this information into two geometrically meaningful parity families. Consequently, $\mathcal{P}_{ab}^{(E)}$ and $\mathcal{P}_{ab}^{(B)}$ should not be viewed as standalone classifiers of the physical origin of a spectral deformation. Mechanisms such as ageing, Faraday rotation, and line-of-sight superposition are primarily identified through their characteristic complex spectral signatures in amplitude and angle. The role of the $P_E/P_B$ decomposition is complementary: it indicates how these signatures are distributed between gradient-like and curl-like structures on the sky. This distinction is useful for interpretation and for CMB-oriented analyses, but it must be used with care since the non-fully-local $E/B$ transform also induces apparent spectral deformations of its own.}

The paper is organized as follows. In Sec.~\ref{sec:EBseparating} we define the map-space $E/B$-separating complex operators and introduce the complex log--Taylor and moment parametrizations used throughout. In Sec.~\ref{sec:predictions_from_physics} we derive analytic predictions for the spectral signatures of several synchrotron mechanisms: spatially varying spectral index, line-of-sight mixing, intrinsic curvature, ageing, and Faraday effects. In Sec.~\ref{sec:comparaison_toy_model} we validate these predictions and compare the induced spectral behaviour across field representations using both a toy model and a realistic \texttt{PySM} synchrotron sky, including finite-channel reconstructions. {Finally, in Sec.~\ref{sec:applications_cmb} we present three simple CMB-oriented applications: distinguishing between spectral models defined in $P$ or in $(P_E,P_B)$, cleaning with $E/B$-separated ILCs, and constructing masks from $B$-family foreground tracers.}

\section{\textit{E}/\textit{B}-separating complex linear operators and spectral properties}
\label{sec:EBseparating}

In this section we introduce the formalism used to describe the spectral properties of polarized emission in map space, especially when decomposed into $E$- and $B$-separated fields (before later linking parameters to physics in Sec.~\ref{sec:predictions_from_physics} and testing them on a toy model and a \texttt{PySM} sky in Sec.~\ref{sec:comparaison_toy_model}). We first present in Sec.~\ref{sec:EBseparating_complex_op} the complex linear operators employed throughout this work. We then detail in Sec.~\ref{sec:spectral_parametrizations} the spectral parametrizations that can be applied to the various fields. In Sec.~\ref{sec:consequences_linearity} we examine the consequences of linearity for the \textit{E}/\textit{B}-projected spectral parameters. Finally, in Sec.~\ref{sec:conceptual_advantages}, we summarize the conceptual advantages and limitations of the different fields.

\subsection{\textit{E}/\textit{B}-separating complex linear operators}
\label{sec:EBseparating_complex_op}

For convenience, we represent the linear polarization at each sky direction by the complex spin-2 field
\begin{equation}
    \underline{P} \equiv Q + i\,U,
\end{equation}
which contains the same Stokes information as the 
matrix formulation of the tensor $\mathcal{P}_{ab}$, while being simpler to manipulate and interpret. The modulus of this complex field gives the polarization amplitude $P$, and half of its argument gives the polarization angle $\psi$. In this work, any complex quantity is denoted by an underline.

A variety of linear operators can be constructed to separate $\underline{P}$ into its $E$- and $B$-mode components.

First, we define the spin-preserving \textit{E}/\textit{B} projectors that implement the decomposition introduced in Eq.~\ref{eq:decompo_eb}.
\begin{equation}
    \underline{P}_E = Q_E+iU_E = \underline{L}_E[\underline{P}], 
    \qquad
    \underline{P}_B = Q_B+iU_B = \underline{L}_B[\underline{P}],
\end{equation}
where $\underline{P}_E$ and $\underline{P}_B$ are the complex spin-2 field version of $\mathcal{P}_{ab}^{(E)}$ and $\mathcal{P}_{ab}^{(B)}$, just as $\underline{P}$ is for $\mathcal{P}_{ab}$. Again, $\underline{P}_E$ and $\underline{P}_B$ differ from $\underline{P}$ only through the selection of $E$- or $B$-like patterns. 

Second, in the more standard construction, one interprets the harmonic coefficients $a_{\ell m}^E$ and $a_{\ell m}^B$ as those of scalar fields, respectively called standard $E$ and $B$ scalar fields. We combine these into the complex scalar field
\begin{equation}
    \label{eq:S_complex_field}
    \underline{S} \equiv E + i\,B = \underline{L}_S[\underline{P}],
\end{equation}
which contains the full information of $\underline{P}$ but is not itself a polarization field.

On the full sphere, $\underline{{L}}_E$ and $\underline{{L}}_B$ have nice properties: they are \emph{orthogonal projection operators} satisfying $\underline{{L}}_X^2 = \underline{{L}}_X$, $\underline{{L}}_X \underline{{L}}_Y = 0$ and $\underline{{L}}_X + \underline{{L}}_Y = \mathbb{I}$ ($X, Y\in E, B$), as well as being \emph{linear}, \textit{i.e.} $\underline{{L}}_X(\underline{P}_1+\lambda\underline{P}_2)= \underline{{L}}_X(\underline{P}_1)+\lambda\,\underline{{L}}_X(\underline{P}_2)$ for two complex polarization fields $\underline{P}_1$ and $\underline{P}_2$ and a uniform scalar $\lambda$. Their explicit expressions, both in harmonic space and in their real-space non-fully-local forms, along with the typical shape of their kernels, are given in Appendix~\ref{app:convolution_kernel}.

Eq.~\ref{eq:decompo_eb} rewritten in terms of complex spin-2 fields yields
\begin{equation}
    \label{eq:closure_complex}
    \underline{P} = \underline{P}_E + \underline{P}_B ,
\end{equation}
at all frequencies. Hereafter, we refer to this as the (complex) \emph{closure relation}. No such closure property exists for the scalar fields $(E,B)$. 

\subsection{Spectral parametrizations of a polarized signal}
\label{sec:spectral_parametrizations}

All the fields introduced so far can be observed at several different frequencies $\nu$, which can be used to provide a spectral description for each of them. To this end, we choose a pivot (or "reference") frequency $\nu_0$ and define
\begin{equation}
    \label{eq:log_freq_s}
    s \equiv \log\!\left(\frac{\nu}{\nu_0}\right).
\end{equation}
We now describe different options for modelling the spectral behaviour of a generic field, which may be real, $X_\nu$, or complex, $\underline{X}_\nu$ (the dependence on $\nu$ is hereafter denoted by a lower index).
In what follows, $\underline{X}_\nu$ may stand for any of the complex fields $\underline{P}$, $\underline{P}_E$, $\underline{P}_B$, or $\underline{S}$.

\subsubsection{Real log–Taylor expansion}

While the exact spectral dependence of polarized synchrotron emission is unknown and might be complex (as we will emphasize later on), one expects it not to deviate strongly from a power law, see, \textit{e.g.}, \cite{Pacholczyk:1970}.

A commonly used way to parametrize the frequency dependence of a quasi-power-law real-valued spectral energy distribution (SED) is to expand the logarithm of the field (see, e.g., \citealt{Rybicki:2004hfl} in the context of synchrotron emission), yielding a line-of-sight-dependent ``tilt'' and ``curvature'':
\begin{equation}
    X_\nu
    =
    A
    \exp\!\left(
        \beta\, s
        + \tfrac{1}{2}\gamma\, s^2
        + \cdots
    \right),
\end{equation}
where $A$ is an amplitude, $\beta$ is a spectral tilt, and $\gamma$ is a curvature parameter. If applied to the polarization amplitude $P$, \textit{e.g.}, \cite{Planck:2018yye, Galloway:2022zoo, Adak:2025hiz, Rizzieri:2025mwo}, this parametrization implicitly assumes a polarization angle that is constant with frequency, which need not hold in general.

One might instead apply this expansion to $Q$ and $U$ (see \citealt{delaHoz:2023yvz} for a comparison with $P$ in a practical case), or even to $E$ and $B$ individually. Although this is straightforward numerically, it is conceptually unsatisfactory for at least two reasons: (i) the expansion does not allow for a sign change in $X_\nu$, which frequently occurs for non–positive-definite fields such as $Q$, $U$, $E$, or $B$; and (ii) the two fields should vary coherently according to the underlying polarization angle~$\psi$, reflecting the geometrical relationship that unifies them.

\subsubsection{Complex log–Taylor expansion}

To overcome the limitations of a real log-Taylor expansion, it is natural to consider directly the spectral behaviour of the \emph{complex} field $\underline{X}_\nu$ itself. As already emphasized in Sec.~\ref{sec:EBseparating_complex_op}, this choice follows from the fact that the linear polarization field is fundamentally a spin-2 object, whose physically meaningful degrees of freedom are not the real components $(Q,U)$ taken separately, but their joint amplitude and orientation in the complex plane. Treating $Q$ and $U$ as independent real fields implicitly breaks this geometric structure and makes it difficult to describe coherent frequency evolution of the polarization angle, sign changes, or rotations in a robust and coordinate–independent manner.

By contrast, working with complex fields preserves the intrinsic geometry of the polarization maps and allows amplitude and angle variations to be encoded simultaneously and consistently. In particular, a complex spectral expansion provides a minimal and well–defined way to capture both changes in polarization intensity and frequency–dependent rotations of the polarization direction within a single set of parameters. Although such complex parametrizations have been implicitly used in specific contexts, their systematic application to map–space spectral modelling of polarized foregrounds is not standard, and constitutes a central ingredient of the present work.

Hence, we introduce the "complexified" version of the log–Taylor expansion, which reads
\begin{equation}
    \label{eq:complex_log_taylor}
    \underline{X}_\nu = \underline{A} \exp\!\left(     \underline{\beta}\, s     + \tfrac{1}{2}\underline{\gamma}\, s^2     + \cdots \right),
\end{equation}
where $\underline{A}$ is a complex amplitude, $\underline{\beta}$ a complex spectral tilt, and $\underline{\gamma}$ a complex curvature. Denoting by $\underline{\kappa}_n$ the $n$-th log–Taylor spectral parameter ($\underline{\kappa}_0 = \log\underline{A}$, $\underline{\kappa}_1 = \underline{\beta}$, $\underline{\kappa}_2 = \underline{\gamma}$), we can obtain it as
\begin{equation}
    \label{eq:logtaylor_prediction}
    \underline{\kappa}_n = \left.\frac{d^n}{ds^n}\log \underline{X}_\nu\right|_{\nu_0}.
\end{equation}

Under the restrictive (and somewhat unrealistic) assumption that the polarization angle is strictly constant with frequency, the real log–Taylor coefficients can be identified with the real parts of the complex ones. When the polarization angle $\psi$ varies with frequency, the imaginary parts of the log–Taylor parameters are directly related to the successive derivatives of $\psi$:
\begin{align}
\begin{cases}
\log A = \Re(\log \underline{A}) = \log|\underline{A}|, \\
\beta  = \Re(\underline{\beta}), \\
\gamma = \Re(\underline{\gamma}),
\end{cases}
\qquad\text{and}\qquad
\begin{cases}
    \psi                 = \tfrac{1}{2}\,\Im(\log \underline{A}), \\
    \dfrac{d\psi}{ds}    = \tfrac{1}{2}\,\Im(\underline{\beta}), \\
    \dfrac{d^2\psi}{ds^2} = \tfrac{1}{2}\,\Im(\underline{\gamma}).
\end{cases}
\end{align}

\subsubsection{Complex moment expansion}
\label{sec:complex_moment_expansion}

Working with complex log-Taylor parameters for low-frequency polarized emission is a novel viewpoint and, as we emphasize below, it can be closely linked to the underlying physics. A related complex formalism has previously been proposed through the spin-moment expansion \citep{Vacher:2022xdw,Vacher:2022mvr}, which generalizes the original moment expansion proposed by \cite{Chluba:2017rtj} to polarized signals using complex numbers. This formalism has also been applied to foreground modelling \citep{Vacher:2024adb} and to study the emission properties of thermal dust with the Planck-HFI data \citep{Guillet:2025tur}. More generally, moment expansion has been widely applied by the CMB community in various forms and spaces (pixels, needlets and power-spectra) in order to perform component separation for recent ground- and space-based missions (see, \textit{e.g.}, \citealt{Ichiki:2018glx,Azzoni:2020hpw,Mangilli:2019opl,Vacher:2021heq,Remazeilles:2020rqw,Azzoni:2022otq,Wolz:2023lzb,Carones:2024urc, Liu:2025hmw}).

The spin-moment expansion of a complex field $\underline{X}_\nu$ requires expanding around a reference power law:
\begin{equation}
    \label{eq:moment_expansion}
    \underline{X}_\nu=\left(\frac{\nu}{\nu_0}\right)^{\overline{\beta}}\left(    \underline{w}_0    + \underline{w}_1 s    + \frac{\underline{w}_2}{2}\, s^2    + \cdots\right),
\end{equation}
where $\overline{\beta}$ is a fixed, spatially uniform real reference spectral index chosen to represent a typical mean-sky value\footnote{{Another option would be to choose a pixel-dependent reference index $\overline{\beta}(\mathbf{n})$, which could itself be real or complex. In this work, we keep the reference uniform so that the spatial variability of the frequency scaling is entirely encoded by the moments, for reasons discussed at the end of this subsection.}}, and $\underline{w}_n$ is a complex moment of order $n$ (which is a map of complex numbers). These moments can be obtained from
\begin{align}
\label{eq:complex_moment_prediction}
    \underline w_n = \left.\frac{d^n}{ds^n}\left[\underline X_\nu\,\exp(-\overline \beta s)\right]\right|_{\nu_0}.
\end{align}

This expansion was first introduced in order to model the astrophysical signal resulting from the averaging of non-linear SEDs, and we will argue here that its range of application is much broader. We return to the corresponding physical predictions for the moments in Sec.~\ref{sec:predictions_from_physics}.

\subsubsection{Relations between complex log–Taylor parameters and moments}
\label{sec:relations_logtaylor_moments}

Expanding Eqs.~\ref{eq:logtaylor_prediction} and \ref{eq:moment_expansion} around $s=0$, equating the two representations, matching orders, and solving for the log–Taylor parameters yields 
\begin{align}
\label{eq:moments_2_taylor}
\begin{cases}
\underline{A} = \underline{w}_0, \\[6pt]
\displaystyle 
\underline{\beta}
= \overline{\beta}
  + \dfrac{\underline{w}_1}{\underline{w}_0}, \\[12pt]
\displaystyle
\underline{\gamma}
= 
      \dfrac{\underline{w}_2}{\underline{w}_0}
      - \left(\dfrac{\underline{w}_1}{\underline{w}_0}\right)^{\!2} 
  .
\end{cases}
\end{align}
Inverting these relations, we obtain for the moments
\begin{align}
\label{eq:w_versus_logtaylor}
\begin{cases}
\underline{w}_0 = \underline{A}, \\[6pt]
\displaystyle
\underline{w}_1
= \underline{A}\,(\underline{\beta} - \overline{\beta}), \\[10pt]
\displaystyle
\underline{w}_2
= \underline{A}\left[(\underline{\beta} - \overline{\beta})^2+\underline{\gamma}\right].
\end{cases}
\end{align}
Thus, the moment expansion and the log–Taylor expansion provide mathematically equivalent, but conceptually distinct, spectral descriptions of a complex field. Furthermore, a curvature in the log-Taylor expansion can be reabsorbed as a second order moment, an observation already made in \cite{Remazeilles:2020rqw}. 

For both parametrizations, the real and imaginary parts of the spectral parameters can be related to the frequency evolution of the amplitude and angle of the corresponding complex field. To first order in $s$, the log-amplitude and the angle of $\underline{X}_\nu$ can be expressed as
\begin{align}
\label{eq:first_order_evol_amplitude}
    &\log |\underline X_\nu| \simeq \log A+\Re\!\left(\underline{\beta}\right) s \simeq \log |\underline w_0|+\overline{\beta}s+\Re\!\left(\frac{\underline w_1}{\underline w_0}\right)s,\\
    \label{eq:first_order_evol_angle}
    &\psi_\nu \simeq \psi_{\nu_0}+\tfrac{1}{2}\Im\!\left(\underline{\beta}\right) s\simeq \psi_{\nu_0}+\tfrac{1}{2}\Im\!\left(\frac{\underline w_1}{\underline w_0}\right) s,
\end{align}
consistently with Eq.~22 of \cite{Vacher:2022mvr}. Equation~\ref{eq:first_order_evol_angle} shows that, at first order, the stability of the polarization angle across frequency is directly encoded in the imaginary part of $\underline{w}_1/\underline{w}_0$.

{It is useful to distinguish two related but distinct notions of spectral complexity. The first is \emph{functional spectral complexity}: the local, i.e. per-pixel, frequency scaling is not well described by a pure power law and requires higher-order parameters, such as the curvature $\underline{\gamma}$ in Eq.~\ref{eq:complex_log_taylor} or higher-order moments in Eq.~\ref{eq:moment_expansion}. The second is the \emph{spatial variability of spectral parameters}: a unique and simple functional form for the frequency scaling may describe the emission everywhere locally, while its effective parameters vary strongly across the sky. Both aspects are relevant for foreground modelling, because spatially varying spectral parameters increase the number of degrees of freedom required by a component-separation model, even when each individual line of sight is spectrally simple.}

{This distinction also clarifies the different roles of the log--Taylor and the moment expansion we use in this specific analysis. The complex log--Taylor parameters of Eq.~\ref{eq:complex_log_taylor} describe the local functional shape of the frequency scaling around $\nu_0$: for instance, $\underline{\gamma}=0$ for a local pure power law, independently of how the spectral index varies elsewhere on the sky. By contrast, because the moment expansion in Eq.~\ref{eq:moment_expansion} is written around a single full-sky reference index $\overline{\beta}$, its higher-order terms encode both genuine departures from a local power law and spatial departures of the local tilt from $\overline{\beta}$. This is explicit in Eq.~\ref{eq:moments_2_taylor}: even when $\underline{\gamma}=0$, a non-zero $(\underline{\beta}-\overline{\beta})$ generates higher-order moments. If one wished the moments to isolate only local functional complexity, one could instead choose a line-of-sight-dependent reference index $\overline{\beta}(\mathbf{n})$; however, this would break the full-sky linearity of Eq.~\ref{eq:moment_expansion}, which is essential for our forthcoming discussions. 

Moreover, such a local-reference moment expansion would be largely equivalent to the log--Taylor expansion itself: around the pivot, once the local reference tilt has absorbed the first-order slope, the normalized higher-order moments reduce to the corresponding local log--Taylor coefficients, $\underline{\Tilde{w}}_n/\underline{w}_0=\underline{\kappa}_n$ (the tilde denotes the use of a per-pixel reference $\overline{\beta}$). We therefore keep a uniform $\overline{\beta}$ throughout. With this choice, the moments provide the natural variables for predicting and projecting spectral behaviour under the $E/B$ operators, while the log--Taylor parameters provide the most direct local description of the frequency scaling shape.}

\subsection{Consequences of linearity for the \textit{E}/\textit{B}-projected spectral parameters}
\label{sec:consequences_linearity}

Because $\underline{L}_E$, $\underline{L}_B$, and $\underline{L}_S$ are linear operators acting on complex fields,
they commute with the formation of linear combinations in the moment expansion.
Therefore, for all $n\ge 0$,
\begin{equation}
\label{eq:project_moments}
    \underline{w}_{n,E} = \underline{L}_E[\underline{w}_{n}],
    \qquad
    \underline{w}_{n,B} = \underline{L}_B[\underline{w}_{n}],
    \qquad
    \underline{w}_{n,S} = \underline{L}_S[\underline{w}_{n}].
\end{equation}
Thanks to these simple relations, all the formalism introduced so far can be directly translated to the $\underline{P}_E$ and $\underline{P}_B$ fields. For instance, Eq.~\ref{eq:first_order_evol_angle} simply applies to $\underline{P}_E$ and $\underline{P}_B$, and the imaginary part of these relationships tells how the complex spin-2 $E$/$B$ fields turn in the complex plane with frequency. Another important consequence is that the closure relation at each frequency also applies to $E$ and $B$ decomposed complex moments:
\begin{equation}
\label{eq:closure_moments}
    \underline{w}_{n} = \underline{w}_{n,E} + \underline{w}_{n,B}.
\end{equation}

This leads to more detailed relations for the Taylor spectral parameters, which can be derived straightforwardly using Eq.~\ref{eq:moments_2_taylor}. The log-Taylor parameters for $\underline{P}$ can be expressed as a function of those of $\underline{P}_E$ and $\underline{P}_B$:
\begin{align}
\begin{cases}
\label{eq:relations_P_from_PEPB}
    \displaystyle
    P = \big| \underline{A}_E + \underline{A}_B \big|
      = \sqrt{P_E^2 + P_B^2 + 2 P_E P_B \cos\!\big[2(\psi_E-\psi_B)\big]},\\[6pt]
    \displaystyle
    \psi \;=\; \tfrac{1}{2}\arg\!\big(\underline{A}_E+\underline{A}_B\big)
    =\tfrac{1}{2}\arctan\!\Bigg(
        \dfrac{P_E\sin 2\psi_E + P_B\sin 2\psi_B}
             {P_E\cos 2\psi_E + P_B\cos 2\psi_B}
    \Bigg),\\[6pt]
    \displaystyle
    \beta \;=\; \Re\!\left[
    \frac{\underline{A}_E\,\underline{\beta}_E + \underline{A}_B\,\underline{\beta}_B}
         {\underline{A}_E + \underline{A}_B}
    \right],\\[6pt]
    \displaystyle
    \frac{d\psi}{ds} \;=\; \tfrac{1}{2}\,\Im\!\left[
    \frac{\underline{A}_E\,\underline{\beta}_E + \underline{A}_B\,\underline{\beta}_B}
         {\underline{A}_E + \underline{A}_B}
    \right],\\[6pt]
    \displaystyle
    \gamma \;=\; \Re\!\Bigg[
    \frac{\underline{A}_E\!\big(\underline{\beta}_E^2 + \underline{\gamma}_E\big)
          + \underline{A}_B\!\big(\underline{\beta}_B^2 + \underline{\gamma}_B\big)}
         {\underline{A}_E + \underline{A}_B}
    \;-\; \underline{\beta}^2\Bigg],\\[6pt]
    \displaystyle
    \frac{d^2\psi}{ds^2} \;=\; \tfrac{1}{2}\Im\!\Bigg[
    \frac{\underline{A}_E\!\big(\underline{\beta}_E^2 + \underline{\gamma}_E\big)
          + \underline{A}_B\!\big(\underline{\beta}_B^2 + \underline{\gamma}_B\big)}
         {\underline{A}_E + \underline{A}_B}
    \;-\; \underline{\beta}^2
    \Bigg].
\end{cases}
\end{align}
Developing moduli and arguments in these relations quickly leads to cumbersome expressions. Above, we have only displayed the explicit results for the polarization amplitude and angle, which remain tractable and we consistently recover, for the polarization angle, Eq.~3.2 of \cite{Liu:2018oqp}. An important consequence of the amplitude relation is that $(P_E-P_B)^2\leq P^2\leq (P_E+P_B)^2$, restricting the $(P_E/P, P_B/P)$ plane to
\begin{equation}
\label{eq:plane_restriction}
    \begin{cases}
      P_B/P\leq P_E/P+1,\\
      P_B/P\geq P_E/P-1,\\
      P_B/P\geq -P_E/P+1.
    \end{cases}  
\end{equation}

Once again, no analogous identity exists for $(E,B)$ or for $\underline{S}$. Even though spectral expansions can be inferred in $(E,B,\underline{S})$, only the spin-preserving maps retain spectral parameters that can be interpreted directly in terms of the frequency dependence of polarization amplitudes and angles.

\subsection{Conceptual advantages of the \texorpdfstring{$\bm{\underline{P}_{\NoCaseChange{\mathrm{E}}}}$}{PE} and \texorpdfstring{$\bm{\underline{P}_{\NoCaseChange{\mathrm{B}}}}$}{PB} decomposition}
\label{sec:conceptual_advantages}

The spin-preserving fields $\underline{P}_E$ and $\underline{P}_B$ enjoy several conceptual advantages over the scalar fields $(E,B)$ or the complex scalar $\underline{S}=E+iB$. By construction they retain the geometric nature of the underlying polarization field: they remain spin-2 quantities with amplitudes and polarization angles that may (with due caution) be interpreted locally in map space. Although the meaning of these angles requires care (they are the angles after \textit{E}/\textit{B} filtering, not necessarily those of the underlying emission components) their interpretation remains considerably more direct than that of scalar fields obtained through the spin-2 $\rightarrow$ spin-0 conversion.

{The practical distinction between the spin-preserving $B$-family field $(Q_B,U_B)$ and the scalar $B$ map is therefore not that one contains more information than the other. On the full sky they are related by linear transformations. The difference is one of representation. The field $(Q_B,U_B)$ remains a spin-2 polarization field: it has an amplitude $P_B$, an angle $\psi_B$, and can be modelled with complex spectral parameters that retain the same interpretation as those of $P$. By contrast, the scalar $B$ map is sign-changing and its absolute value $|B|$ is a non-linear quantity whose spectral parameters are harder to interpret (and ultimately as shown in Sec.~\ref{sec:applications_cmb}, this distinction matters in concrete applications for CMB science, \textit{e.g.}, masking strategy).}

One might worry that $\underline{P}_E$ and $\underline{P}_B$ are somewhat ``coordinate dependent'' whereas $E$ and $B$ are not.  However, this is analogous to the familiar distinction between the spin–2 field $\underline{P}=Q+iU$ and the scalar temperature $T$: $\underline{P}_E$ and $\underline{P}_B$ are spin–2 fields that transform with a phase under local rotations of the polarization basis, while $E$ and $B$ are scalar, basis–invariant quantities built from them. While the coordinates of these spin-2 objects in a specific basis are obviously not basis independent, $\underline{P}_E$ and $\underline{P}_B$ taken as proper spin-2 objects are just as coordinate independent as $\underline{P}$, $E$, $B$ or $T$ are. 
 
A further advantage is the closure relation, which holds identically at all frequencies and at every order in any spectral expansion. This allows a direct, term-by-term comparison between the full field and its gradient- and curl-like contributions. In particular, differences between the moments of $\underline{P}$, $\underline{P}_E$, and $\underline{P}_B$ admit an immediate physical interpretation in terms of the relative spectral behaviour of the corresponding spatial patterns.

A familiar diagnostic in map space is the ratio of amplitudes $|\underline{P}_E|/|\underline{P}_B|$, which quantifies, along each line of sight, the balance between the \textit{E}-like and \textit{B}-like contributions to the total polarization pattern. The moment expansion naturally generalizes this idea to any spectral order. Ratios such as $|\underline{w}_{n,E}|/|\underline{w}_{n,B}|$ provide direction- and order-dependent measures of how the \textit{E}/\textit{B} balance varies across frequency. For example, a relative increase of $|\underline{w}_{1,B}|$ compared to $|\underline{w}_{1,E}|$ immediately indicates that curl-like structures respond more strongly to spectral-index variations than gradient-like ones. Higher-order ratios $|\underline{w}_{n,E}|/|\underline{w}_{n,B}|$ (for $n\ge2$) in turn probe the curvature of this behaviour, distinguishing between frequency-dependent morphological changes driven by coherent structures, turbulence, or Faraday effects. These quantities have simple interpretations only when the fields remain spin-2, which is no longer the case for $(E,B)$ or for the scalar $\underline{S}$.

By contrast, although the scalar field $\underline{S}=E+iB$ contains the same information as $\underline{P}$, the interpretation of its spectral parameters is less direct: its amplitude and phase are not related to observable polarization angles or magnitudes, but to scalar quantities whose connection to the underlying geometry of the magnetic fields is non-fully-local. For this reason, moments such as $\underline{w}_{n,S}$ lack the intuitive map-space meaning of their spin-2 counterparts.

For all these reasons, the decomposition into $\underline{P}_E$ and $\underline{P}_B$ provides, from a conceptual standpoint, a particularly natural framework for studying the frequency dependence of the polarization field. We will illustrate this in more detail with concrete examples in Sec.~\ref{sec:comparaison_toy_model}.

\section{Physical mechanisms generating synchrotron spectral moments}
\label{sec:predictions_from_physics}

In this section we summarize the main physical mechanisms that generate non-zero spectral moments $\underline{w}_n$ of the complex polarization field $\underline{P}$ and of its spin-preserving projections $\underline{P}_E$ and $\underline{P}_B$. Specifically, after deriving useful equations for the multiplicative corrections to an underlying spectrum in Sec.~\ref{sec:general_considerations}, we discuss five families of mechanisms that produce non-zero higher-order spectral parameters:
\begin{itemize}
    \item[(i)] line-of-sight superposition of components with distinct power-law spectra (see Sec.~\ref{sec:superposition});
    \item[(ii)] spatially varying spectral indices $\beta(\mathbf{n})$ for a single emitting component per line of sight (see Sec.~\ref{sec:spatial_beta});
    \item[(iii)] line-of-sight superposition of components with distinct curved spectra (see Sec.~\ref{sec:curvature});
    \item[(iv)] synchrotron ageing, which produces a smooth spectral break (see Sec.~\ref{sec:ageing});
    \item[(v)] Faraday rotation and depolarization, which introduce both amplitude and phase effects (see Sec.~\ref{sec:faraday}).
\end{itemize}
We focus on the essential formulae and their physical interpretation, and refer the reader to the cited literature for detailed derivations. We conclude in Sec.~\ref{sec:coexistence_EB} with a brief discussion of the coexistence of \textit{E}- and \textit{B}-dominated structures and the impact of the non-fully local nature of the \textit{E}/\textit{B} operators.

\subsection{General considerations}
\label{sec:general_considerations}
To organize the predictions for spectral parameters, we express departures from a reference power law as multiplicative corrections. For a generic complex field $\underline{X}_\nu$ we write
\begin{align}
    \label{eq:correction_factor}
    \underline{X}_\nu
    &=
    \underline{X}_{0}
    \left(\frac{\nu}{\nu_0}\right)^{\overline{\beta}}
    \,\underline{C}(\nu),\\
    &=\sum_i \underline{X}_{i,0}
    \left(\frac{\nu}{\nu_0}\right)^{\beta_i}
    \prod_k \underline{C}_k^{(i)}(\nu).
\end{align}
where $\underline{C}$ is a map-space frequency-dependent correction field such that $\underline{C}(\nu_0)=1$ by construction. More generally, the signal can be described in terms of several emitting components $i$ present along the line of sight, each affected by several multiplicative physical effects $k$, denoted $\underline{C}_k^{(i)}(\nu)$. In terms of $s$ (introduced in Eq.~\ref{eq:log_freq_s}), this becomes $\underline{X}_\nu=\underline{X}_0\,\exp(\overline{\beta}s)\,\underline{C}(s)$.

The complex log–Taylor parameters of $\underline{X}_\nu$ can then be written as
\begin{equation}
    \label{eq:logtaylor_prediction_in_practice}
    \underline{\kappa}_n
    =
    \left.\frac{d^n}{ds^n}\left[\log \underline{X}_\nu\right]\right|_{s=0}
    =
    \frac{d^n}{ds^n}\left[\log \underline X_{\nu_0} + \overline \beta s+\log \underline{C}(s)\right|_{s=0}.
\end{equation}
In the specific case of a single emitting component present along the line of sight affected by several multiplicative physical effects $k$, the first three log-Taylor parameters read
\begin{align}
\underline{A} &= \underline{X}_0,\\
\underline{\beta} &= \overline{\beta} + \sum_k \left.\frac{d}{ds}\log \underline{C}_k(s)\right|_{0},\\
\underline{\kappa}_n &= \sum_k \underline{\kappa}_{k,n}
= \sum_k \left.\frac{d^n}{ds^n}\log \underline{C}_k(s)\right|_{0},\qquad n\ge 2.
\end{align}

Moreover, in the general case, the formula for the complex moments follows from Eq.~\ref{eq:complex_moment_prediction} as
\begin{equation}
    \label{eq:moment_prediction_in_practice}
    \underline{w}_n
    =
    \left.\frac{d^n}{ds^n}\left[\underline{X}_\nu\,e^{-\overline{\beta}s}\right]\right|_{s=0}
    =
    \underline{w}_0
    \left.\frac{d^n}{ds^n}\underline{C}(s)\right|_{s=0}.
\end{equation}
In the case of several emitting components $i$ present along the line of sight, each affected by several multiplicative physical effects $k$, the first three complex moments read
\begin{align}
\label{eq:w0_combination}
\underline{w}_0 &= \sum_i \underline{X}_{i,0},
\\
\label{eq:w1_combination}
\underline{w}_1 &= \sum_i \underline{X}_{i,0}\,\left.\frac{d\log C^{(i)}}{ds}\right|_{0}
= \sum_i \underline{X}_{i,0}\left[(\beta_i-\overline{\beta})+\left.\sum_k\frac{d\log \underline{C}_k^{(i)}}{ds}\right|_{0}\right],
\\
\label{eq:w2_combination}
\underline{w}_2 &= \sum_i \underline{X}_{i,0}\left[\left.\frac{d\log C^{(i)}}{ds}\right|_{0}^{\!2} + \left.\frac{d^2\log C^{(i)}}{ds^2}\right|_{0}\right] \nonumber\\
&=\sum_i \underline{X}_{i,0}\left\{\left[(\beta_i-\overline{\beta})+\left.\sum_k\frac{d\log \underline{C}_k^{(i)}}{ds}\right|_{0}\right]^2+\left.\sum_k\frac{d^2\log \underline{C}_k^{(i)}}{ds^2}\right|_{0}\right\}.
\end{align}
with $\underline{w}_0=\underline{X}_0$ since $\underline{C}_0=1$. 

As a trivial illustration, consider $N$ emitting components on the line of sight, all sharing the same spectral index $\beta_i=\overline{\beta}$ and no additional effects. Then
\begin{align}
\underline{P}_\nu &=
\sum_{i=1}^N\underline{P}_{i,0}\left(\frac{\nu}{\nu_0}\right)^{\overline \beta}
=\underline{P}_0\left(\frac{\nu}{\nu_0}\right)^{\overline \beta}\underline{C}_{\overline \beta},\\
\underline{C}_{\overline \beta} &= \sum_{i=1}^N\frac{\underline{P}_{i,0}}{\underline{P}_0},
\end{align}
which is independent of frequency. In this case, the equations above trivially imply that all higher-order spectral moments vanish, $\underline{w}_n=0$ for $n\ge1$, and the log–Taylor parameters have $\underline{\kappa}_n=0$ for $n\ge2$.

\begin{figure}
    \centering
    \includegraphics[width=0.98\linewidth]{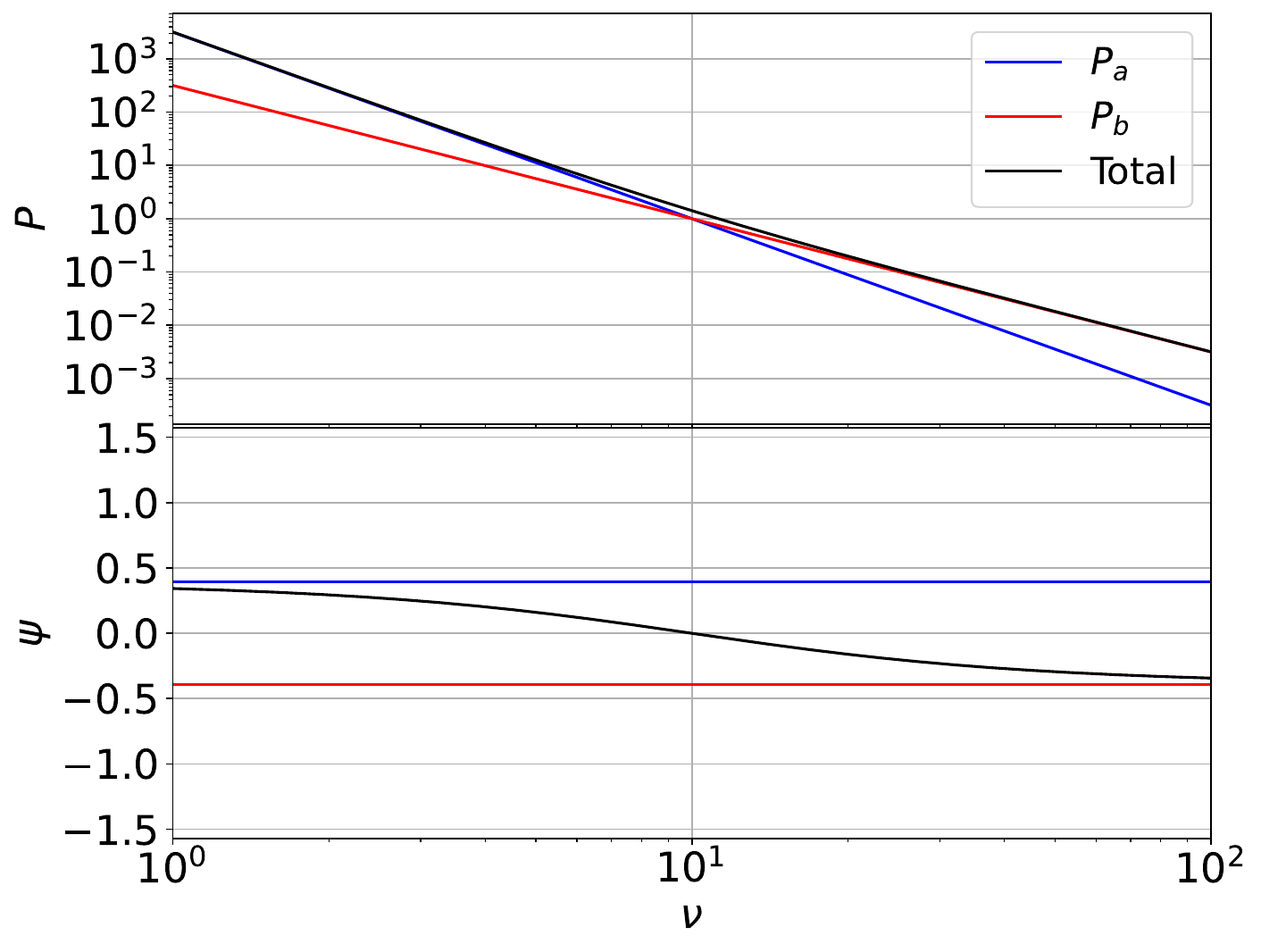}
    \caption{Example of the power law superposition effect (Sec.~\ref{sec:superposition}): Eq.~\ref{eq:superposition_equation} with two components, $\underline{P}_\nu = \underline{P}_{a, \nu} + \underline{P}_{b, \nu}$ with $\underline{P}_{a, 0} = e^{2i\pi/8}$, $\beta_a = -3.5$, $\underline{P}_{b, 0} = e^{-2i\pi/8}$ and $\beta_b = -2.5$. \textit{Upper panel}: SED, \textit{lower panel}: polarization angle, both as functions of frequency. The blue curves describe the baseline polarization spectrum, which is modified into the total black curves by the superimposition effect (whose correction $\underline{C}_\textrm{LoS}(\nu)$ is shown in dashed-red).
    }
    \label{fig:superposition}
\end{figure}

\subsection{Line-of-sight superposition of distinct power laws}
\label{sec:superposition}

We now consider multiple components along the line-of-sight (LoS hereafter) with different spectral indices:
\begin{align}
\label{eq:superposition_equation}
&\underline{P}_\nu
=
\sum_{i=1}^N\underline{P}_{i,0}
\left(\frac{\nu}{\nu_0}\right)^{\beta_i}
=\underline{P}_0\left(\frac{\nu}{\nu_0}\right)^{\overline \beta}\underline{C}_\mathrm{LoS}(\nu), \mathrm{\ \ \ where \ \ \ }
\nonumber\\
&\underline{C}_\mathrm{LoS}(\nu)
=
\sum_{i=1}^N\frac{\underline{P}_{i,0}}{\underline{P}_0}\left(\frac{\nu}{\nu_0}\right)^{\beta_i-\overline \beta}.
\end{align}
Substituting $\underline{C}_\mathrm{LoS}$ into Eq.~\ref{eq:moment_prediction_in_practice} yields (see also \citealt{Vacher:2022mvr})
\begin{equation}
\label{eq:prediction_moments_superposition}
\underline{w}_n=\sum_i \underline{P}_{i,0}\,(\beta_i-\overline\beta)^{n}.
\end{equation}
Because the sum of power laws is not itself a power law in general, higher-order log–Taylor terms are generated. Using either Eq.~\ref{eq:logtaylor_prediction_in_practice} or Eq.~\ref{eq:moments_2_taylor}, 
and defining complex weighted averages 
\begin{equation}
\label{eq:complex_w_average}
\underline{\langle} X_i\underline{\rangle} \equiv \frac{\sum_i \underline{P}_{i,0} X_i}{\sum_j \underline{P}_{j,0}},
\end{equation}
one finds 
\begin{align}
\underline{\beta}
&=
\underline{\langle} \beta_i\underline{\rangle},\\
\underline{\gamma}
&=
  \underline{\langle} \beta_i^2\underline{\rangle}-\underline{\langle} \beta_i\underline{\rangle}^2,
\end{align}
and similarly for higher orders. Hence, superposition of power laws induces curvature that is equal to the complex-weighted variance of the spectral indices of the complex power laws along the LoS. In general, for arbitrary amplitudes and angles $\underline{P}_{i,0}$, these coefficients are complex. The imaginary parts encode a frequency-dependent polarization angle. Using Eq.~\ref{eq:first_order_evol_angle}, the first-order angle evolution around $s=0$ is
\begin{equation}
    \label{eq:freq_dependence_of_angle_for_superposition}
    \psi_\nu \simeq \psi_{\nu_0}
    +\tfrac{1}{2}\Im\left(\frac{\sum_i \underline{P}_{i,0}\,(\beta_i - \overline{\beta})}{\sum_i \underline{P}_{i,0}}\right) s.
\end{equation}
This frequency-dependent angle induces intrinsic $E\leftrightarrow B$ mixing. At zeroth order, the $E/B$ ratio is that of the total field
$\underline{P}_0=\sum_i\underline{P}_{i,0}$. At higher order however,
\begin{equation}
\frac{|\underline{w}_{n,E}|}{|\underline{w}_{n,B}|} = \frac{\left|\sum_i \underline{P}_{i,E,0}\,(\beta_i-\overline\beta)^{n}\right|}{\left|\sum_i \underline{P}_{i,B,0}\,(\beta_i-\overline\beta)^{n}\right|}
\end{equation}
characterizes the frequency-dependent redistribution between \textit{E}-like and \textit{B}-like structures. There is no simpler generic expression unless we know the whole map. A non-zero imaginary part of $\underline{w}_1$ drives a change in the \textit{E}/\textit{B} balance with frequency, so that
$|\underline{w}_{1,E}|/|\underline{w}_{1,B}|$ can differ significantly from $|\underline{P}_{0,E}|/|\underline{P}_{0,B}|$.

One can also assess the relative stability of the polarization angle for different fields, \textit{e.g.}, by comparing $\Im(\underline{w}_{1,E}/\underline{w}_{0,E})$ with $\Im(\underline{w}_1/\underline{w}_0)$ as a measure of how strongly $P_E$ and $P$ respond to line-of-sight mixing.

{We show in Fig.~\ref{fig:superposition} the effect of superposing two components along the line of sight with different $\beta$ and $\psi$. For two components with two complex power-laws $P_a$ and $P_b$, the modulus contains an interference term,
\begin{equation}
|\underline{P}_a+\underline{P}_b|^2
=
P_{a,0}^2e^{2\beta_as}+P_{b,0}^2e^{2\beta_bs}+2P_aP_be^{(\beta_a+\beta_b)s}\cos(2\Delta\psi),
\end{equation}
so the curvature of the amplitude is not determined only by the scalar sum of the two power laws. In the illustrative case shown here, $\psi_a=\pi/8$ and $\psi_b=-\pi/8$, hence $\Delta\psi=\pi/4$ and the resulting SED (amplitude) is positively curved; for larger angular separations, of order $\Delta\psi\simeq \pi/2$, the destructive interference term can make the amplitude of the sum fall below the scalar sum, and individual frequency branches can display negative effective curvature. In any case, the SED (\textit{upper panel}) is affected by higher-order complexity around the pivot scale, and the polarization angle evolves from $\psi_a$ at low frequency to $\psi_b$ at high frequency (\textit{lower panel}) (as for the next examples of this section, the absolute values and units of polarization do not matter for the discussion).}

\subsection{Spatially varying \texorpdfstring{$\bm{\beta}$}{beta} (single component per LoS)}
\label{sec:spatial_beta}

Spatial variations of the synchrotron spectral index $\beta(\mathbf{n})$ induce a direction-dependent deviation from the reference index $\overline\beta$. For a single component per pixel,
\begin{align}
&\underline{P}_\nu
=
\underline{P}_0\left(\frac{\nu}{\nu_0}\right)^{\beta}
=
\underline{P}_0\left(\frac{\nu}{\nu_0}\right)^{\overline \beta}\,C_{\beta}(\nu),\mathrm{\ \ \ where\ \ \ }
C_{\beta}(\nu)
=
\left(\frac{\nu}{\nu_0}\right)^{\beta-\overline{\beta}}.
\end{align}
Using Eq.~\ref{eq:moment_prediction_in_practice}, one obtains the hierarchy (see, \textit{e.g.}, \citealt{Vacher:2022mvr})
\begin{equation}
\underline{w}_n=\underline{P}_0\,(\beta-\overline\beta)^n,
\end{equation}
which is the $N=1$ case of Eq.~\ref{eq:prediction_moments_superposition} above. Because the spectrum in each pixel remains a pure power law, the log–Taylor curvature $\underline{\gamma}$ and all higher-order log–Taylor parameters vanish locally: $\underline{\kappa}_n=0$ for $n\ge 2$, and $\underline{w}_n/\underline{w}_0$ is purely real. At the level of $\underline{P}$ itself there is therefore no intrinsic curvature and no frequency evolution of the local polarization angle.

For the \textit{E}/\textit{B} decomposition, however, the situation is more subtle because $\underline L_E$ and $\underline L_B$ are non-fully-local operators on the sphere.  On a discrete sky, let pixel indices $p$ and $q$ correspond to line-of-sight directions $\mathbf{n}$ and $\mathbf{n}'$, respectively.  In Appendix \ref{app:convolution_kernel} we give the map-space expressions of the convolution kernels that implement $\underline{L}_E$ and $\underline{L}_B$.  Using Eqs.~\ref{eq:PE_kernel_geom} and \ref{eq:PB_kernel_geom} and substituting the spatially varying power law $\underline{P}_{\nu,q}=\underline{P}_{0,q}(\nu/\nu_0)^{\beta(q)}$ yields
\begin{align}
\underline{P}_{E,\nu,p}
&=
\sum_q
\Big[
    \underline{\mathcal{K}}^{(+)}_{E,pq}\,\underline{P}_{0,q}
  + \underline{\mathcal{K}}^{(-)}_{E,pq}\,\underline{P}_{0,q}^*
\Big]
\left(\frac{\nu}{\nu_0}\right)^{\beta(q)},
\label{eq:PE_beta_kernel}
\\
\underline{P}_{B,\nu,p}
&=
\sum_q
\Big[
    \underline{\mathcal{K}}^{(+)}_{B,pq}\,\underline{P}_{0,q}
  + \underline{\mathcal{K}}^{(-)}_{B,pq}\,\underline{P}_{0,q}^*
\Big]
\left(\frac{\nu}{\nu_0}\right)^{\beta(q)},
\label{eq:PEPB_beta_kernel}
\end{align}
where the kernels $\underline{\mathcal{K}}^{(\pm)}_{E/B,pq}$ are frequency-independent and depend only on geometry (Euler angles between $\mathbf{n}$ and $\mathbf{n}'$) and on the radial functions $\mathcal{K}_{\mathcal{I}},\mathcal{K}_{\mathcal{D}}$ from \citet{Rotti:2018pzi} (see Appendix \ref{app:convolution_kernel}). Eqs.~\ref{eq:PE_beta_kernel} and \ref{eq:PEPB_beta_kernel} show that, for a fixed pixel $p$ in $\underline{P}_E$ or $\underline{P}_B$, the spectrum is a weighted superposition of power laws with different indices $\beta(q)${, in complete analogy with line-of-sight mixing (see Sec.~\ref{sec:superposition}). The two mechanisms are physically connected, but it is useful to keep them separate because one is intrinsic to the line of sight whereas the other is induced by angular mixing on the sphere. In practice, the distinction can become blurred when structures with different spectral properties are separated by angular scales smaller than the instrumental beam: after beam convolution, they are effectively observed as a single line of sight containing several superimposed spectral components.}

Formally one still has the projected moments Eq.~\ref{eq:project_moments} but since $\underline{w}_n(q)\propto \underline{P}_{0,q}\big(\beta(q)-\overline\beta\big)^n$ contains a spatially varying scalar factor, the \textit{E}/\textit{B} projectors mix contributions from regions with different $\beta(q)$.  As a consequence, even though $\underline{P}_\nu$ is locally a pure power law, the projected fields $\underline{P}_{E,\nu}$ and $\underline{P}_{B,\nu}$ can exhibit non-trivial higher-order moments (effective spectral curvature in $E$ and $B$), frequency-dependent changes of the local polarization angle in the $E$ and $B$ maps, and hence a frequency-dependent $E\leftrightarrow B$ balance. Only in the limiting case where $\beta(q)$ is nearly constant on the angular scales to which $\underline L_E$ and $\underline L_B$ are most sensitive does one recover an approximately constant ratio $|\underline{w}_{n,E}|/|\underline{w}_{n,B}|$.  

More generally, any effect that mixes power laws with different~$\beta$, and in particular any effect that mixes pixels on the sphere, will tend to induce additional spectral complexity. This is the case for the \textit{E}/\textit{B} transform, but it would also arise for instrumental effects that mix neighbouring pixels, such as a finite-width beam (and especially when the beam size is larger than the typical scale of the $\beta$ variations).

Since the illustration of the \textit{E}/\textit{B} transform loss-of-locality effect requires the knowledge of the whole field, we defer it to Sec.~\ref{sec:comparaison_toy_model}.

\subsection{Intrinsic curvature and its superposition on the LoS}
\label{sec:curvature}

\begin{figure}
    \centering
    \includegraphics[width=0.98\linewidth]{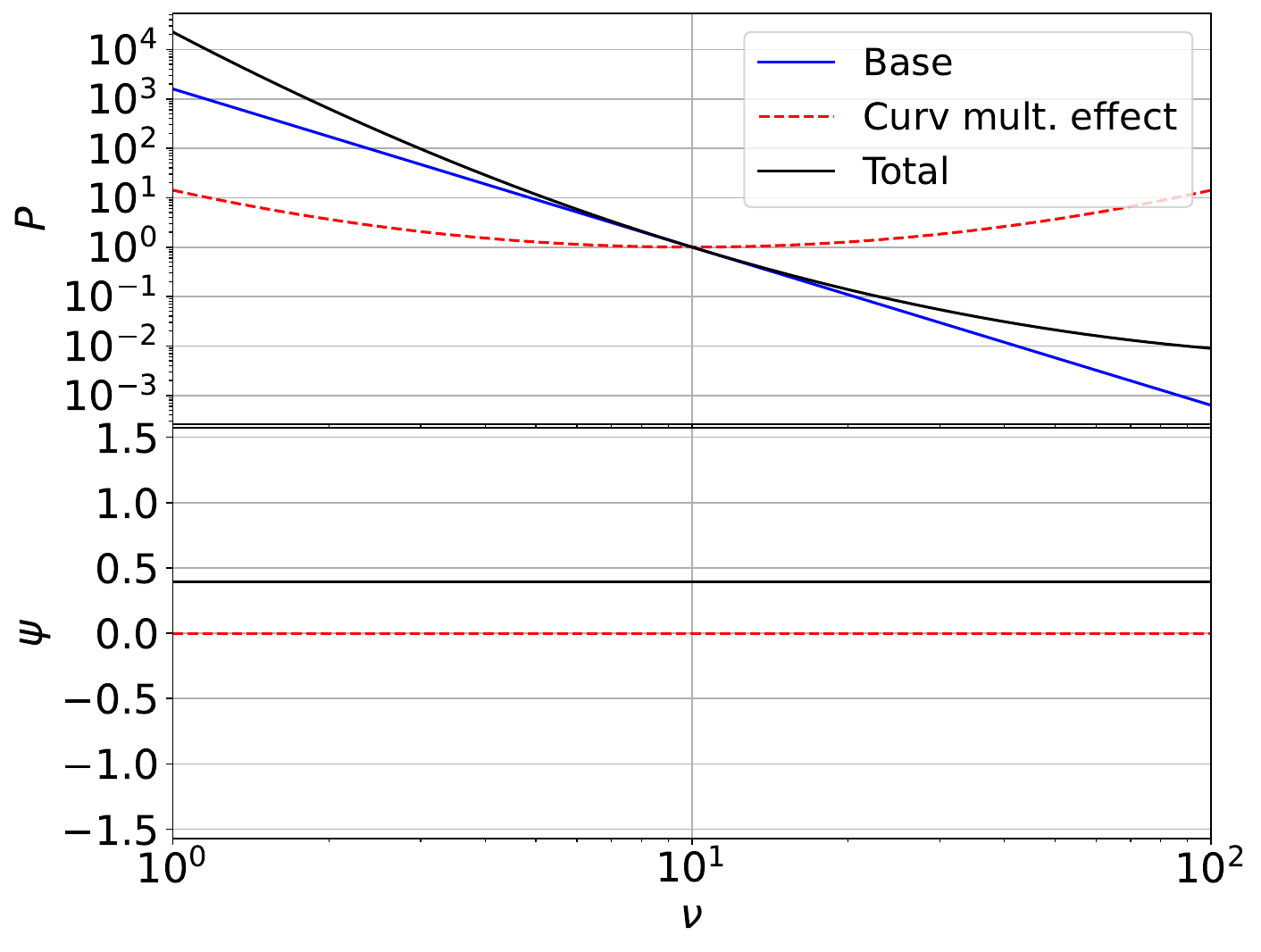}
    \caption{Example of a single-component intrinsic positive curvature effect (Sec.~\ref{sec:curvature}): Eq.~\ref{eq:curvature_equation} with $\underline{P}_{0}=e^{2i\pi/8}$, $\overline{\beta}=-3.2$ and $\gamma = 1$. The panels and curves are otherwise similar to the ones in Fig.~\ref{fig:superposition}.}
    \label{fig:curvature}
\end{figure}

We now generalize the LoS superposition case presented in Sec.~\ref{sec:superposition} to components with intrinsic curvature. Consider several components $i$ along the line of sight with intrinsic log–Taylor curvature $\gamma_i$:
\begin{align}
\label{eq:curvature_equation}
&\underline{P}_\nu
= \sum_{i=1}^N \underline{P}_{i,0}\,
  \exp\!\big[\beta_i s + \tfrac{1}{2}\gamma_i s^2\big]
= \underline{P}_0\left(\frac{\nu}{\nu_0}\right)^{\overline \beta}\underline{C}_\mathrm{\gamma LoS}(\nu),\mathrm{\ \ \ where \ \ \ }
\nonumber\\
&\underline{C}_\mathrm{\gamma LoS}(\nu)
=\sum_{i=1}^N \frac{\underline{P}_{i,0}}{\underline{P}_{0}}\,
\exp\!\big[(\beta_i-\overline\beta)s + \tfrac{1}{2}\gamma_i s^2\big].
\end{align}
Using Eq.~\ref{eq:moment_prediction_in_practice} one finds
\begin{align}
\underline{w}_0
&= \sum_i \underline{P}_{i,0},\\
\underline{w}_1
&= \sum_i \underline{P}_{i,0}\,(\beta_i-\overline\beta),
\\[4pt]
\underline{w}_2
&= \sum_i \underline{P}_{i,0}\big[(\beta_i-\overline\beta)^2 + \gamma_i\big].
\end{align}
and using again the complex weighted average defined in Eq.~\ref{eq:complex_w_average},
\begin{align}
\underline{\beta}
&= \left.\frac{d\log\underline{P}_\nu}{ds}\right|_{0}
 = \underline{\langle} \beta\underline{\rangle},\\
\underline{\gamma}
&= 
   \left.\frac{d^2\log\underline{P}_\nu}{ds^2}\right|_{0}
 = \underline{\langle} \gamma\underline{\rangle}
 + \Big(\underline{\langle} \beta^2\underline{\rangle} - \underline{\langle} \beta\underline{\rangle}^2\Big),
\end{align}
\textit{i.e.}\ the effective curvature is the sum of the complex-weighted average of the intrinsic curvatures plus the complex-weighted variance of the spectral indices.

We illustrate in Fig.~\ref{fig:curvature} the effect of curvature for the given (large) positive value of $\gamma=1$: the SED (amplitude) becomes curved (\textit{upper panel}) without affecting the fiducial input polarization angle (\textit{lower panel}).

\subsection{Synchrotron ageing}
\label{sec:ageing}

The classical relation between a power-law electron energy distribution and the optically thin synchrotron spectrum is given in \citet{Pacholczyk:1970}; for $N(E)\;\propto\; E^{-p}$, the synchrotron flux density follows
\begin{equation}
S_\nu \;\propto\; \nu^{\alpha_{\rm inj}},\quad\textrm{where}\quad
\alpha_{\rm inj} = -\frac{p-1}{2}.
\end{equation}
The global picture is that radiative losses progressively deplete high-energy electrons and produce a break in the spectrum at a characteristic frequency $\nu_b$. More precise and complicated treatments of spectral ageing are given in \citet{Murgia_Ageing}. For simplicity, we assume a simple smooth broken-power-law correction factor (relative to a high-frequency reference), given by
\begin{align}
\underline{C}_\mathrm{age}^{\rm(orig)}(\nu)
=
\left(\frac{\nu}{\nu_b}\right)^{\Delta\alpha}
 \left[1+\left(\frac{\nu}{\nu_b}\right)^{1/\sigma}\right]^{-\sigma\,\Delta\alpha} \in \mathbb{R},
\end{align}
where $\Delta\alpha$ is the steepening of the spectral index and $\sigma$ controls the sharpness of the break. This form is real-valued and thus does not affect polarization angles directly.

To ensure that the local complex spectral index at $\nu_0$ is exactly $\overline{\beta}$, we renormalize this factor as
\begin{align}
\label{eq:ageing_equation}
&\underline{P}_\nu=\underline{P}_0\left(\frac{\nu}{\nu_0}\right)^{\overline \beta}{C}_\mathrm{age}(\nu),\mathrm{\ \ \ where \ \ \ } \\
&C_{\rm age}(\nu)
=
\frac{\displaystyle
  \left(\frac{\nu}{\nu_b}\right)^{\Delta\alpha}
  \Big[1+(\nu/\nu_b)^{1/\sigma}\Big]^{-\sigma\Delta\alpha}}
{\displaystyle
  \left(\frac{\nu_0}{\nu_b}\right)^{\Delta\alpha}
  \Big[1+(\nu_0/\nu_b)^{1/\sigma}\Big]^{-\sigma\Delta\alpha}}
\;
\left(\frac{\nu}{\nu_0}\right)^{-\frac{\Delta\alpha}{1+(\nu_0/\nu_b)^{1/\sigma}}}.\nonumber
\end{align}
Defining $a \equiv (\nu_0/\nu_b)^{1/\sigma}$ and using $s$ (see Eq.~\ref{eq:log_freq_s}), this can be written compactly as
\begin{equation}
    \log C_{\rm age}(s)
= \Delta\alpha\frac{a}{1+a}\,s
- \Delta\alpha\,\sigma\,
\log\!\frac{1+a\,e^{s/\sigma}}{1+a}.
\end{equation}

The log–Taylor coefficients $\underline{\kappa}_n$ are obtained by differentiating this expression. Using Eq.~\ref{eq:logtaylor_prediction_in_practice}, the first two orders are
\begin{align}
\beta &= \overline{\beta} + \left.\frac{d\log C_{\rm age}}{ds}\right|_{0}
      = \overline{\beta},\\[6pt]
\gamma &= \left.\frac{d^2\log C_{\rm age}}{ds^2}\right|_{0}
      = -\,\frac{\Delta\alpha}{\sigma}\,\frac{a}{(1+a)^2}.
\end{align}
{Importantly the curvature due to ageing is negative.} 

By construction $(d\log C_{\rm age}/ds)_{s=0}=0$, so the local slope at $\nu_0$ equals $\overline\beta$, and all $\underline{\kappa}_n$ are real. Applying Eq.~\ref{eq:moment_prediction_in_practice} (with $\underline{w}_0=\underline{P}_0$) yields
\begin{align}
\underline{w}_1^{\rm(age)} &= \underline{w}_0\,C'_{\rm age}(0) = 0,\\
\underline{w}_2^{\rm(age)} &= \underline{w}_0\,C''_{\rm age}(0)
= -\,\underline{w}_0\;\frac{\Delta\alpha}{\sigma}\,\frac{a}{(1+a)^2}.
\end{align}
The leading non-zero contribution is therefore a purely real second-order moment: ageing produces curvature in the amplitude spectrum, but no frequency-dependent angle evolution (no imaginary moments). Consequently,
\begin{align}
\frac{|\underline{w}_{n,E}^{\rm(age)}|}{|\underline{w}_{n,B}^{\rm(age)}|}
\approx
\frac{|\underline{P}_{0,E}|}{|\underline{P}_{0,B}|}
\qquad (n\ge 0),
\end{align}
\textit{i.e.}\ $E/B$ ratios are preserved at all orders.

\begin{figure}
    \centering
    \includegraphics[width=0.98\linewidth]{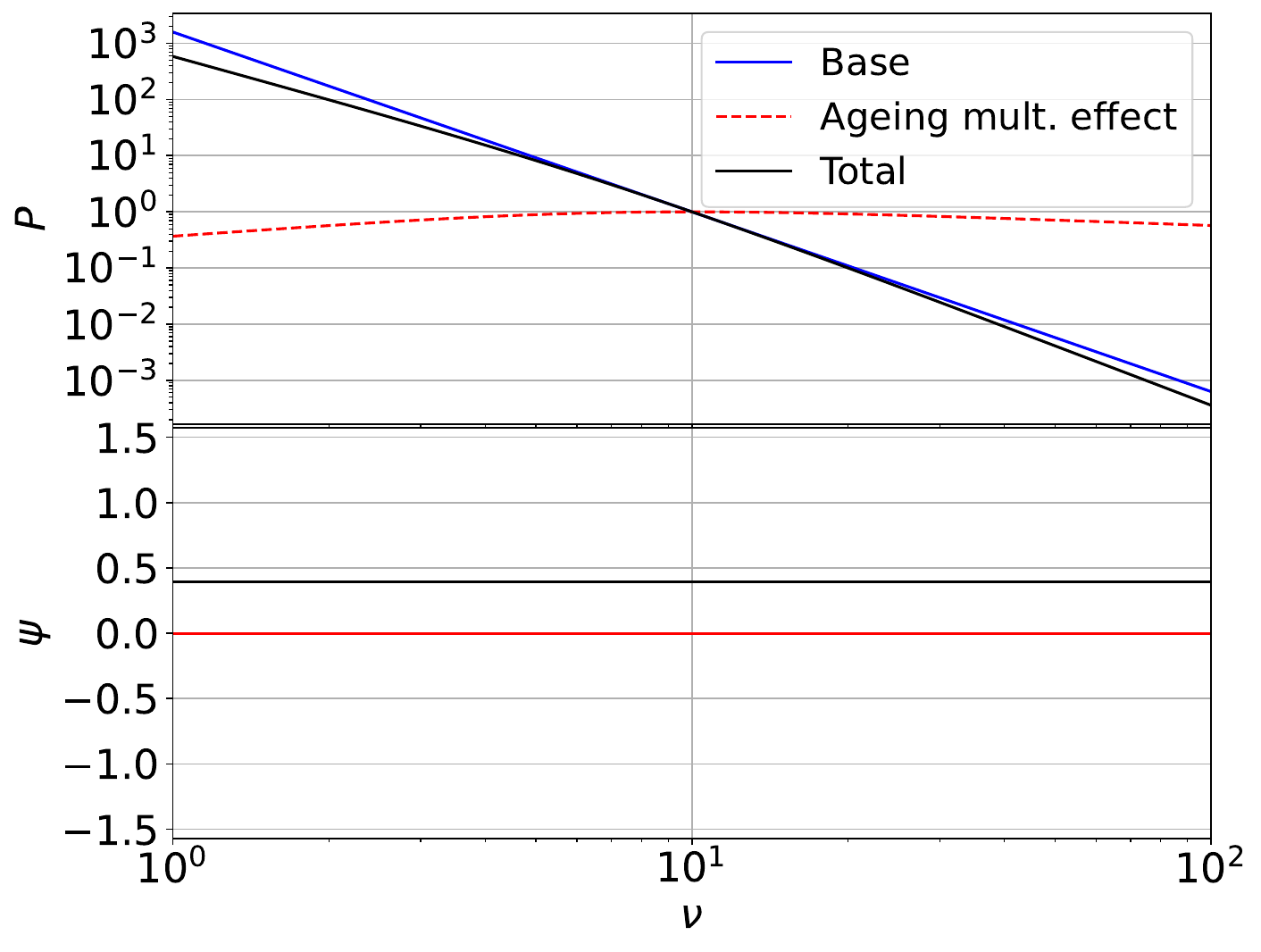}
    \caption{Example of the ageing effect (Sec.~\ref{sec:ageing}): Eq.~\ref{eq:ageing_equation} with $\overline{\beta}=-3.2$, $\sigma=0.5$, $\Delta\alpha=1$ and $\nu_b = 7\,\mathrm{GHz}$. The panels and curves are otherwise similar to the ones in Fig.~\ref{fig:superposition}.}
    \label{fig:ageing}
\end{figure}

We illustrate in Fig.~\ref{fig:ageing} the specific example of an aged synchrotron (with given ageing parameters that are chosen to emphasize the effect), showing that ageing does not affect the angle (\textit{lower panel}) and adds spectral complexity in the SED amplitude (\textit{upper panel}).

\subsection{Faraday rotation effects}
\label{sec:faraday}

At low frequency the observed complex linear polarization
$\underline{P}_\nu$ is further modified by Faraday rotation and depolarization, which we encode in a complex propagation factor $\underline{C}_{\rm FR}(\lambda)$, with $\lambda=c/\nu$. These effects arise from magneto-ionic material both co-spatial with the synchrotron-emitting region (internal Faraday rotation) and located in distinct foreground screens (external Faraday rotation).

For a uniform slab where synchrotron emission and Faraday rotation are co-spatial, \citet[][Eq.~34]{Sokoloff_Faraday} showed that the complex polarization is multiplied by
\begin{equation} 
\underline{C}_{\rm int}(\lambda)
=\frac{1-\exp[-\underline{S}(\lambda)]}{\underline{S}(\lambda)},
\qquad
\underline{S}(\lambda)=2\,\lambda^{4}\,\sigma_{\rm RM, int}^{2}-2\,i\,\lambda^{2}\,\mathrm{RM}_{\rm int},
\label{eq:Cint}
\end{equation}
where $\mathrm{RM}_{\rm int}=K\int n_e B_\parallel\,dz$ is the internal Faraday depth in units of Rotation Measure (RM) and $\sigma_{\rm RM,int}$ is the rms internal RM produced by small-scale magnetic fluctuations. The imaginary part of $\underline{C}_\mathrm{int}$ is responsible for differential Faraday rotation (phase), while the real part produces internal depolarization ($\sim\exp[-{\rm const}\,\lambda^4]$ in the strong-fluctuation limit).

A distinct external magneto-ionic screen contributes multiplicatively, see \citet[Eqs.~23 and 25]{Burn:1966} or \citet[Eq.~B3]{Sokoloff_Faraday}:
\begin{equation}
\underline{C}_{\rm ext}(\lambda)
=\exp\left(-2\lambda^{4}\sigma_{\rm RM,ext}^{2}+2i\lambda^{2}\mathrm{RM}_{\rm ext}\right)\;
\label{eq:Cext}
\end{equation}
with mean screen RM $\mathrm{RM}_{\rm ext}$ and rms fluctuations
$\sigma_{\rm RM,ext}$. The real part gives depolarization, while the imaginary part gives rotation.

The total Faraday factor is then
\begin{equation}
\underline{C}_{\rm FR}^{\rm(orig)}(\lambda)
=
\underline{C}_{\rm int}(\lambda)\,\underline{C}_{\rm ext}(\lambda).
\end{equation}
Writing
\begin{equation}
\underline{C}_{\rm int}(\lambda)
= A_{\rm int}(\lambda)\,\exp\!\bigl[i\,\phi_{\rm int}(\lambda)\bigr],
\textrm{\ and\ }
\underline{C}_{\rm ext}(\lambda)
= A_{\rm ext}(\lambda)\,\exp\!\bigl[i\,\phi_{\rm ext}(\lambda)\bigr],
\end{equation}
with
\begin{align}
A_{\rm ext}(\lambda)
&= \exp\!\bigl[-2\lambda^{4}\sigma_{\rm RM,ext}^{2}\bigr],\qquad
\phi_{\rm ext}(\lambda)
= 2\lambda^{2}\mathrm{RM}_{\rm ext},\\
A_{\rm int}(\lambda) &= |\underline{C}_{\rm int}(\lambda)|,\qquad
\phi_{\rm int}(\lambda) = \arg\underline{C}_{\rm int}(\lambda),
\end{align}
the total amplitude and phase are 
\begin{align}
A_{\rm FR}(\lambda)
= A_{\rm int}(\lambda)\,A_{\rm ext}(\lambda), \textrm{\ and\ }
\phi_{\rm FR}(\lambda)
= \phi_{\rm int}(\lambda) + \phi_{\rm ext}(\lambda),
\end{align}
and the full correction factor is
\begin{equation}
\underline{C}_{\rm FR}^{\rm(orig)}(\lambda)
= A_{\rm FR}(\lambda)\,
  \exp\!\bigl[i\,\phi_{\rm FR}(\lambda)\bigr].
\end{equation}
An effective (freq.-dependent) rotation measure may be defined as
\begin{equation}
\mathrm{RM}_{\rm eff}(\lambda^2)
\equiv \frac{1}{2}\,\frac{d\phi_{\rm FR}}{d\lambda^{2}}
= \mathrm{RM}_{\rm ext} + \frac{1}{2}\,\frac{d\phi_{\rm int}}{d\lambda^{2}},
\end{equation}
so that in the absence of internal structure $\mathrm{RM}_{\rm eff}\to\mathrm{RM}_{\rm ext}$, while in general both internal and external contributions affect the total rotation and depolarization.

To ensure that the local spectral index at $\nu_0$ equals $\overline{\beta}$, we again introduce a renormalized Faraday factor $\underline{C}_{\rm FR}(s)$ via
\begin{align}
\label{eq:faraday_equation}
&\underline{P}_\nu
=\underline{P}_0\left(\frac{\nu}{\nu_0}\right)^{\overline \beta}{\underline{C}}_\mathrm{FR}(s),\mathrm{\ \ \ where \ \ \ } \\
&\underline{C}_{\rm FR}(s)
=
\frac{\underline{C}_{\rm FR}^{\rm(orig)}(s)}
     {\underline{C}_{\rm FR}^{\rm(orig)}(0)}\,
\exp\!\left[-\,s\,\left.\frac{d}{ds}\ln\underline{C}_{\rm FR}^{\rm(orig)}(s)\right|_{s=0}\right].\nonumber
\end{align}
With this choice, applying Eq.~\ref{eq:logtaylor_prediction_in_practice} gives
\begin{align}
\beta &= \overline{\beta} + 
\left.\frac{d\log \underline{C}_{\rm FR}}{ds}\right|_{0}
= \overline{\beta},
\end{align}
while the curvature is in general complex:
\begin{align}
\underline{\gamma} &= \left.\frac{d^2\log \underline{C}_{\rm FR}}{ds^2}\right|_{0}\\ &= 
\underline{S}''_0\,f(\underline{S}_0)
+\underline{S}_0'^2
                    f'(\underline{S}_0)
\\&-32\,\sigma_{\rm RM,ext}^{2}c^4/\nu_0^4+8i\mathrm{RM}_{\rm ext}\,c^2/\nu_0^2
,
\end{align}
with
\begin{equation}
f(U)= \frac{e^{-U}}{1-e^{-U}} - \frac{1}{U},\textrm{\ and\ } f'(U) =  -\frac{e^{-U}}{(1-e^{-U})^2} + \frac{1}{U^2},
\end{equation}
and
\begin{align}
\underline{S}_0 &= 2\sigma_{\rm RM,int}^{2}(c/\nu_0)^{4}-2i\,\mathrm{RM}_{\rm int}(c/\nu_0)^{2},\\[3pt]
\underline{S}_0'&= -8\sigma_{\rm RM,int}^{2}(c/\nu_0)^{4}+4i\,\mathrm{RM}_{\rm int}(c/\nu_0)^{2},\\[3pt]
\underline{S}_0''&= 32\sigma_{\rm RM,int}^{2}(c/\nu_0)^{4}-8i\,\mathrm{RM}_{\rm int}(c/\nu_0)^{2}.
\end{align}
Here, $\Re(\underline{\gamma})$ encodes the Faraday-induced curvature of the amplitude spectrum (depolarization), while $\Im(\underline{\gamma})$ encodes the frequency-dependent rotation of the polarization angle. {As in the case of spectral ageing, the effective curvature induced by Faraday rotation is expected to be negative.}

\begin{figure}
    \centering
    \includegraphics[width=0.98\linewidth]{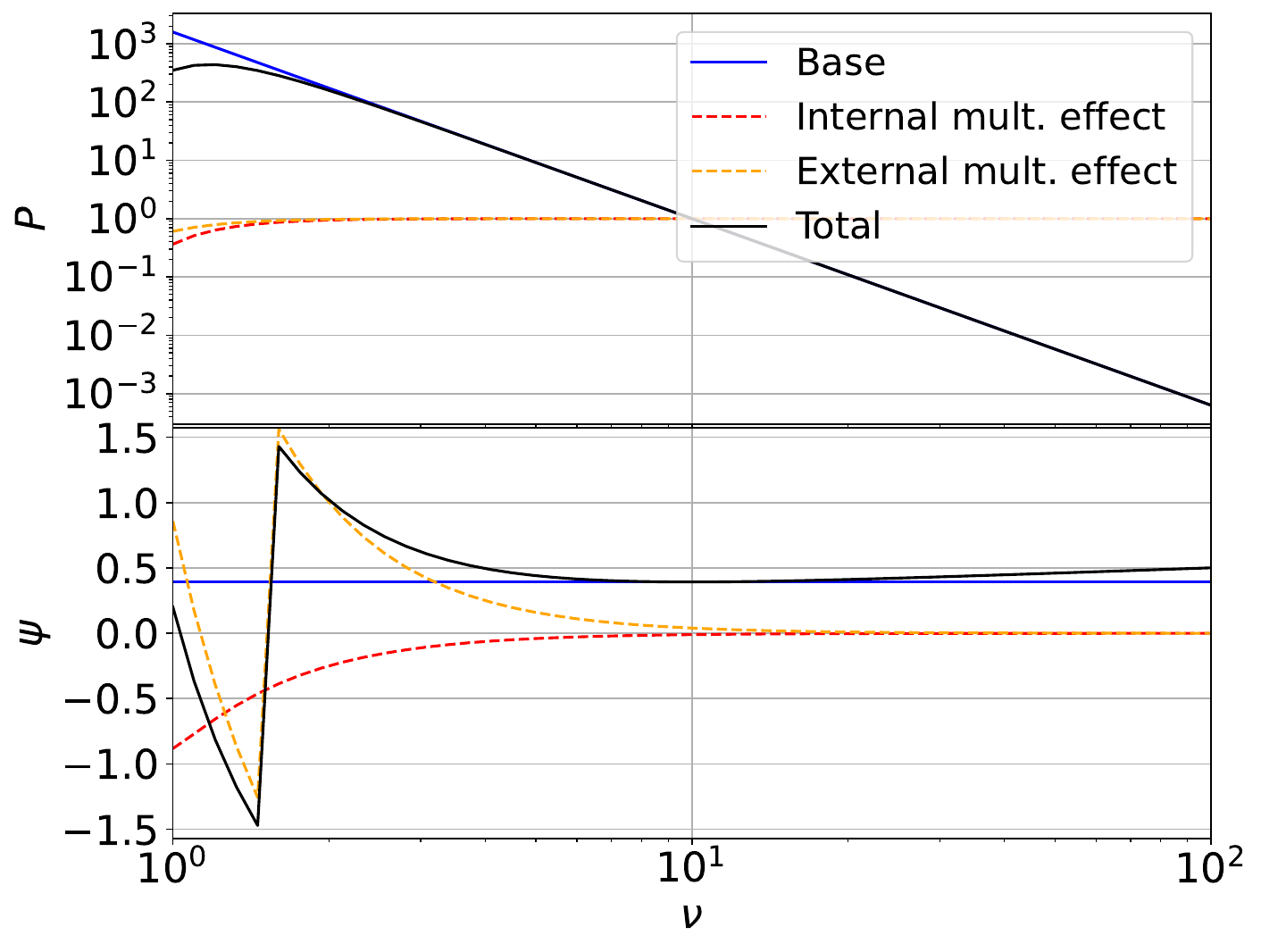}
    \caption{Example of the combination of internal and external Faraday effects (Sec.~\ref{sec:faraday}): Eq.~\ref{eq:faraday_equation} with $\underline{P}_{0}=e^{2i\pi/8}$, $\overline\beta=-3.2$, $\sigma_\mathrm{RM,int}=\sigma_\mathrm{RM,ext}=0.5/c^2$, $\mathrm{RM}_\mathrm{int}=-2/c^2$ and  $\mathrm{RM}_\mathrm{ext}=4/c^2$, with RMs expressed in units of $\mathrm{GHz}^{2}/c^{2}$. The panels and curves are otherwise similar to the ones in Fig.~\ref{fig:superposition} (for this particular illustration, (i) polarization angles are left wrapped along $\nu$ and are between $[-\pi/2, \pi/2]$, (ii) two distinct correction curves are represented, in red the internal Faraday effect and in orange the external Faraday effect).
    }
    \label{fig:faraday}
\end{figure}

From Eq.~\ref{eq:moment_prediction_in_practice} (with $\underline{w}_0=\underline{P}_0$), the Faraday-induced moments are
\begin{align}
\underline{w}_1^{\rm(FR)} &= \underline{w}_0\,
\underline{C}'_{\rm FR}(0) = 0,\\[4pt]
\underline{w}_2^{\rm(FR)} &= \underline{w}_0\,
\underline{C}''_{\rm FR}(0)=\underline{w}_0\,
\underline{\gamma},
\end{align}
which are also generically complex. 

To see this more explicitly, consider the pure-rotation limit. The Faraday phase can then be written in terms of the effective rotation measure as
\begin{equation}
\Phi(\mathbf{n}) \equiv 2\,(c/\nu_0)^2\,\mathrm{RM}_{\rm eff}(\mathbf{n})
= \overline{\Phi} + \delta\Phi(\mathbf{n}),
\end{equation}
with $\overline{\Phi}$ the sky-mean value and $\delta\Phi$ the spatial
fluctuations. A spatially uniform RM (\textit{i.e.}\ $\delta\Phi=0$) preserves the
zeroth-order $E/B$ balance: for instance, in the specific case in which $\underline{P}_{0,B}=0$ then no $B$ is generated by Faraday rotation alone. Instead, spatial variations $\delta\Phi$ generate $\textit{E}\leftrightarrow\textit{B}$ mixing at first order in a small-rotation expansion (still in the specific example of $\underline{P}_{0,B}=0$):
\begin{align}
\underline{w}_{1,E}^{\rm(FR)} \;\propto\; \overline{\Phi}\,\underline{w}_{0,E},\mathrm{\ \ and\ \ }\underline{w}_{1,B}^{\rm(FR)} \;\propto\; \underline{L}_B\!\big[\delta\Phi\,\underline{w}_0\big],
\end{align}
so that
\begin{equation}
    \frac{|\underline{w}_{1,E}^{\rm(FR)}|}{|\underline{w}_{1,B}^{\rm(FR)}|}
    \;\sim\;
    \frac{\big|\overline{\Phi}\,\underline{w}_{0,E}\big|}
         {\big|\underline{L}_B[\delta\Phi\,\underline{w}_0]\big|}.
\end{equation}
An increase of $|\underline{w}_{1,B}^{\rm(FR)}|/|\underline{w}_{1,E}^{\rm(FR)}|$
therefore signals frequency-dependent angle rotation sourced by spatial
variations of $\mathrm{RM}_{\rm eff}(\mathbf{n})$.

Fig.~\ref{fig:faraday} illustrates, for a specific choice of internal and external Faraday parameters, the impact of these effects on the spectral dependence of polarization. Depolarization manifests as a suppression of the SED amplitude at low frequencies (upper panel), while Faraday rotation produces a significant frequency-dependent rotation of the polarization angle (lower panel), which approximately follows an inverse-square scaling with frequency\footnote{We note that, over most of the parameter space of this model, the onset of noticeable rotation occurs at higher frequencies than that of significant depolarization.}.

\begin{table*}
{
\centering
\caption{Summary of the main spectral signatures discussed in Sec.~\ref{sec:predictions_from_physics}. The entries describe the intrinsic spectral behaviour associated with each physical mechanism. Additional spectral deformations may appear in the projected fields $(\underline{P}_E,\underline{P}_B)$ as a consequence of the non-fully-local nature of the map-space $\textit{E}/\textit{B}$ decomposition. These deformations are not intrinsic to the underlying mechanism itself and are generally expected to be localised near transitions between regions with different spectral or morphological properties. Cases in which the complexity arises only through this projection effect are labelled as ``secondary'' in the table.}
\label{tab:spectral_signatures}
\renewcommand{\arraystretch}{1.18}
\begin{tabular}{p{0.24\textwidth} c p{0.14\textwidth} p{0.12\textwidth} p{0.17\textwidth} p{0.17\textwidth}}
\hline
Mechanism & Sec. &
Curvature in $\underline{P}$ &
Evol. of $\psi$ in $\underline{P}$ &
Curvature in $(\underline{P}_E,\underline{P}_B)$ &
Evol. of $\psi$ in $(\underline{P}_E,\underline{P}_B)$ \\
\hline
Spatially varying $(A, \beta, \psi)(\mathbf{n})$ with single component per LoS &
\ref{sec:spatial_beta} &
No &
No &
Secondary &
Secondary \\

LoS superposition of distinct power laws, with mixed parity content &
\ref{sec:superposition} &
Yes, with either sign &
Yes &
Yes &
Yes \\

Synchrotron ageing &
\ref{sec:ageing} &
Yes, negative &
No &
Yes &
Secondary \\

Faraday rotation and depolarization &
\ref{sec:faraday} &
Yes, negative &
Yes &
Yes &
Yes \\

Coexistence of parity-specific emitters with different SEDs &
\ref{sec:coexistence_EB} &
Yes, with either sign &
Yes &
Secondary &
Secondary \\
\hline
\end{tabular}}
\end{table*}

\subsection{Coexistence of \textit{E}- and \textit{B}-dominated emitters}
\label{sec:coexistence_EB}

As a final physical ingredient, and one of the main motivations for working in $\textit{E}$- and $\textit{B}$-family fields, we consider the coexistence of structures with different parity content and different spectral behaviour. The simplest case is that of two components projected onto the same sky region: an approximately $\textit{E}$-dominated component $\underline{P}_a$ with spectral index $\beta_a$, and an approximately $\textit{B}$-dominated component $\underline{P}_b$ with spectral index $\beta_b$. At first order in the fixed-reference moment expansion,
\begin{equation}
    \underline{w}_1
    \simeq
    \underline{P}_a(\beta_a-\overline{\beta})
    +
    \underline{P}_b(\beta_b-\overline{\beta}) .
\end{equation}
If the two templates are sufficiently well separated by the spin-preserving projectors, then
\begin{equation}
    \underline{w}_{1,E}
    \simeq
    \underline{P}_a(\beta_a-\overline{\beta}),
    \qquad
    \underline{w}_{1,B}
    \simeq
    \underline{P}_b(\beta_b-\overline{\beta}) ,
\end{equation}
and therefore
\begin{equation}
    \frac{|\underline{w}_{1,E}|}{|\underline{w}_{1,B}|}
    \simeq
    \frac{|\underline{P}_a|}{|\underline{P}_b|}
    \frac{|\beta_a-\overline{\beta}|}{|\beta_b-\overline{\beta}|}.
\end{equation}
Thus the first spectral moment traces not only the relative amplitudes of the two structures, but also their relative spectral slopes.

{This example also illustrates why $\underline{P}_E$ and $\underline{P}_B$ can be spectrally simpler than the total field $\underline{P}$. If the two parity components are individually close to power laws but have different indices, their sum $\underline{P}=\underline{P}_E+\underline{P}_B$ is not, in general, a rigid-angle power law. Through the relations Eqs.~\ref{eq:relations_P_from_PEPB} (and even if one sets $\Im({\underline{\beta}_E})=\Im({\underline{\beta}_B})=\underline{\gamma}_E=\underline{\gamma}_B=0$ in these relations), the addition of two simple parity components can generate apparent curvature and frequency-dependent polarization-angle rotation in the total field $\underline{P}$. Conversely, analysing the two parity families separately can partially disentangle the spectral behaviour of the underlying structures.

This situation is physically natural rather than artificial. Synchrotron maps can contain several projected structures, with different magnetic-field geometries, cosmic-ray electron populations, ageing histories, or Faraday screens, that contribute with different $\textit{E}/\textit{B}$ balance. The spin-preserving decomposition does not identify the physical mechanism by itself, nor does it add information beyond $(Q,U)$. Its role is instead to reorganize the spectral signatures into parity families, thereby showing whether a given deformation is associated mainly with coherent $\textit{E}$-like structures, with $\textit{B}$-like structures, or with a mixture of both.}

The preceding argument is intentionally idealized. As discussed in Sec.~\ref{sec:spatial_beta} and Appendix~\ref{app:convolution_kernel}, the operators $\underline{L}_E$, $\underline{L}_B$, and $\underline{L}_S$ are non-fully local convolutions on the sphere. The projected fields $\underline{P}_E(\mathbf{n})$ and $\underline{P}_B(\mathbf{n})$ therefore receive contributions from neighbouring structures with potentially different spectral properties. This partially blurs the simple picture of perfectly separable ``$\textit{E}$-emitters'' and ``$\textit{B}$-emitters'': the same transform that can disentangle co-spatial structures with different parity content can also induce additional spectral deformations by mixing nearby regions.

Whether the gain from parity separation outweighs this non-fully-local mixing is therefore a quantitative, sky-dependent question. It depends on how well the relevant structures are separated by parity, how different their SEDs are, how large the $\textit{E}$-to-$\textit{B}$ imbalance is, and how strongly neighbouring regions with different spectra are mixed by the kernels. This balance cannot be assessed from the formalism alone, and motivates the explicit diagnostics and applications developed below.

{
\subsection{Summary of spectral signatures and role of the \texorpdfstring{$\textit{E}/\textit{B}$}{E/B} decomposition}
\label{sec:summary_spectral_signatures}

The mechanisms discussed in this section leave different measurable signatures in the complex spectral parameters. Table~\ref{tab:spectral_signatures} summarizes whether each mechanism generates amplitude curvature in $\underline{P}$, frequency evolution of the polarization angle, and corresponding effects in the spin-preserving fields $(\underline{P}_E,\underline{P}_B)$. The table should be read as a statement about spectral properties. It does not provide a unique physical classification: in a realistic sky, several emitting structures can coexist, each affected by different spectral effects and each having its own $\textit{E}/\textit{B}$ parity balance.

Two points are particularly important. First, amplitude curvature and polarization-angle evolution in full polarization are distinct and important observables. In the simple models considered here, ageing produces negative real curvature without angle evolution, while Faraday depolarization also tends to produce negative amplitude curvature but can additionally rotate the polarization angle. By contrast, line-of-sight superposition naturally produces curvature of either sign and can rotate the net polarization angle when the mixed components have different angles and spectral indices. Therefore, observing a robust positive curvature in the total polarization amplitude $P$ would strongly point towards line-of-sight superposition, rather than towards ageing or Faraday depolarization alone.

Second, the mechanisms discussed above show why a rigid-angle power law in $\underline{P}$ should not be expected to be the most accurate or physically complete model in general. Several physical effects can curve the synchrotron amplitude, rotate the polarization angle $\psi$ with frequency, and therefore change the $\textit{E}/\textit{B}$ balance across frequency. At the same time, Sec.~\ref{sec:coexistence_EB} shows that the separated fields $(\underline{P}_E,\underline{P}_B)$ can themselves be naturally close to simple SEDs when structures with different spectral properties also have different parity content. In that case, the total field $\underline{P}=\underline{P}_E+\underline{P}_B$ inherits additional spectral complexity from recombining simpler parity components. Thus, the relevant modelling question is not whether $\underline{P}$ or $(\underline{P}_E,\underline{P}_B)$ is intrinsically preferable, but which representation provides the simpler and more physical description for the sky region under consideration.

This last case is the key motivation for modelling directly in parity-separated spin-2 fields. It is not guaranteed to occur everywhere, because the non-fully local projection can itself generate spectral complexity by mixing neighbouring regions. The useful regime is therefore the one in which the gain from separating structures with different parity content and SEDs exceeds the additional complexity induced by the projection. The toy-model and PySM diagnostics of Sec.~\ref{sec:comparaison_toy_model} and the CMB-oriented applications of Sec.~\ref{sec:applications_cmb} are designed to test precisely this balance.}

\section{Verification of predictions and interpretability of the fields}
\label{sec:comparaison_toy_model}

Because the spectral complexity induced in $\underline{P}_E$, $\underline{P}_B$ and $\underline{S}$ depends in a non-fully-local way on the full-sky morphology, quantitative predictions for these transformed fields can only be made once a concrete sky model is specified. In this section we therefore move from the general mechanisms discussed in Sec.~\ref{sec:predictions_from_physics} to explicit examples. In Sec.~\ref{sec:simple_toy_model}, we construct a controlled toy model that allows us to verify the superposition, ageing and Faraday predictions derived previously, and to investigate how these effects impact the interpretability of $\underline{P}$, $\underline{P}_E$, $\underline{P}_B$ and $\underline{S}$. This sets the stage for the more realistic \texttt{PySM}-based analysis presented in Sec.~\ref{sec:more_realistic}.

\subsection{Simple toy model}
\label{sec:simple_toy_model}

We first consider a simple eight-component toy model with varying \textit{E}/\textit{B} balance and spectral properties. The components are chosen to span a range of morphologies (loops, circular sources, stochastic-angle structures) and physical mechanisms (pure spectral power laws, intrinsic spectral curvature, ageing, Faraday effects). We present its morphology at the reference frequency in Sec.~\ref{sec:central_freq_tm} and analyse its spectral properties in Sec.~\ref{sec:spec_pred_validation}, thereby validating the prediction equations derived in Sec.~\ref{sec:predictions_from_physics}. We then use the toy model in Sec.~\ref{sec:spectral_conclusion_tm} to illustrate how spectral conclusions can be drawn from the various \textit{E}/\textit{B}-separated fields, before showing in Sec.~\ref{ssec:finite_channels_performance} the complementarity of the two types of spectral expansion considered in this work.

\subsubsection{Central-frequency morphology}
\label{sec:central_freq_tm}

\begin{table}
\centering
\begin{tabular}{l | c c c c}
\hline
 & $E/B$ & $A$ & $\beta$ & Other prop. \\
\hline
(1) $E$ loop                       & $+\infty$          & 1    & -3.4 & --  \\
(2) $B$ loop                       & $0$                & 1    & -2.6 & --  \\
(3) Random-angle galaxy      & rand  (5$^\circ$)             & 1    & -3.0 & --  \\
(4) Circular source              & $0$                & 1    & -3.2 & Aged \\
(5) Circular source              & 1                  & 1    & -3.2 & $\gamma>0$\\
(6) Circular source              & $+\infty$          & 1    & -3.2 & Faraday \\
(7) Stochastic-angle stripe      & rand  (20$^\circ$)             & 1    & -2.8 & --  \\
(8) Random-angle bg.  & rand  (20$^\circ$)             & 0.01 & -3.0 & --  \\
\hline
\end{tabular}
\caption{Components included in the toy model. The numbering corresponds to the labels in the polarization–amplitude panel of Fig.~\ref{fig:toy_model_maps} (top row, centre). The first column indicates the relative contribution of each feature to $E$ and $B$, while the second gives the relative total polarization amplitude, whose unit does not matter for the discussion. In the spectral analysis of Sec.~\ref{sec:spec_pred_validation}, these two morphological columns are defined at the pivot frequency of 10 GHz, together with the two spectral columns used to extrapolate the model to lower and higher frequencies: the third column is the spectral index at 10 GHz and the fourth column specifies whether a more complicated spectral property is considered beyond the simple power law ($\gamma>0$ meaning positively curved).} 
\label{tab:table_components_tm}
\end{table}

\begin{figure*}
\centering
    \includegraphics[width=0.86\linewidth]{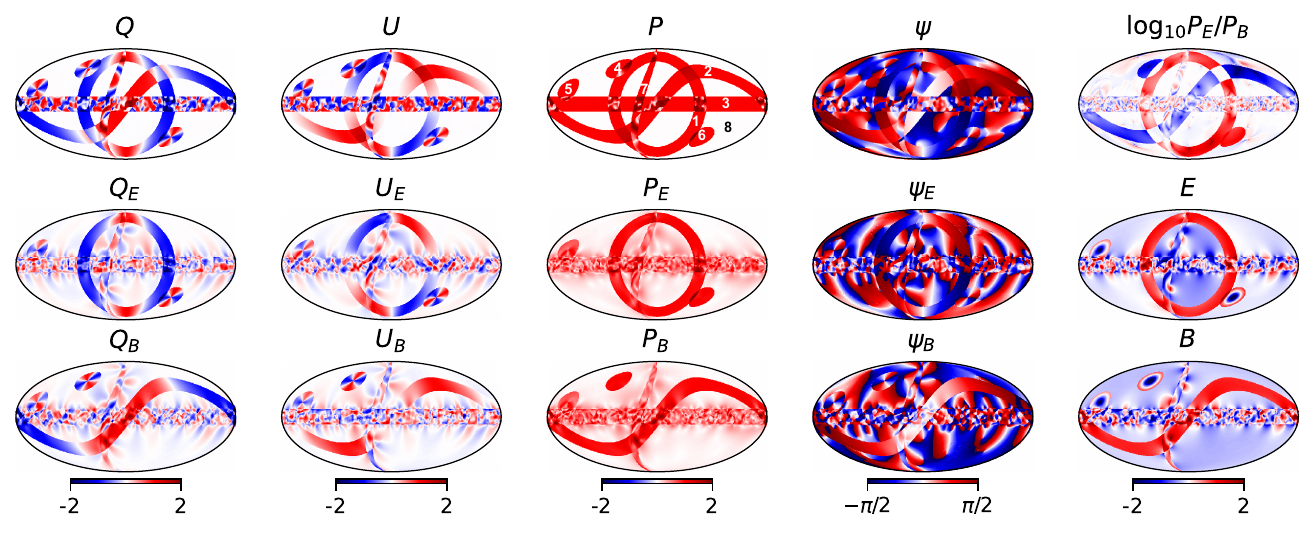}
    \caption{Toy-model polarization sky at a single frequency. The numbered structures in the total polarization amplitude panel (top row, centre) correspond to the components defined in Table~\ref{tab:table_components_tm}, each chosen to illustrate a distinct $E/B$ balance and morphology. The first row shows the full Stokes fields $(Q,U)$ together with the resulting total polarization amplitude $P$ and angle $\psi$. The second and third rows display the corresponding $E$- and $B$-family fields $(Q_E,U_E,P_E,\psi_E)$ and $(Q_B,U_B,P_B,\psi_B)$ obtained through the map-space spin-2 \textit{E}/\textit{B} decomposition. The upper-right panel shows the local ratio $P_E/P_B$, while the rightmost panels of the second and third rows display the standard scalar $E$ and $B$ maps. {In this toy model, the different structures are cleanly separated by the $E/B$ decomposition: coherent radial or tangential patterns project predominantly into the $E$ family, while curl-like structures project into the $B$ family. The spin-preserving amplitudes $P_E$ and $P_B$ can therefore be interpreted as parity-selected contributions to the total polarization amplitude, remaining localized around regions where polarized emission with the corresponding parity is present.}}
    \label{fig:toy_model_maps}
\end{figure*}

The properties of the eight components of the toy model are summarized in Table~\ref{tab:table_components_tm}. In order to obtain a band-limited toy model, for which harmonic-space transformations can be applied consistently, we first construct the morphology in pixel space and then apply a harmonic-space cut, removing all modes above $\ell_{\mathrm{max}} = 3N_{\mathrm{side}} - 1$, where $N_{\mathrm{side}} = 512$ is the \texttt{HEALPix} resolution parameter of the toy model\footnote{This procedure may induce a very small loss of locality, and hence a negligible amount of associated spectral complexity, but it is required to ensure the consistency of the harmonic transforms and, in particular, to preserve the closure relation.}. The obtained $(Q, U, P, \psi)$ morphology is illustrated in the first row of Fig.~\ref{fig:toy_model_maps}, where we have overlaid the identification number of each component on the $P$ map. The second and third rows display the corresponding $E$- and $B$-family fields $(Q_E,U_E,P_E,\psi_E)$ and $(Q_B,U_B,P_B,\psi_B)$ obtained through the map-space \textit{E}/\textit{B} decomposition introduced in Sec.~\ref{sec:EBseparating}. Finally, the last column of Fig.~\ref{fig:toy_model_maps} shows the $P_E$ to $P_B$ ratio as well as the $E$ and $B$ scalar maps (from upper panel to lower panel).

Several key features are immediately apparent. Purely $E$-type structures (components 1 and 6) disappear from the $B$-family panels, while purely $B$-type structures (component 2) vanish from the $E$-family fields. Sources with mixed morphology or stochastic angles (components 3, 5, 7, 8) appear in both families, as expected. The map of the local ratio $P_E/P_B$ (top-right panel) clearly separates radial-like ($E$-dominated) from solenoidal ($B$-dominated) patterns and highlights regions where a single component dominates the local \textit{E}/\textit{B} balance. This matches the conceptual picture of Sec.~\ref{sec:EBseparating}: $\underline{P}_E$ and $\underline{P}_B$ retain the spin-2 nature of the field  (contrary to scalar $E$, $B$, $\underline{S}$) while efficiently isolating gradient- and curl-like contributions (contrary to $\underline{P}$).

\begin{figure}
    \centering
    \includegraphics[width=\linewidth]{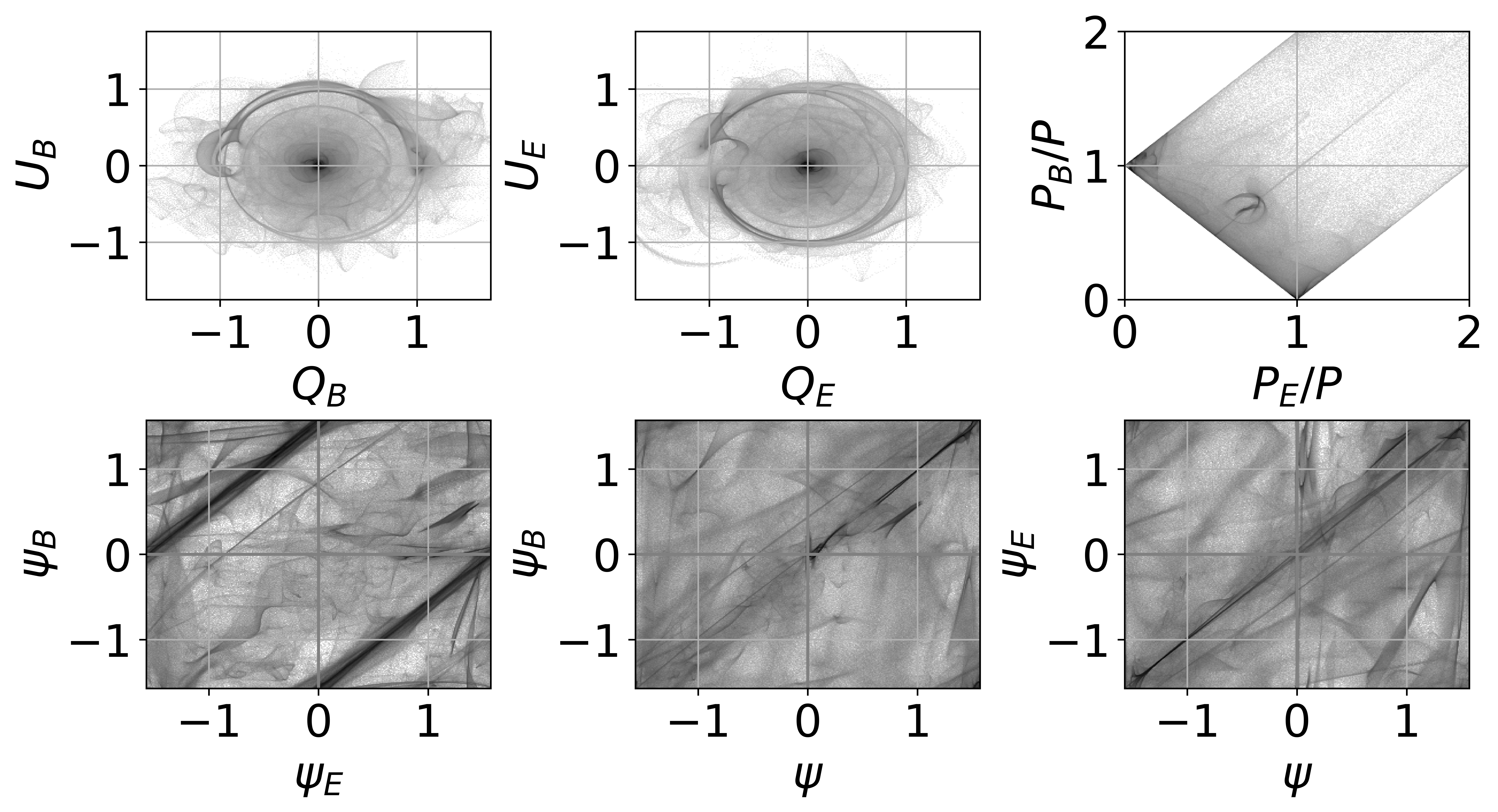}
    \caption{Two-dimensional distributions of several pairs of map-level quantities derived from the toy model. The first row shows the $(Q_B,U_B)$, $(Q_E,U_E)$ and $(P_E/P,\,P_B/P)$ planes, while the second row displays the planes $(\psi_E,\,\psi_B)$, $(\psi,\,\psi_B)$, and $(\psi,\,\psi_E)$, all evaluated at the reference frequency of 10 GHz. The grey-scale density displays the pixel counts in logarithmic scale, in each parameter space.}
    \label{fig:toy_model_2D}
\end{figure}

Fig.~\ref{fig:toy_model_2D} complements this map-space view with two-dimensional distributions. The $(Q_B,U_B)$ and $(Q_E,U_E)$ planes (\textit{upper-left} and \textit{upper-middle}) show how the \textit{E}/\textit{B} decomposition projects each pixel into its $E$-family or $B$-family Stokes components. Purely $E$- or $B$-type structures cluster in compact regions of their respective planes, while mixed or stochastic components populate broader domains. The $(P_E/P,\,P_B/P)$ plane (\textit{upper-right}) directly visualizes the constraint $(P_E-P_B)^2 \leq P^2 \leq (P_E+P_B)^2$ derived in Eq.~\ref{eq:plane_restriction}: the pixel distribution is confined to the allowed region, with $E$-dominated pixels lying near the $(P_E/P, P_B/P)\simeq (1,0)$ corner and $B$-dominated pixels near $(P_E/P, P_B/P)\simeq (0,1)$. The angular (\textit{lower}) planes $(\psi_E,\psi_B)$, $(\psi,\psi_B)$, and $(\psi,\psi_E)$ reveal that regions containing a single dominant morphological component lie close to the diagonal, whereas pixels combining multiple structures with different orientations fill larger areas of these plots.

Taken together, Figs.~\ref{fig:toy_model_maps} and \ref{fig:toy_model_2D} illustrate how the \textit{E}/\textit{B} decomposition reorganizes the polarization field both in map space and in the $(Q,U)$, $(P_E/P,P_B/P)$ and angle planes. They also provide a concrete playground in which to apply the spectral predictions derived in Sec.~\ref{sec:predictions_from_physics}.

\begin{figure*}
    \centering
    \includegraphics[width=\linewidth]{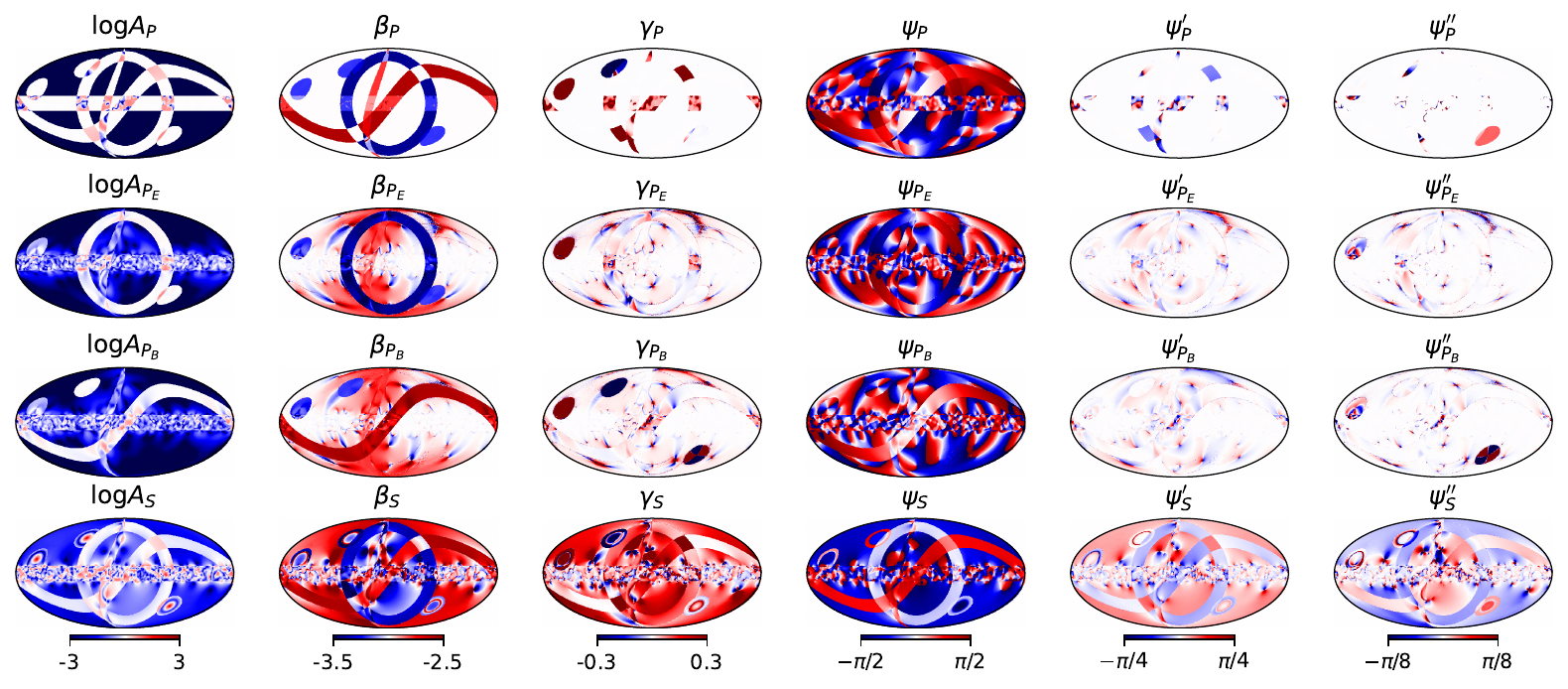}
    \caption{Maps of complex log--Taylor parameters for the toy model, shown for $\underline{P}$ (top row), $\underline{P}_E$ (second row), $\underline{P}_B$ (third row), and $\underline{S}$ (bottom row). The columns display, from left to right, $\log A_X$, $\psi_X$, $\beta_X$, $\psi'_X$, $\gamma_X$, and $\psi''_X$, where primes denote derivatives with respect to $s$ (\textit{cf.} Eq.~\ref{eq:log_freq_s}). The colour scale of a given column is common to all rows and is shown below the figure. Angular quantities are given in radians. {The toy model is globally somewhat simpler in $\underline{P}$ than in $(\underline{P}_E,\underline{P}_B)$ within the single-component background regions, reflecting the non-fully local nature of the \textit{E}/\textit{B} projection. However, the opposite behaviour is observed where the \textit{E}- and \textit{B}-dominated loops overlap: in these regions, $\underline{P}$ becomes more spectrally complex than either $\underline{P}_E$ or $\underline{P}_B$, illustrating the coexistence effect discussed in Sec.~\ref{sec:coexistence_EB}.}}
    \label{fig:all_fields_logtaylor_TM}
\end{figure*}

\subsubsection{Verifying spectral predictions}
\label{sec:spec_pred_validation}

We now extend the toy model to multiple frequencies by assigning specific spectral behaviours to the components, as indicated in the last column of Table~\ref{tab:table_components_tm}. The circular source (4) is affected by synchrotron ageing with the parameters used for Fig.~\ref{fig:ageing}: $\overline{\beta}=-3.2$, $\sigma=0.5$, $\Delta\alpha=1$ and $\nu_b = 7~\mathrm{GHz}$. The circular source (5) has intrinsic curvature with $\overline{\beta}=-3.2$ and $\gamma=1$ (Fig.~\ref{fig:curvature}). The circular source (6) is affected by internal and external Faraday effects with $\overline\beta=-3.2$, $\sigma_\mathrm{RM,int}=\sigma_\mathrm{RM,ext}=0.5/c^2$, $\mathrm{RM}_\mathrm{int}=-2/c^2$ and $\mathrm{RM}_\mathrm{ext}=4/c^2$ (Fig.~\ref{fig:faraday}). The remaining components follow simple power laws with the spectral indices listed in Table~\ref{tab:table_components_tm}.

Using the combination rules of Eqs.~\ref{eq:w0_combination}--\ref{eq:w2_combination}, the moments of the total polarization field in this eight-component model are
\begin{align}
\label{eq:predi_tm_w0}
\underline{w}_0 &= \sum_{i=1}^{8} \underline{A}_i,\\
\label{eq:predi_tm_w1}
\underline{w}_1 &= \sum_{i=1}^{8} \underline{A}_i\,\Delta\beta_i,\\
\label{eq:predi_tm_w2}
\underline{w}_2 &= \sum_{i=1}^{8} \underline{A}_i
\Big[
   \Delta\beta_i^2
 + \gamma_i
 + C^{\rm(age)}_{2,i}
 + C^{\rm(FR)}_{2,i}
\Big],
\end{align}
where $\underline{A}_i$ is the complex amplitude of component $i$ at $\nu_0$, $\Delta\beta_i\equiv \beta_i-\overline{\beta}$, and $C^{\rm(age)}_{2,i}$ and $C^{\rm(FR)}_{2,i}$ are the second-order (real and complex) contributions from ageing and Faraday effects as given in Secs.~\ref{sec:ageing} and \ref{sec:faraday}. The effective log–Taylor coefficients of the total field then follow from Eq.~\ref{eq:moments_2_taylor}.

We compare these analytic predictions with parameters fitted directly to simulated maps at three frequencies\footnote{Hereafter, we fix the reference $\overline\beta$ of the moment expansion to -3, close to the sky-averaged $\beta$ by construction of all sky models studied hereafter. We verified that as long as this reference is varied only slightly, the reliability of the predictions is not affected and the performance of the moment extrapolation is only weakly affected.}. To isolate purely numerical effects, we first use three very closely spaced channels, $\nu_a=9.999~\mathrm{GHz}$, $\nu_b=\nu_0=10~\mathrm{GHz}$ and $\nu_c=10.001~\mathrm{GHz}$\footnote{As one expects, using three frequencies to fit a model with only three complex parameters limits the validity of the fit away from $\nu_0$, especially in regions where higher-order spectral parameters are non-negligible (\textit{e.g.}, in the vicinity of the curved or Faraday-active sources). A low-order expansion around $s=0$ provides an accurate local description, but cannot perfectly reproduce spectral properties far from the pivot. To illustrate this, more realistically spaced frequency channels will be considered in Sec.~\ref{ssec:finite_channels_performance}.}. 

In this regime the moments and log–Taylor parameters fitted on $\underline{P}$ agree with those predicted from Eqs.~\ref{eq:predi_tm_w0}--\ref{eq:predi_tm_w2} and Eq.~\ref{eq:moments_2_taylor} to better than $10^{-5}$ relative accuracy. We illustrate this comparison in map space in Appendix \ref{sec:app_verif_rel}. This confirms that the expressions derived in Sec.~\ref{sec:predictions_from_physics} correctly capture the behaviour of the complex field under line-of-sight superposition, intrinsic curvature, ageing and Faraday rotation, when evaluated near the pivot frequency.

We repeat the same exercise for the $E$- and $B$-family fields and for the scalar field $\underline{S}$. One approach is to fit log–Taylor parameters directly to $\underline{P}_E$, $\underline{P}_B$ or $\underline{S}$ at the three frequencies. Alternatively, we can (i) predict the $\underline{P}$ moments using Eqs.~\ref{eq:predi_tm_w0}--\ref{eq:predi_tm_w2}, (ii) map these to moments for $\underline{P}_E$, $\underline{P}_B$ and $\underline{S}$ using the linear relations of Eq.~\ref{eq:project_moments}, and (iii) convert the resulting moments to log–Taylor parameters via Eq.~\ref{eq:moments_2_taylor}. The two procedures agree to within $10^{-4}$ relative accuracy, thereby validating the combination of (a) the physical predictions for the $\underline{P}$ moments, (b) the moment–log–Taylor conversion, and (c) the linear $E$/$B$ projection relations established in Sec.~\ref{sec:EBseparating}.

\begin{figure*}
    \centering
    \includegraphics[width=\linewidth]{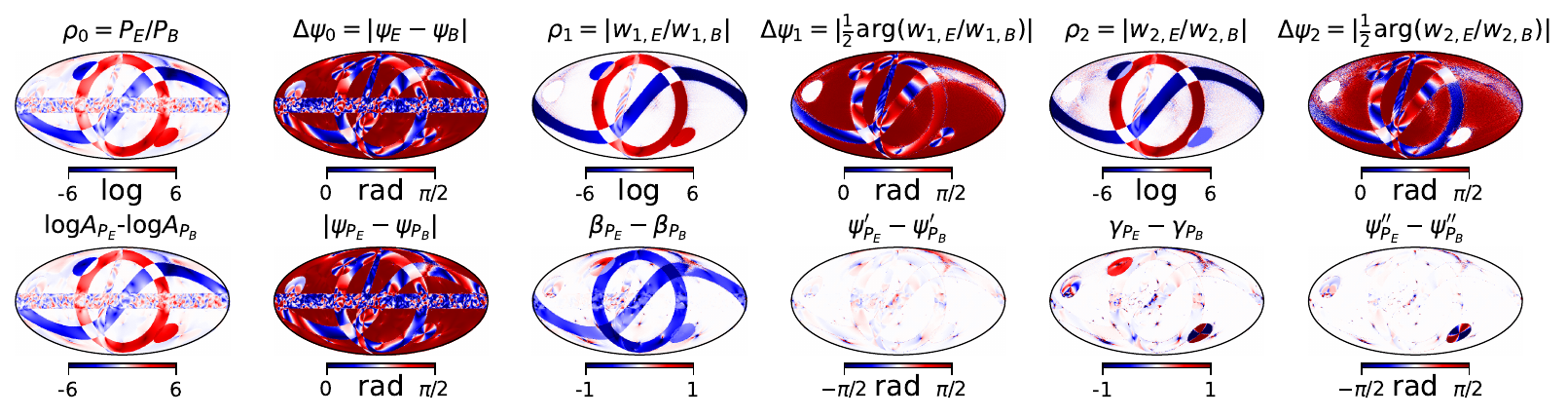}
    \caption{Generalized \textit{E}/\textit{B} diagnostics for the toy model. \textit{Upper panels}: maps of $\rho_n=|\underline{w}_{n,E}/\underline{w}_{n,B}|$ and $|\Delta\psi_n|=\tfrac{1}{2}|\arg(\underline{w}_{n,E}/\underline{w}_{n,B})|$ for $n=0,1,2$. \textit{Lower panels}: differences between $E$ and $B$ log–Taylor parameters, namely $\log A_{P_E}-\log A_{P_B}$, $|\psi_{P_E}-\psi_{P_B}|$, $\beta_{P_E}-\beta_{P_B}$, $\psi'_{P_E}-\psi'_{P_B}$, $\gamma_{P_E}-\gamma_{P_B}$, and $\psi''_{P_E}-\psi''_{P_B}$.}
    \label{fig:ratio_EB}
\end{figure*}

\subsubsection{Spectral conclusions on the toy model in $\underline{P}$, $\underline{P}_E$, $\underline{P}_B$ and $\underline{S}$}
\label{sec:spectral_conclusion_tm}

We now use the multi-frequency toy model to compare the spectral behaviour of the different fields $\underline{P}$, $\underline{P}_E$, $\underline{P}_B$ and $\underline{S}$. This serves three purposes: it illustrates the physical mechanisms discussed in Sec.~\ref{sec:predictions_from_physics}, proposes various diagnostics that are meaningful to apply to data, and anticipates the tendencies that will reappear in the more realistic \texttt{PySM} simulations. We focus on four aspects: (i) the stability of the polarization angle (Sec.~\ref{sec:angle_stability}), (ii) generalized \textit{E}/\textit{B} ratios at different spectral orders (Sec.~\ref{sec:generalized_EB}), (iii) direct comparisons of log–Taylor parameters between $E$ and $B$ (Sec.~\ref{sec:logtaylor_params}), and (iv) simple diagnostics of spectral complexity (Sec.~\ref{sec:spectral_complex}). All quantities below can be computed either from the moments $\underline{w}_n$ or from the associated complex log–Taylor coefficients and are illustrated in Figs.~\ref{fig:all_fields_logtaylor_TM} and \ref{fig:ratio_EB}.

\paragraph{Angle stability in $\bm{\underline{P}}$, $\bm{\underline{P}_E}$ and $\bm{\underline{P}_B}$}
\label{sec:angle_stability}

As discussed in Sec.~\ref{sec:complex_moment_expansion}, the imaginary parts of the complex log–Taylor coefficients control the evolution of the polarization angle with frequency: $\tfrac{1}{2}\Im(\underline{\beta})$ and $\tfrac{1}{2}\Im(\underline{\gamma})$ correspond to the first and second derivatives of $\psi$ with respect to $s$. In practice, all angles reported hereafter are unwrapped (using the \texttt{unwrap} function of \texttt{numpy}), which also ensures a robust determination of the frequency evolution of the angles.

In the total polarization $\underline{P}$ (top row of Fig.~\ref{fig:all_fields_logtaylor_TM}), the angle derivatives $\psi'_P$ and $\psi''_P$ are close to zero over most of the sky. Significant deviations occur only where components with different intrinsic angles and non-trivial spectra overlap, notably around the Faraday-active circular source (6) and, more mildly, near intersections of the large $E$ and $B$ loops (1 and 2) with the stochastic-angle structures (3, 7, 8). This behaviour directly reflects the mechanisms of Sec.~\ref{sec:superposition} and \ref{sec:faraday}: angle evolution in $\underline{P}$ arises either from line-of-sight mixing of misaligned power laws or from Faraday rotation.

By contrast, the projections $\underline{P}_E$ and $\underline{P}_B$ (second and third rows) present much richer patterns in $\psi'_X$ and $\psi''_X$. The $E$ projection enhances angle variations along the large $E$ loop and the stochastic galactic band, while the $B$ projection concentrates strong derivatives along the $B$ loop and around the aged and Faraday circular sources (4–6). Even when $\psi'_P$ is small, the derivatives of $\psi_{P_E}$ and $\psi_{P_B}$ can be substantial. This illustrates a key point anticipated in Sec.~\ref{sec:predictions_from_physics}: the map-space \textit{E}/\textit{B} transform redistributes the same physical effects into different parity channels through non-fully-local kernels, so a modest frequency dependence of the total angle can correspond to a much more significant angle evolution in $\underline{P}_E$ and $\underline{P}_B$.

The scalar field $\underline{S}$ (\textit{lower panels}) shows yet another behaviour: its angle and angle derivatives vary smoothly on large scales set by the loops and band but are rarely close to zero. This is consistent with the expectation that $\underline{S}$ combines amplitude and angle in a non-linear way, and therefore tends to exhibit enhanced spectral complexity.

{Angle stability provides a first diagnostic of \textit{spectral complexity}. A field with small absolute values of $\psi'_X$ and $\psi''_X$ can be accurately extrapolated using simple rigid-angle (i.e.\ amplitude-only) models, and also indicates a morphology that remains stable across frequencies, making such fields well suited for modelling the physics of polarized emission over the sky.}

\paragraph{Generalized \textit{E}/\textit{B} ratios}
\label{sec:generalized_EB}

To quantify how the \textit{E}/\textit{B} balance evolves with spectral order, we introduce
\begin{equation}
\rho_n = \left|\frac{\underline{w}_{n,E}}{\underline{w}_{n,B}}\right|,
\qquad
\Delta \psi_n = \frac{1}{2}\arg\left[\frac{\underline{w}_{n,E}}{\underline{w}_{n,B}}\right],
\end{equation}
where $\underline{w}_{n,E}$ and $\underline{w}_{n,B}$ are the $n$-th moments of $\underline{P}_E$ and $\underline{P}_B$. The ratio $\rho_n$ measures the amplitude balance between $E$ and $B$ at order $n$, while $\Delta\psi_n$ measures the relative phase. When $\Delta\psi_n\simeq 0$, the \textit{E} and \textit{B} contributions at order $n$ share a similar angle evolution; when $|\Delta\psi_n|\sim \pi/2$, their contributions are nearly orthogonal in the complex plane. In that sense, $\{\rho_n,\Delta\psi_n\}$ provide a natural generalization of the familiar map-level ratio $\underline{P}_E/\underline{P}_B$ discussed in Sec.~\ref{sec:EBseparating}.

Fig.~\ref{fig:ratio_EB} shows $\rho_n$ and $|\Delta\psi_n|$ for $n=0,1,2$. For $n=0$, $\rho_0$ reproduces the single-frequency $E/B$ amplitude ratio. The large $E$ loop (1) appears as strongly \textit{E}-dominated, the large $B$ loop (2) as strongly \textit{B}-dominated, while the stochastic-angle components (3, 7, 8) and the intrinsically \textit{E}/\textit{B}-balanced circular source (5) populate intermediate values. The corresponding angle difference $|\Delta\psi_0| = |\psi_E-\psi_B|$ is small only where a single component dominates; elsewhere it is close to $\pi/2$, reflecting the fact that $E$ and $B$ spin-2 objects in this model are mostly perpendicular in the background region.

{This behaviour can be understood from the combination of the closure relation Eq.~\ref{eq:closure_complex} and the non-fully-local nature of the $E/B$ projectors.  By construction, the pure fields obey $\underline{P}=\underline{P}_E+\underline{P}_B$ at every pixel.  In the background region of the toy model the ``true'' polarization $\underline{P}$ is very small (only the weak stochastic component is present), whereas the non-fully-local $E/B$ transform spreads the signal from the bright structures over a much wider area.  The tails of this response are therefore carried almost entirely by $\underline{P}_E$ and $\underline{P}_B$, which must satisfy $\underline{P}_E(\mathbf{n})\simeq-\underline{P}_B(\mathbf{n})$ wherever $\underline{P}(\mathbf{n})\simeq 0$.  In terms of Stokes parameters this implies $Q_B\simeq -Q_E$ and $U_B\simeq -U_E$, and hence $2\psi_B\simeq 2\psi_E+\pi$, \textit{i.e.}\ $\psi_B\simeq\psi_E+\pi/2$. The near-orthogonality of the $E$- and $B$-polarization orientations in the background is thus not an independent physical feature of the sky, but the natural way in which the non-fully-local $E/B$ decomposition enforces $\underline{P}=\underline{P}_E+\underline{P}_B$ in regions where the input polarization is intrinsically weak.}

At $n=1$, $\rho_1$ compares the first-order spectral gradients of $E$ and $B$. Regions where components with distinct spectral indices overlap (\textit{e.g.}, intersections of loops, galaxy and stripe) stand out more clearly in $\rho_1$ than in $\rho_0$, indicating that first-order departures from a power law are preferentially carried by either gradient-like or curl-like structures depending on the local composition. This is precisely the behaviour anticipated in Sec.~\ref{sec:spatial_beta} and \ref{sec:superposition}, where spatially varying $\beta$ and line-of-sight superposition were shown to generate non-zero $\underline{w}_1$. The corresponding $|\Delta\psi_1|$ shows that, even when the static \textit{E}/\textit{B} patterns are asymmetric, the angle evolution of the $E$ and $B$ gradients can remain relatively aligned in some regions, but is strongly decorrelated in others, especially around the Faraday source (6) where complex $\underline{\gamma}$ is expected.

For $n=2$, $\rho_2$ is dominated by two curved components: (4), the aged \textit{B}-like circular source, and (6), the Faraday-rotated \textit{E}-like source. By contrast, $\rho_2$ vanishes for the intrinsically curved but \textit{E}/\textit{B}-balanced source (5). The rest of the sky, where the spectra are close to power laws, contributes little to $\rho_2$, apart from the spectrally steep \textit{B} loop and the spectrally soft \textit{E} loop, both relative to $\overline{\beta}$. These contribute to the second-order moment through the $(\underline{\beta}-\overline{\beta})^2$ term (see Eq.~\ref{eq:w_versus_logtaylor}).

\paragraph{Log–Taylor parameters per field}
\label{sec:logtaylor_params}

The \textit{bottom row} of Fig.~\ref{fig:ratio_EB} summarizes the same information directly in terms of differences between $E$- and $B$-mode log–Taylor parameters: $\log A_{P_E} - \log A_{P_B}$, $|\psi_{P_E} - \psi_{P_B}|$, $\beta_{P_E} - \beta_{P_B}$, $\psi'_{P_E} - \psi'_{P_B}$, $\gamma_{P_E} - \gamma_{P_B}$ and $\psi''_{P_E} - \psi''_{P_B}$. The first two maps essentially reproduce $\rho_0$ and $|\Delta\psi_0|$. The difference $\beta_{P_E} - \beta_{P_B}$ highlights where effective spectral indices differ between $E$ and $B$: large positive and negative values are found on the pure $E$ and $B$ loops (1 and 2), while the stochastic background (8) shows values close to zero, consistent with nearly symmetric \textit{E}/\textit{B} contributions.

The difference in angle gradients, $\psi'_{P_E} - \psi'_{P_B}$, is generally modest, confirming that to first order the angle evolution in $E$ and $B$ is often similar, except around regions where several mechanisms combine, such as the Faraday-active source (6) and overlap zones of multiple components. The curvature differences $\gamma_{P_E} - \gamma_{P_B}$ and $\psi''_{P_E} - \psi''_{P_B}$ isolate precisely the regions where second-order behaviour differs between the two parities;  they are dominated by the $B$ aged and $E$ Faraday rotated circular sources (4–6).

\paragraph{Spectral complexity}
\label{sec:spectral_complex}

In this section, we use the log--Taylor parameters to characterize spectral complexity, in the sense discussed in Sec.~\ref{sec:relations_logtaylor_moments}: departures from a simple local power law, and the spatial variability of the effective low-order spectral parameters. In practice, this can be determined directly from the spatial structure and amplitude of $\underline{\beta}_X$ and $\underline{\gamma}_X$ in Fig.~\ref{fig:all_fields_logtaylor_TM}.

{For this particular toy model, the total polarization field $\underline{P}$ is globally somewhat simpler than its spin-preserving $E$- and $B$-family projections. The maps of $\beta_P$, $\gamma_P$, $\psi'_P$ and $\psi''_P$ contain large regions that are nearly uniform or close to zero, whereas their $\underline{P}_E$ and $\underline{P}_B$ equivalents display more small-scale structure, in particular along the Galactic band and in regions where several components overlap. This behaviour is expected from the non-locality of the $E/B$ projection: as discussed in Sec.~\ref{sec:spatial_beta}, the projected fields effectively combine emission from neighbouring regions with different spectral behaviours, thereby generating additional effective moments even when the original field is locally simpler.}

{The more important lesson, however, is that this ordering is not universal. The toy model deliberately contains regions where an $\textit{E}$-dominated emitter and a $\textit{B}$-dominated emitter overlap on the sky while having different spectral indices. At the crossings of the $\textit{E}$ and $\textit{B}$ loops, the full field $\underline{P}$ is more spectrally complex than either $\underline{P}_E$ or $\underline{P}_B$ taken separately. This is precisely the coexistence effect described in Sec.~\ref{sec:coexistence_EB}: when components with different SEDs also have different parity content, combining them into $\underline{P}$ mixes their spectra, whereas the separated fields partially disentangle them. In these regions, $\underline{P}_E$ and $\underline{P}_B$ are visibly simpler than $\underline{P}$ in terms of $\gamma$, $\psi'$, and $\psi''$.

This effect is central to the motivation of the paper. It shows that analysing spectral behaviour in $\underline{P}_E$ and $\underline{P}_B$ is not merely a formal exercise: in physically plausible situations, the parity-separated spin-2 fields can be the more economical variables to model. The CMB-oriented tests of Sec.~\ref{sec:applications_cmb} are designed to exploit exactly this possibility.}

Finally, the scalar field $\underline{S}=E+iB$ behaves differently. The maps of $\beta_S$ and $\gamma_S$ in the bottom row of Fig.~\ref{fig:all_fields_logtaylor_TM} show strong variations and ring-like structures around most components, and the corresponding phase derivatives are rarely close to zero. This is consistent with the conceptual discussion of Sec.~\ref{sec:EBseparating_complex_op}: although $\underline{S}$ contains the same information as $\underline{P}$ on the full sky, its amplitude and phase are not a polarization amplitude and angle. Its spectral parameters are therefore harder to interpret physically, and in this toy model they are also more structured than those of the spin-2 fields.

These conclusions are drawn from an idealized and highly structured toy sky, so the quantitative ordering between $\underline{P}$, $\underline{P}_E$, $\underline{P}_B$, and $\underline{S}$ will depend on the morphology and physical content of the actual foregrounds. The robust conclusion is instead methodological. Beside validating the theoretical framework of Sec.~\ref{sec:predictions_from_physics}, the toy model shows that the field with the simplest spectral description is sky-dependent. In practice, diagnostics such as the present log--Taylor comparison, or the angle-stability test described in Sec.~\ref{sec:angle_stability}, can be applied to data to determine whether the total polarization field or the $E/B$-separated spin-2 fields provide the more economical and physically transparent spectral representation.

\subsubsection{Performance with a finite number of frequency channels}
\label{ssec:finite_channels_performance}

In order to assess how well the different spectral parametrizations recover the underlying frequency dependence when only a few channels are available, we designed the following test on the toy model.  We select three frequencies, $\nu_0=10$ (the pivot) and $\nu=\{5,15\}$, and evaluate the full toy model at these three frequencies to obtain the ``truth'' maps for the complex polarization fields $\underline{P}$, $\underline{P}_E$, $\underline{P}_B$, which are the focus of this discussion.

For each field $\underline{X}\in\{\underline{P},\underline{P}_E,\underline{P}_B\}$, we then use the map-space predictions for its second-order complex log-Taylor parameters or complex moments (predictions that we have derived in Sec.~\ref{sec:predictions_from_physics} and validated in Sec.~\ref{sec:spec_pred_validation} in the near-frequency setup) to extrapolate the pivot-frequency value $\underline{X}_{\nu_0}$ to $\nu=\{5,15\}$ (with SEDs Eqs.~\ref{eq:complex_log_taylor} or \ref{eq:moment_expansion} \textit{resp.} for the log-Taylor or moment expansions).  This yields approximate model maps $\underline{X}^{\rm model}_\nu$ based on (i) a truncated log-Taylor expansion and (ii) a truncated moment expansion, each retaining three spectral parameters per pixel. 
\begin{figure}
    \centering
    \includegraphics[width=\linewidth]{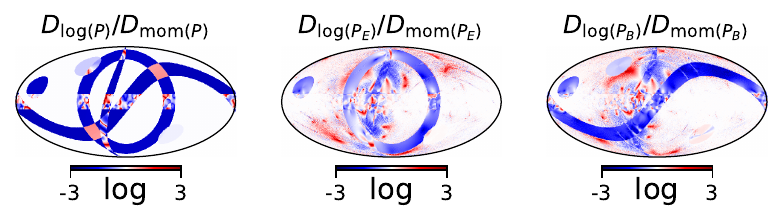}
    \caption{Maps of the decimal logarithm of the ratio between the log-Taylor and moment reconstruction errors, evaluated at $\nu=\{5,10,15\}\,\mathrm{GHz}$ with pivot $\nu_0=10\,\mathrm{GHz}$.  From left to right the panels show $\log_{10}\!\big[D_{\log(P)}/D_{\mathrm{mom}(P)}\big]$, $\log_{10}\!\big[D_{\log(P_E)}/D_{\mathrm{mom}(P_E)}\big]$ and $\log_{10}\!\big[D_{\log(P_B)}/D_{\mathrm{mom}(P_B)}\big]$, where $D_X$ is defined by Eq.~\ref{eq:D_complex}.}
    \label{fig:wide_lin_ratioD}
\end{figure}

The quality of these reconstructions is quantified using the criterion defined by 
Eq.~(25) of \citet{Chluba:2017rtj}, that we generalize to complex polarization fields. For a given complex field $\underline{X}_\nu$ and its approximation $\underline{X}^{\rm model}_\nu$, we define
\begin{equation}
\label{eq:D_complex}
D_{\underline{X}}
=
\sqrt{
\frac{1}{N_{\rm ch}}
\sum_{i=1}^{N_{\rm ch}}
\left|
\frac{\underline{X}^{\rm model}_{\nu_i}}
     {\underline{X}^{\rm input}_{\nu_i}}
-1
\right|^2
},
\end{equation}
where the sum runs over the three frequency channels and the modulus is in the complex plane.  Applied to the second-order log-Taylor and moment models in the three fields, this yields per-pixel maps $D_{\log(P)}$ and $D_{\mathrm{mom}(P)}$, $D_{\log(P_E)}$ and $D_{\mathrm{mom}(P_E)}$, and $D_{\log(P_B)}$ and $D_{\mathrm{mom}(P_B)}.$

\begin{figure*}
\centering
    \includegraphics[width=0.86\linewidth]{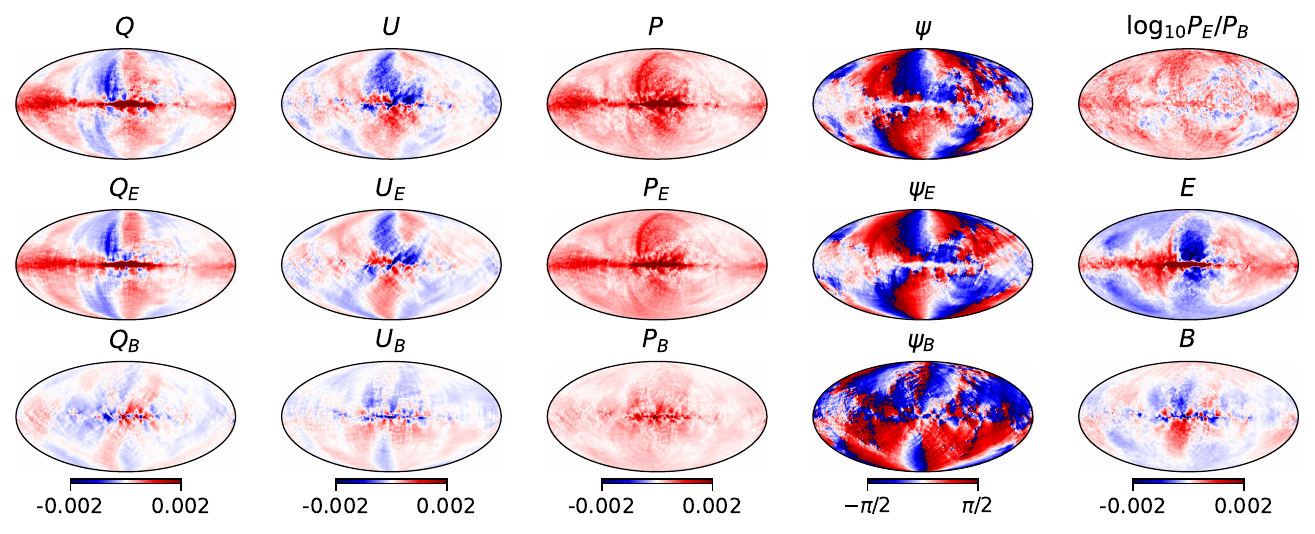}
    \caption{Similar to Fig.~\ref{fig:toy_model_maps}, but with \texttt{PySM s5} at 10 GHz. {Filamentary coherent structures in s5 are primarily $E$-like as seen in $P_E$, while ``blobby'' structures populate $P_B$ (this is in contrast to the corresponding scalar fields where these features are less distinct).}} 
    \label{fig:pysm_morpho_maps}
\end{figure*}

Fig.~\ref{fig:wide_lin_ratioD} displays the decimal logarithm of the ratio between the log-Taylor and moment errors, $\log_{10}\!\left[\frac{D_{\log(P)}}{D_{\mathrm{mom}(P)}}\right]$, $\log_{10}\!\left[\frac{D_{\log(P_E)}}{D_{\mathrm{mom}(P_E)}}\right]$ and $\log_{10}\!\left[\frac{D_{\log(P_B)}}{D_{\mathrm{mom}(P_B)}}\right],$ for the three complex fields.  Blue regions correspond to $D_{\log}<D_{\mathrm{mom}}$, \textit{i.e.}\ a better reconstruction from the log-Taylor parametrization, while red regions indicate the opposite.

For this particular toy model and choice of physical components, the log-Taylor expansion systematically outperforms the whole-sky--referenced moment expansion when both are truncated at second order and constrained by only three frequency channels.  This trend is seen consistently in $\underline{P}$, $\underline{P}_E$ and $\underline{P}_B$, suggesting that, in this regime, the complex log-Taylor description is better suited to capture the detailed spectral behaviour of the polarized sky than the corresponding low-order moment expansion around a single reference index $\overline{\beta}$.

This result would differ if one were to define a reference spectral index $\overline{\beta}$ on a pixel-by-pixel basis (for instance as the mean spectral index of the components along the line of sight). Such a choice, however, would break the full-sky linearity of the moment expansion, which is a key property exploited throughout this work. The quantitative outcome is also sensitive to the details of the toy model, in particular to the number of components along the line of sight, as well as to the choice and spacing of the frequency channels. For example, \citet{Chluba:2017rtj} showed that, when more spectral parameters and more observing frequencies are available, the (real) moment expansion converges more rapidly than the (real) log-Taylor expansion. Although the setup considered here differs significantly, our results are not in tension with theirs: they also obtain that, in regimes with a limited number of spectral parameters and frequency channels, the real log-Taylor expansion can indeed perform comparably to, or even slightly better than, the real moment expansion.

\subsection{A more five{morphologically} realistic sky model}
\label{sec:more_realistic}

We now turn to a more realistic synchrotron sky, using the \texttt{s5} model of \texttt{PySM3} (\citealt{Thorne_2017, Zonca_2021, Pan-ExperimentGalacticScienceGroup:2025vcd}, denoted simply \texttt{PySM} hereafter), evaluated at a pivot frequency of 10\,GHz. We first recall the morphology of the \texttt{s5} model in Sec.~\ref{sec:s5_maps_central_freq}, with particular emphasis on the \textit{E}/\textit{B}-separated fields.

In this baseline model the polarization is, by construction, a rigid-angle power law in $P$: at each pixel, $Q_\nu$, $U_\nu$ and $P_\nu$ share the same spectral index and exhibit no intrinsic curvature. In the language of Sec.~\ref{sec:predictions_from_physics}, the sky is a pure power law in $\underline{P}$ with $\underline{\gamma}=0$ and all higher-order log--Taylor parameters vanishing. A log--Taylor fit around a pivot frequency should therefore return a single (real) spectral tilt and vanishing higher-order and angle-evolution parameters.

Any non-zero moments or curvature found in transformed fields must thus originate from the non-fully-local \textit{E}/\textit{B} operators rather than from the input SEDs. Because $\underline{L}_E$ and $\underline{L}_B$ are non-fully-local convolutions on the sphere, they mix structures with different morphology and orientation, exactly as discussed in Secs.~\ref{sec:EBseparating} and \ref{sec:superposition}. As a result, $P_E$, $P_B$ and $S$ are not guaranteed to remain perfect power laws even when $P$ is. This constitutes spectral complexity induced by the \textit{E}/\textit{B} transform itself, which we illustrate in Sec.~\ref{sec:spec_complexity_in_pysm_hist}. We then demonstrate in Sec.~\ref{sec:finite_freq_channels_pysm} the complementarity of the two spectral expansions when only a limited number of frequency channels is available.

\subsubsection{Central frequency \textit{E}/\textit{B}-separated maps (\texttt{PySM s5})}
\label{sec:s5_maps_central_freq}

As a complement to the toy-model illustration of Sec.~\ref{sec:central_freq_tm}, we now show the morphology of the main polarization fields for the \texttt{PySM s5} model at 10 GHz.

Fig.~\ref{fig:pysm_morpho_maps} displays full-sky maps of $(Q,U)$, the polarization amplitude $P$, the polarization angle $\psi$, and their \textit{E}/\textit{B}-separated counterparts, including $P_E$, $P_B$, $\psi_E$, $\psi_B$, and the scalar fields $E$ and $B$. While the Stokes maps encode the full information, their morphology is strongly dependent on the chosen coordinate frame and does not readily isolate physically distinct structures (and even less morphologically distinct parity components). In contrast, the \textit{E}/\textit{B}-separated spin-2 fields provide a direct and geometrically meaningful decomposition of the signal.

{In particular, coherent large-scale features such as loops and filamentary structures, which dominate the total polarized intensity $P$, are largely captured by $P_E$ and are strongly suppressed in $P_B$. The $E$-family fields therefore appear well suited to isolate astrophysical loop-like structures and to address Galactic-science questions. Conversely, the $B$-family field, from which coherent $E$-mode structures have been filtered out, is of particular interest for tensor-to-scalar ratio searches. 
}

\begin{figure*}
    \centering
    \includegraphics[width=0.68\linewidth]{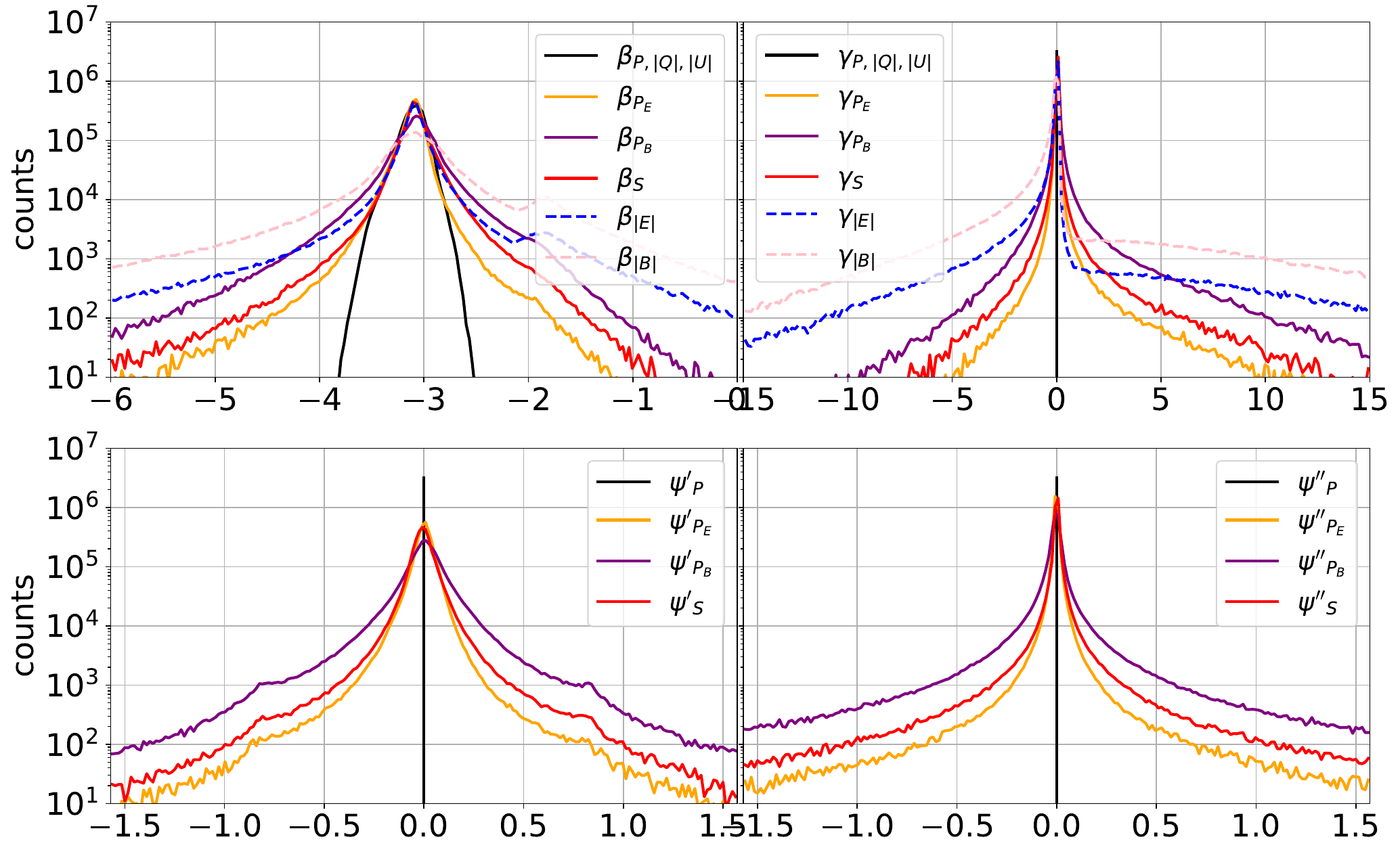}
    \caption{Pixel histograms (logarithmic $y$-axis) of fitted spectral parameters in the \texttt{PySM s5} model. Top-left: spectral tilt $\beta_X$. Top-right: curvature $\gamma_X$. Bottom-left: first angle derivative $\psi'_X\equiv d\psi_X/ds$. Bottom-right: second angle derivative $\psi''_X\equiv d^2\psi_X/ds^2$. Curves are shown for $X\in\{P,\ P_E,\ P_B,\ S,\ |E|,\ |B|\}$; the $P$ curve also overlaps the corresponding distributions for $|Q|$ and $|U|$ in this model. The vertical line marks zero in the $\gamma$, $\psi'$ and $\psi''$ panels. {We see that, although the input \texttt{s5} sky is spectrally rigid in $P$, non-fully-local $E/B$-derived fields acquire effective spatially varying spectral parameters, with the largest deformations appearing in $|E|$ and $|B|$.}}
    \label{fig:summary_spectral_s5}
\end{figure*}

The resulting, more fragmented, $\textit{B}$-family morphology, illustrated by $P_B$ in Fig.~\ref{fig:pysm_morpho_maps}, may appear unfamiliar from a traditional CMB-analysis perspective. However, standard map-level operations can be applied directly to this $\textit{B}$-family foreground field, including local component separation, mask construction using a $P_B$ tracer, and power-spectrum estimation from $(Q_B,U_B)$. {Some of these possibilities are explored in idealized form in Sec.~\ref{sec:applications_cmb}.}

{The complementary nature of the two parity families is also clearly visible in the $\log_{10}(P_E/P_B)$ ratio, which highlights regions dominated by one component or the other. This type of separation cannot be achieved at the level of $Q$ and $U$, whose individual morphologies mix parity contributions and depend on the choice of reference frame. Once again, the $\underline{P}_E$/$\underline{P}_B$ decomposition provides a clearer association between observed structures and their underlying parity-family morphology.
}

five{We do not claim that the visual distinction between coherent filamentary structures in $P_E$ and more patchy structures in $P_B$ is universal. 
The morphology of each family depends on the sky. In the toy model, the input components were deliberately constructed to have controlled $\textit{E}/\textit{B}$ content, so both the spin-preserving fields and the scalar $\textit{E}/\textit{B}$ maps identify the same patterns. The \texttt{PySM s5} map is different: its more realistic morphology contains extended loop-like and filamentary structures, which project predominantly into the $\textit{E}$ family, while the remaining $\textit{B}$-family map is more fragmented.

This behaviour is also plausible for polarized thermal dust, although we do not analyse dust in this paper. In the diffuse interstellar medium, elongated dust structures tend to align with the local magnetic field, and aligned aspherical grains emit polarized radiation whose orientation traces that field. Such coherent alignments naturally generate radial or tangential polarization patterns around filaments, and therefore predominantly $\textit{E}$-family emission, as observed in \textit{Planck} dust maps \citep[see, e.g., Fig.~7 of][]{Liu:2018oqp}. More generally, Fig.~\ref{fig:pysm_morpho_maps} illustrates that realistic foregrounds can contain multiple structures with distinct parity content. If these structures also have distinct SEDs, the total field $\underline{P}$ can be spectrally more complex than $(\underline{P}_E,\underline{P}_B)$, as discussed in Sec.~\ref{sec:coexistence_EB} and illustrated in Sec.~\ref{sec:spectral_complex}.}

five{This possibility, already anticipated conceptually in previous sections, motivates a central objective of this work: constructing spectral analyses that operate on fields which are able to isolate physically distinct foreground components. Such representations are potentially easier to model, more stable across frequency, and therefore particularly attractive for Galactic studies and CMB foreground mitigation.}

\subsubsection{{Spatial variability and spectral complexity of spectral parameters induced by the $E/B$ transform}}

\label{sec:spec_complexity_in_pysm_hist}

{Importantly, the two distinct notions of functional spectral complexity and spatial variability that we have introduced in Sec.~\ref{sec:relations_logtaylor_moments} are not independent. In this section, we will show that, even for a simple rigid-angle power law in $P$, non-zero higher-order spectral parameters can appear in any $E/B$-transformed field. This effect arises from the interplay between the non-local nature of the transform and the spatial variability of $\beta$, as described in more detail in Sec.~\ref{sec:spatial_beta}.}

The full-sky maps of the complex log--Taylor coefficients for the \texttt{PySM s5} model around 10\,GHz are presented in Appendix~\ref{sec:log_taylor_pysm} and briefly discussed there in analogy with the toy-model case (see Fig.~\ref{fig:all_fields_logtaylor_pysm}). Here, we focus instead on the pixel distributions of these parameters, shown in Fig.~\ref{fig:summary_spectral_s5}. This provides a compact way to quantify both the spatial variability of the fitted spectral parameters and the functional spectral complexity induced by the $\textit{E}/\textit{B}$ transforms.

{Fig.~\ref{fig:summary_spectral_s5} contains four histograms. The top-left panel shows the fitted spectral tilt $\beta_X$, the top-right panel the fitted curvature $\gamma_X$, the bottom-left panel the first derivative of the polarization angle with respect to $s=\log(\nu/\nu_0)$, $\psi'_X$, and the bottom-right panel the second derivative, $\psi''_X$. The curves correspond to $X\in\{P,\ P_E,\ P_B,\ S,\ |E|,\ |B|\}$, with the $P$ curve also coinciding with the histograms obtained from $|Q|$ and $|U|$ in this model. In the $\gamma$, $\psi'$ and $\psi''$ panels, the vertical black line marks zero, which is the expected value for a rigid-angle power law.}

{The behaviour of the curves follows a clear hierarchy. The distribution of $\beta_P$ is narrowly peaked around the input \texttt{s5} value, while $\beta_{P_E}$, $\beta_{P_B}$ and $\beta_S$ have progressively broader tails, indicating increasing spatial variability of the effective spectral index after the $E/B$ transform. The broadest $\beta$ distributions are those of $|E|$ and $|B|$. The same ordering is visible in the curvature panel: $\gamma_P$ is concentrated at zero, whereas $\gamma_{P_E}$, $\gamma_{P_B}$ and $\gamma_S$ extend to non-zero values, with $|E|$ and especially $|B|$ showing the widest tails. The angle-derivative panels show an analogous effect. While $\psi'_P$ and $\psi''_P$ vanish identically in the input model, the transformed fields acquire spatially varying non-zero values, with the dispersion increasing from $P_E$ to $S$ and then to $P_B$.}

{This figure therefore illustrates explicitly the link between the two notions of spectral complexity introduced above. The \texttt{s5} model is functionally simple in $P$: locally, its frequency scaling is a rigid-angle power law, with zero curvature and no frequency evolution of the polarization angle. However, $\beta_P$ and $\psi$ vary across the sky. As discussed in Sec.~\ref{sec:spatial_beta}, the non-fully-local $E/B$ transform mixes neighbouring sky directions. Consequently, each transformed pixel receives contributions from regions with different values of $\beta_P$ and different polarization angles. From the point of view of the transformed field, this angular mixing behaves like an effective superposition of different local power laws and angles. It therefore generates non-zero higher-order log--Taylor parameters, such as $\gamma_X$, $\psi'_X$ and $\psi''_X$, even though these parameters are absent in the original field $P$.}

{In this precise sense, Fig.~\ref{fig:summary_spectral_s5} shows both effects at once: spatial variability of the effective parameters, through the widths of the distributions, and induced functional spectral complexity, through the appearance of non-zero curvature and angle-derivative parameters in the transformed fields. The spin-preserving fields $P_E$ and $P_B$ are not free from this effect: they also inherit non-zero effective curvature and angle evolution. However, the induced distributions are narrower than for the spin-0 scalar amplitudes $|E|$ and $|B|$. Thus, the conclusion is not that $P_E$ and $P_B$ remain exact power laws while $E$ and $B$ do not, but rather that the same non-fully-local mechanism produces milder and more interpretable induced deformations in the spin-2 fields than in the scalar-amplitude fields.}

Second, and crucially for practical use, these conclusions are insensitive to the truncation scale of the \textit{E}/\textit{B} transform. Although not shown here, we have repeated the analysis for $\ell_{\max}=95$ and  $\ell_{\max}=11$ (instead of $\ell_{\max}=1535$) and found essentially unchanged distributions. If the deformations were dominated by the finite support of the real-space kernels (``effective beam'' or kernel-truncation effects; see \citealt{Rotti:2018pzi}), one would expect them to widen when decreasing $\ell_{\max}$. This is not observed: the lack of full locality associated with the kernel width is subdominant compared to the intrinsic geometric differences between the $E$- and $B$-like projections of the polarization tensor. In other words, the extra spectral complexity of $P_E$, $P_B$, $S$, and especially of $|E|$ and $|B|$, is mainly a consequence of the \textit{E}/\textit{B} geometry in map space, not an artefact of the implementation.

\subsubsection{Performance with a finite number of channels (\texttt{PySM s5})}
\label{sec:finite_freq_channels_pysm}

\begin{figure}
    \centering
    \includegraphics[width=\linewidth]{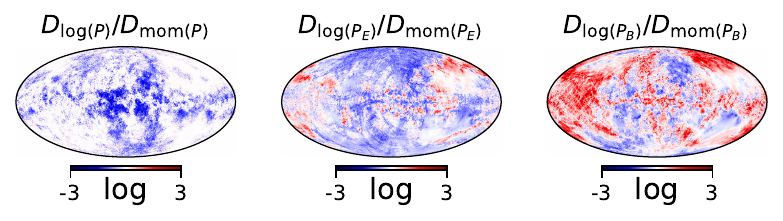}
    \caption{Same as Fig.~\ref{fig:wide_lin_ratioD} but for the \texttt{PySM s5} model. {The figure shows that, for realistic morphology and simple spectral properties, the preferred low-order parametrization can depend on the field: log--Taylor performs better over most of $P$ and much of $P_E$, whereas the moment truncation is often favoured in $P_B$.}}
    \label{fig:wide_lin_ratioD_pysm}
\end{figure}

We now repeat, using the \texttt{PySM s5} synchrotron model, the same three-channel reconstruction test introduced for the toy model in Sec.~\ref{sec:comparaison_toy_model}.  We evaluate the ``truth'' maps at $\nu=\{5,10,15\}\,\mathrm{GHz}$ with pivot $\nu_0=10\,\mathrm{GHz}$, and for each complex field $\underline{X}\in\{\underline{P},\underline{P}_E,\underline{P}_B\}$ we build two second-order reconstructions: (i) a truncated complex log-Taylor model and (ii) a truncated complex moment model (each retaining three complex spectral parameters per pixel; cf.\ Eqs.~\ref{eq:complex_log_taylor} and \ref{eq:moment_expansion}).  The reconstruction quality is quantified by the complex generalization of the diagnostic $D_{\underline{X}}$ defined in Eq.~\ref{eq:D_complex}.  As in the toy-model case, we compare the two parametrizations through the ratio of errors, $\log_{10}\!\big[D_{\log(X)}/D_{\mathrm{mom}(X)}\big]$, shown in Fig.~\ref{fig:wide_lin_ratioD_pysm}.  Blue regions correspond to $D_{\log}<D_{\mathrm{mom}}$ (log-Taylor performs better), while red regions indicate the opposite.

The three panels of Fig.~\ref{fig:wide_lin_ratioD_pysm} display distinct behaviours.  For the total field $\underline{P}$ (left), the map is predominantly blue-to-white, with comparatively weak spatial contrast.  For $\underline{P}_E$ (middle), blue regions still cover a large fraction of the sky, but extended red patches appear.  For $\underline{P}_B$ (right), the pattern reverses: red regions dominate, with only a limited set of blue structures.

These trends can be interpreted in direct continuity with the toy-model discussion.  In the \texttt{s5} model, the underlying emission is constructed to be spectrally rigid in $\underline{P}$ (and hence in $Q$ and $U$): the true frequency dependence is close to a single power law with a fixed polarization angle.  It is therefore unsurprising that, for $P$, a clear preference is seen for the log-Taylor expansion (blue).

The situation becomes more instructive once the non-fully-local $E/B$ operators are applied.  As established earlier (Sec.~\ref{sec:EBseparating} and illustrated in Sec.~\ref{sec:more_realistic}), even a spectrally simple $\underline{P}$ generically acquires effective spectral deformations in $\underline{P}_E$ and $\underline{P}_B$ because the map-space projections mix morphology in a non-fully-local way.  Fig.~\ref{fig:wide_lin_ratioD_pysm} shows that these induced deformations are captured with different success by the two low-order parametrizations: the log-Taylor truncation remains preferable over large areas for $P_E$, while the moment truncation becomes more favourable for $P_B$ over much of the sky.  Compared to the toy model, where the log-Taylor description tended to dominate more uniformly, the \texttt{s5} case therefore highlights a more field-dependent competition once realistic full-sky morphology and $E/B$-induced deformations are present.

Overall, this test confirms that (i) the ranking between parametrizations is not purely formal but can depend on which derived field is modelled, and (ii) for realistic skies the $E$- and $B$-projected fields need not share the same ``best'' low-order spectral description, even when the underlying $\underline{P}$ is close to a rigid power law.  This motivates carrying both descriptions forward as practical tools, and, crucially, testing them directly on the data in any three-band analysis, rather than assuming \emph{a priori} that one truncation will be uniformly optimal on all parity channels.

\section{Further illustrations for CMB-oriented analyses}
\label{sec:applications_cmb}

{The previous sections have established the formal and physical interpretation of the spin-2 $E/B$-family fields. We now illustrate why this representation can be important for CMB-oriented analyses. The aim is not to build a complete analysis pipeline, but to clarify a sequence of practical choices that arise in map-space analyses. The first step is to determine in which field the foreground is spectrally simplest, or equivalently which limited-parameter model best describes the data (Sec.~\ref{sec:distinguishing_modelling_pysm}). This choice then informs downstream analysis decisions: which parametrization to adopt in sky modelling, in map-space parametric component-separation methods \citep{Stompor:2008sf, Rizzieri:2025mwo}, or in constrained ILC methods \citep{Remazeilles:2020rqw}; in which field an ILC should minimize variance (Sec.~\ref{sec:EB_separated_ILC}); and from which field one should build masks or estimate $BB$ power spectra (Sec.~\ref{sec:masking_PEPB}). These questions show that the $E/B$ transform is not merely a relabelling of the information contained in $Q/U$, in the sense that it can lead to different modelling, cleaning, and masking strategies.}

{Most of the potential gains of using this approach are expected when the foreground sky is not accurately described by a rigid-angle power law in $P$. If the true sky were precisely described by a power law in $\underline{P}$ at every pixel, then there would be little intrinsic advantage in modelling, cleaning, or masking in $E/B$-separated map-space fields, given the unavoidable loss of locality induced by the transform. The examples below therefore compare two limiting situations: the standard \texttt{PySM s5} model, where the synchrotron frequency scaling is rigid in $P$, and a modified version in which the simple power-law behaviour is instead imposed separately in $P_E$ and $P_B$. These idealized cases are not meant to exhaust the possibilities, but rather to show that the optimal map-space strategy depends on where spectral simplicity and residual foreground power reside.}

\subsection{Application 1: Distinguishing between a power law in $P$ and power laws in $(P_E,P_B)$: modelling and extrapolation}
\label{sec:distinguishing_modelling_pysm}

The preceding Sec.~\ref{ssec:finite_channels_performance} addressed a ``within-field'' question: given a fixed complex field $\underline{X}\in\{\underline{P},\underline{P}_E,\underline{P}_B\}$ and only three frequency channels, which low-order spectral parametrization (complex log--Taylor or complex moments) most accurately transports that same field across frequency? We now turn to a complementary and more operational question: \emph{in which field is the foreground spectrally simplest, and therefore most naturally modelled?} In the total polarization field $\underline{P}$, or in its spin-2 \textit{E}/\textit{B}-separated components $(\underline{P}_E,\underline{P}_B)$?

{Previous sections and particularly Sec.~\ref{sec:spec_complexity_in_pysm_hist} showed that the $\textit{E}/\textit{B}$ projections are non-fully local and can redistribute spectral complexity. Thus, a rigid-angle power law in $\underline{P}$ generally induces effective spectral deformations in $\underline{P}_E$ and $\underline{P}_B$. Conversely, Sec.~\ref{sec:coexistence_EB} showed that the opposite situation is physically plausible: if co-spatial emitters have different parity content and different SEDs, then $\underline{P}_E$ and $\underline{P}_B$ can be closer to simple power laws than their sum $\underline{P}$. Real data are expected to lie between these two limiting cases.}

This motivates two limiting hypotheses:
\begin{enumerate}
    \item \emph{Rigid-angle power law in $\underline{P}$.} The Stokes fields $(Q,U)$, or equivalently $P$, share one power-law SED per pixel. The polarization angle is independent of frequency, and any spectral complexity in $\underline{P}_E$ or $\underline{P}_B$ is induced by the non-fully local projection.
    \item \emph{Rigid-angle power laws in $(\underline{P}_E,\underline{P}_B)$.} The two spin-preserving parity fields have independent power-law SEDs. The total field $\underline{P}=\underline{P}_E+\underline{P}_B$ can then have a frequency-dependent polarization angle and non-zero curvature through the changing relative weights of the two parity families, as in the coexistence mechanism of Sec.~\ref{sec:coexistence_EB}.
\end{enumerate}
A practical question is whether a small number of frequency channels can determine which hypothesis is closer to a given sky realization.

To test this, we compare two extreme \texttt{PySM}-based models:
\begin{description}
    \item[\textbf{Standard \texttt{s5}:}] $Q$, $U$, and $P$ share the same power-law SED at each pixel. Any spectral complexity in $P_E$, $P_B$, or $S$ is induced by the $E/B$ transform.
    \item[\textbf{Modified \texttt{s5}-like model:}] $P_E$ and $P_B$ are assigned independent power-law SEDs with indices $\beta_{P_E}$ and $\beta_{P_B}$, while $P$ is allowed to become spectrally complex after recombining the two parity families.
\end{description}
In both cases the morphology at the pivot frequency is kept identical, so that the only difference is how the sky is propagated in frequency.

We use three noiseless frequency channels: two bracketing frequencies, $\nu_{\rm low}=5~{\rm GHz}$ and $\nu_{\rm high}=15~{\rm GHz}$, and one intermediate channel, $\nu_{\rm mid}=10~{\rm GHz}$. For each sky model we compare two extrapolation schemes:
\begin{description}
    \item[\textbf{Hypothesis P:}] fit a single spectral index $\beta_P$ from the low and high frequencies, extrapolate $Q$ and $U$ to $\nu_{\rm mid}$ under the power-law-in-$P$ assumption, and then apply the $E/B$ transform to obtain $P_E$ and $P_B$.
    \item[\textbf{Hypothesis EB:}] first transform each frequency map to $P_E$ and $P_B$, fit independent indices $\beta_{P_E}$ and $\beta_{P_B}$ from the low and high frequencies, extrapolate $P_E$ and $P_B$ to $\nu_{\rm mid}$, and finally recombine to $Q$ and $U$.
\end{description}
For each of the four combinations of sky model and hypothesis, we compute the mean-squared prediction error at the intermediate frequency for $Q$, $U$, $P$, $Q_E$, $U_E$, $P_E$, $Q_B$, $U_B$, and $P_B$\footnote{{This diagnostic is conceptually close to a standard model-comparison exercise, and in real data it should be implemented with a likelihood, $\chi^2$, or Bayesian evidence that accounts for noise and instrumental covariance. Its purpose here is more specific and deliberately simpler: we compare two \emph{field-level} hypotheses, namely whether spectral simplicity is better imposed before or after the spin-preserving $E/B$ projection. This is not equivalent to fitting independent real power laws to $Q$ and $U$, since $Q$ and $U$ are basis-dependent, sign-changing quantities and do not preserve the coherent amplitude--angle structure of the polarization field. The test is therefore best viewed as a \emph{noiseless} proof of principle for deciding which geometrical representation should be used before applying a full statistical model comparison to (noisy) data.}}.

\begin{figure}
    \centering
    \includegraphics[width=\linewidth]{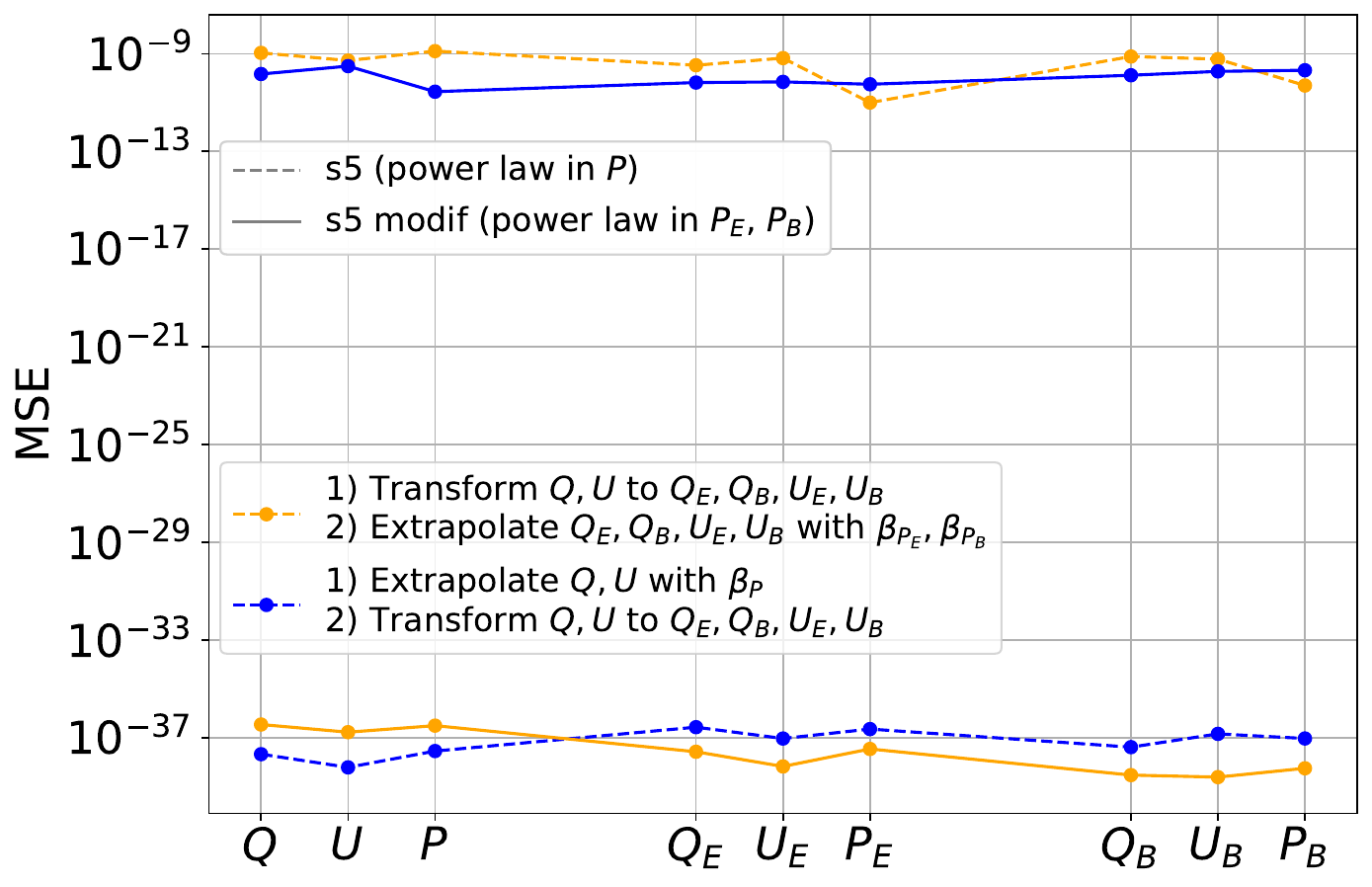}
    \caption{Mean-squared prediction error at the intermediate frequency, $\nu_{\rm mid}=10~{\rm GHz}$, for $Q,U,P,Q_E,U_E,P_E,Q_B,U_B,P_B$, when extrapolating from $\nu_{\rm low}=5~{\rm GHz}$ and $\nu_{\rm high}=15~{\rm GHz}$. Blue curves use Hypothesis P, i.e. a single power law in $P$. Orange curves use Hypothesis EB, i.e. independent power laws in $P_E$ and $P_B$. Dashed lines correspond to the standard \texttt{s5} sky, whose true SED is a power law in $P$. Solid lines correspond to the modified \texttt{s5}-like sky, whose true SED is a power law in $(P_E,P_B)$.}
    \label{fig:prediction_hypotheses}
\end{figure}

Fig.~\ref{fig:prediction_hypotheses} shows the result. The horizontal axis lists the fields in which the prediction error is evaluated, while the vertical axis gives the mean-squared error on a logarithmic scale. For the standard \texttt{s5} sky (dashed curves), the blue curve lies many orders of magnitude below the orange one for all fields: the power-law-in-$P$ hypothesis correctly captures the input model. For the modified sky (solid curves), the situation is reversed: the orange curve lies at the numerical floor, while the blue curve gives much larger errors. Thus the extrapolation scheme that matches the field in which the sky is spectrally simple is selected unambiguously.

Even with only three noiseless channels, the two hypotheses are therefore easily distinguishable. Real observations will of course include noise, bandpass uncertainties, and additional systematics, but the basic diagnostic remains valid: by comparing prediction errors under different parametric assumptions, one can test whether the data favour a description closer to a power law in $P$ or to power laws in $(P_E,P_B)$. We emphasize that the latter should not be interpreted as an expectation that real skies obey exact power laws in $P_E$ and $P_B$.

{This kind of test has direct practical implications. The choice of field in which one imposes spectral simplicity determines where spectral complexity is allowed to reside: in $P$, or in its $E$- and $B$-family components. Applying analogous diagnostics to real multi-frequency data will therefore be an important first step in deciding which map-space model is supported by the sky. The answer need not be unique: different foregrounds, frequency ranges, or sky regions may favour different representations, especially in lines of sight affected by non-standard synchrotron mechanisms, line-of-sight superposition, Faraday effects, or ageing.}

{Such information would be useful for building more realistic frequency-dependent foreground models, for extrapolating foreground templates, and for choosing the parameters used in parametric component-separation methods or constrained ILCs that rely on prior frequency scaling descriptions. In the next two subsections, we show that knowing where spectral complexity resides, and how residual foregrounds are distributed between the two parity families, also informs the choice of field in which to minimize ILC variance and the choice of tracer and target field for masking.}

{
\subsection{Application 2: cleaning the CMB with $E/B$-separated ILCs}
\label{sec:EB_separated_ILC}

We now consider a simple foreground-cleaning exercise based on internal linear combinations (ILCs). The purpose is to ask whether the choice of input field ($P$, $P_E/P_B$, $S$, or scalar $E/B$) matters when the foreground frequency scaling is simple in different representations.

For a generic data vector $d_\nu(p)$ in thermodynamic CMB units, an ILC estimate of the CMB signal can be written as}
\begin{equation}
    \label{eq:ILC}
   \hat s(p)=\sum_{\nu} w^\nu_{\mathcal{D}(p)}\,d_\nu(p),
\end{equation}
{where the weights are estimated over a domain $\mathcal{D}(p)$ containing the pixel $p$. The standard minimum-variance ILC \citep{Tegmark:2003ve,WMAP:2003cmr,Eriksen:2004jg} minimises the variance of $\hat s$ under the unit-response constraint $\sum_\nu w^\nu=1$. This gives
}

\begin{equation}
    \label{eq:weights_ilc}
    w^\nu_{\mathcal{D}} =
    \frac{\sum_{\nu'} (C_{\mathcal{D}}^{-1})^{\nu\nu'}}
         {\sum_{\nu\nu'} (C_{\mathcal{D}}^{-1})^{\nu\nu'}},
\end{equation}
{with empirical frequency--frequency covariance}
\begin{equation}
    \label{eq:domain_covariance}
    C_{\mathcal{D}}^{\nu\nu'}
    =
    \frac{1}{N_{\mathcal{D}}}
    \sum_{p\in\mathcal{D}} d_\nu(p)\,d_{\nu'}^{*}(p),
\end{equation}
{where the complex conjugate is relevant for complex-valued fields and can be dropped for real inputs. In practice we split the sky into $N_{\rm sp}$ \texttt{HEALPix} subpatches and estimate one set of weights per subpatch.

Following the approach of \citet{Fernandez-Cobos:2016iud}, we apply the same variance construction to several real and complex inputs:
\begin{itemize}
    \item $\mathbb{R}P$ILC: real weights from minimizing the variance of the full polarization amplitude $P$;
    \item $\mathbb{R}P_EP_B$ILC: real weights from minimizing the variance of $P_E$ and $P_B$;
    \item $\mathbb{R}S$ILC: real weights from minimizing the variance of $|\underline{S}|=\sqrt{E^2+B^2}$;
    \item $\mathbb{R}EB$ILC: real weights from minimizing the variance of scalar $E$ and $B$;
    \item $\mathbb{C}\underline{P}$ILC: complex weights from minimizing the variance of $\underline{P}$;
    \item $\mathbb{C}\underline{P}_E\underline{P}_B$ILC: complex weights from minimizing the variance of $\underline{P}_E$ and $\underline{P}_B$;
    \item $\mathbb{C}\underline{S}$ILC: complex weights from minimizing the variance of $\underline{S}$.
\end{itemize}

We simulate three noiseless frequency channels at 80, 100, and 120 GHz, containing CMB, thermal dust (\texttt{d0}, which comes with a uniform spectral index of 1.54 and a blackbody temperature of 20 K), and synchrotron. The synchrotron component is either the standard \texttt{s5} model, in which the SED is a rigid-angle power law in $P$, or the modified \texttt{s5}-like model, in which simple power laws are imposed in $P_E$ and $P_B$. To prevent CMB cosmic variance from affecting our conclusions, we apply the methods described below to 100 CMB realizations and average the final results. As a sanity check, we verify that the different ILC configurations recover a CMB $BB$ spectrum whose mean bias is dominated by foreground residuals, with no significant empirical-covariance-induced ILC bias \citep{Delabrouille:2008qd}. For each ILC configuration, we compute the residual foreground $BB$ power spectrum by applying the ILC weights to foreground-only maps, and then compare this residual spectrum to that obtained with the reference $\mathbb{R}P$ILC.}

\begin{figure}
    \centering
    \includegraphics[width=\linewidth]{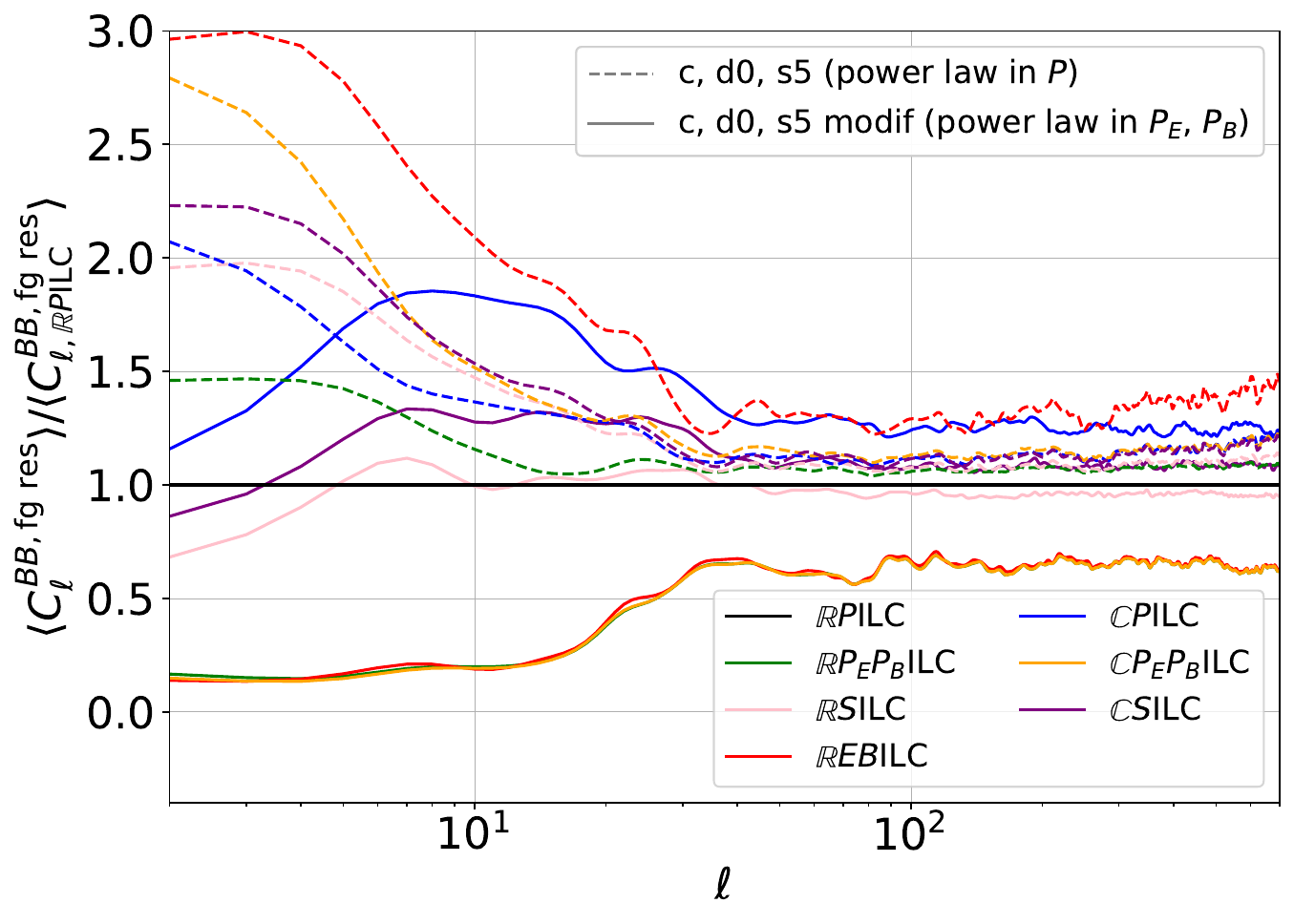}
    \caption{Simulation-averaged foreground residual $BB$ power spectra after several pixel-space ILC configurations, shown as ratios to the mean residual obtained with the reference $\mathbb{R}P$ILC. The horizontal black line at unity is therefore the reference; values below unity indicate a smaller residual foreground bias than $\mathbb{R}P$ILC, and values above unity indicate a larger residual. All ILCs use 12 \texttt{HEALPix} subpatches and three noiseless channels at 80, 100, and 120 GHz, containing CMB, \texttt{d0} dust, and synchrotron. Dashed curves correspond to the standard \texttt{s5} sky, whose synchrotron SED is a power law in $P$. Solid curves correspond to the modified \texttt{s5}-like sky, whose synchrotron SED is a power law in $(P_E,P_B)$. The colours denote the ILC configurations listed in the text. For visualization purposes, the curves have been smoothed with a Gaussian taper in $\ell$ space of $\sigma_\ell=2$. The separated-field ILCs improve the residuals only when the simulated foreground SEDs are simple in $(P_E,P_B)$ (solid lines), illustrating that the optimal cleaning field is sky-dependent.}
    \label{fig:PEPBILC}
\end{figure}

\begin{figure*}
    \centering
    \includegraphics[width=0.9\linewidth]{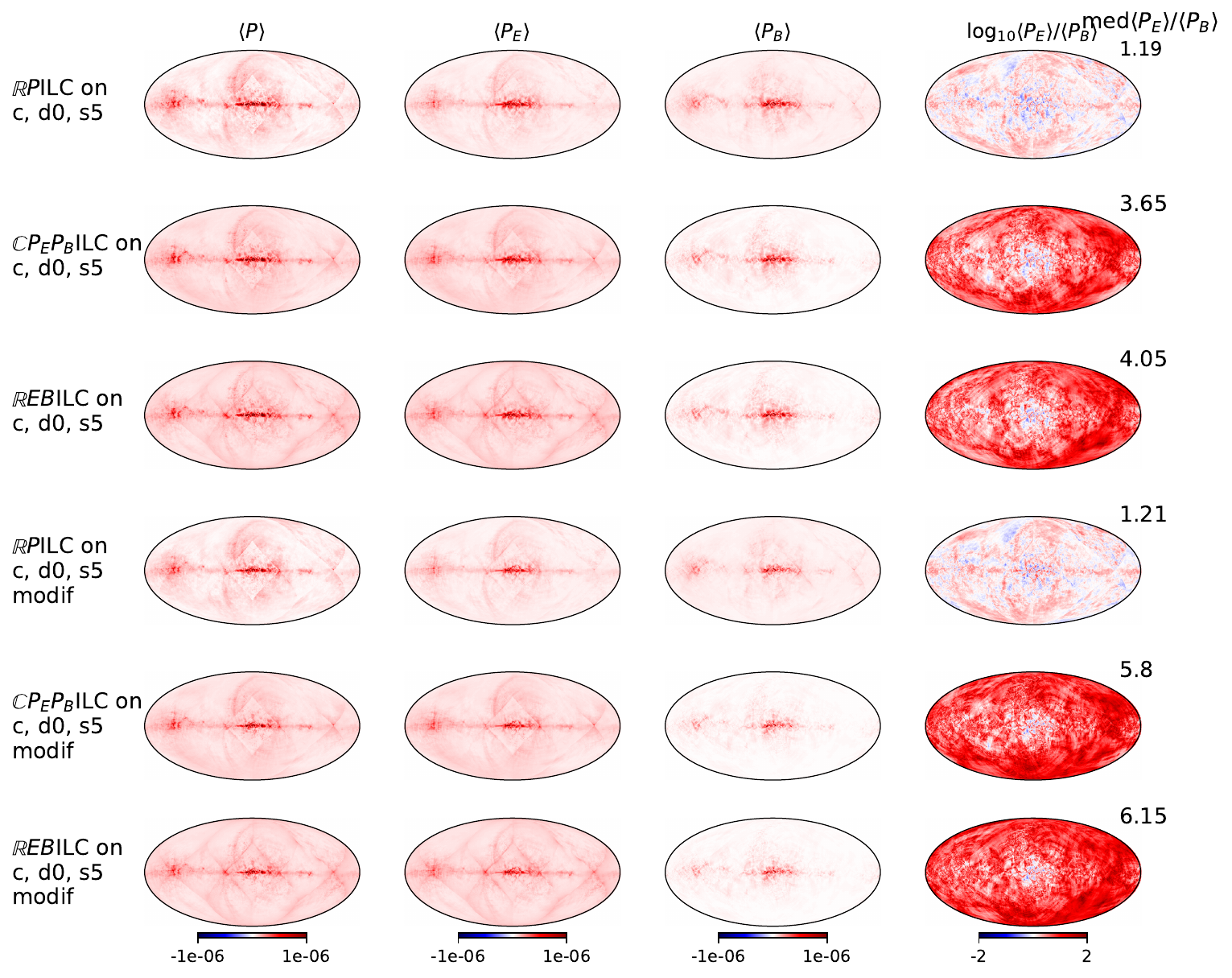}
    \caption{Map-space foreground residuals after ILC cleaning for six representative configurations. Columns show, from left to right, the simulation-averaged residual maps in $\langle P\rangle$, $\langle P_E\rangle$ and $\langle P_B\rangle$, followed by the local ratio $\log_{10}\langle P_E\rangle/\langle P_B\rangle$. The first three rows correspond to $\mathbb{R}P$ILC, $\mathbb{C}\underline{P}_E\underline{P}_B$ILC and $\mathbb{R}EB$ILC applied to CMB+\texttt{d0}+\texttt{s5}. The last three rows show the same three ILC configurations applied to CMB+\texttt{d0}+modified \texttt{s5}. The number shown at the right of each row is the sky median of $\langle P_E\rangle/\langle P_B\rangle$. The figure shows that different ILC choices can yield similar residual $BB$ power while leaving residuals with different map-space parity content: $\mathbb{R}P$ILC residuals are roughly balanced between the two families, whereas the separated-field ILCs leave residuals that are strongly $E$-dominated, especially in the modified \texttt{s5}-like sky.}
    \label{fig:map-space-residuals-postILC}
\end{figure*}

{Fig.~\ref{fig:PEPBILC} displays the ratio
$\langle C_\ell^{BB,{\rm fg\,res}}\rangle /\langle C_{\ell,\mathbb{R}P{\rm ILC}}^{BB,{\rm fg\,res}}\rangle$
as a function of multipole ($\langle.\rangle$ denotes the average over the 100 simulations). Dashed curves correspond to the standard \texttt{s5} sky, and solid curves to the modified \texttt{s5}-like sky. The black line at unity is the reference $\mathbb{R}P$ILC. For the standard \texttt{s5} case, the $P$-based configurations are competitive and the $E/B$-separated ILCs do not provide any improvement; all separated-field variants produce residuals above the reference. In contrast, for the modified model,  the $\mathbb{R}P_EP_B$ILC, $\mathbb{R}EB$ILC and $\mathbb{C}\underline{P}_E\underline{P}_B$ILC configurations all lie below unity at all multipoles, showing reduced foreground residuals relative to $\mathbb{R}P$ILC. On the largest scales, these configurations yield residual levels up to a factor of five lower than the reference case. For the same model, real and complex ILCs based on $\underline{S}$ do not provide a comparable gain: this field is a scalar combination of $E$ and $B$ and effectively re-mixes the two parity families.

The interpretation is consistent with Sec.~\ref{sec:distinguishing_modelling_pysm}. ILCs perform best when their weights act on fields in which the foreground SEDs are comparatively simple. When synchrotron is a rigid-angle power-law in $P$, separating the sky into $E$ and $B$ parities introduces unnecessary effective spectral complexity and does not help. When the simple SEDs instead live in $P_E$ and $P_B$, the separated-field ILCs can exploit this structure and reduce the foreground residuals. This reinforces the main practical message: the usefulness of working in $E/B$-separated fields is sky-dependent and should be tested on the data rather than assumed.

We note that, in this idealized full-sky and noiseless setup, the ILC that minimises the variance separately in the scalar $E$ and $B$ maps performs comparably to the real and complex ILCs applied to the spin-preserving fields: the $\mathbb{R}EB$ILC curve lies at approximately the same level as the $\mathbb{R}P_EP_B$ILC and $\mathbb{C}\underline{P}_E\underline{P}_B$ILC curves. We conclude that, for this minimum-variance ILC test, the difference between scalar and spin-preserving representations does not significantly affect the residual $BB$ power spectrum. We have checked that this remains true for other choices of sky patches. Nevertheless, this difference remains important for interpretation and for subsequent map-level operations: the spin-preserving representation makes it easier to localise, trace, and mask residuals in terms of polarization amplitudes and angles.

To make this distinction explicit, we inspect the post-ILC residuals in map space. Fig.~\ref{fig:map-space-residuals-postILC} shows, for six representative ILC configurations, the residual foreground maps in $\langle P\rangle$, $\langle P_E\rangle$ and $\langle P_B\rangle$, averaged over the 100 simulations, together with the local ratio $\log_{10}\langle P_E\rangle/\langle P_B\rangle$. The first three rows correspond to the standard \texttt{s5} sky, while the last three rows correspond to the modified \texttt{s5}-like sky. Within each group, we compare $\mathbb{R}P$ILC, $\mathbb{C}\underline{P}_E\underline{P}_B$ILC and $\mathbb{R}EB$ILC. 

For the $\mathbb{R}P$ILC rows, the residuals are relatively balanced between the $E$- and $B$-family fields, with median ratios $P_E/P_B\simeq 1.2$. By contrast, the separated-field ILCs leave residuals that are much more strongly concentrated in the $E$ family: the median ratios increase to $\simeq 3.6$--$4.1$ for the standard \texttt{s5} sky and to $\simeq 5.8$--$6.2$ for the modified sky. The $\mathbb{R}EB$ILC rows in particular show larger residual power in $P$ and $P_E$ than the corresponding $\mathbb{C}\underline{P}_E\underline{P}_B$ILC rows, despite their comparable full-sky residual $BB$ spectra in Fig.~\ref{fig:PEPBILC}. This illustrates why power-spectrum residuals alone do not fully describe the structure of the cleaned maps: two ILC configurations can leave similar full-sky $BB$ residuals while producing different parity content in map space.

This is relevant to the upcoming final section in which we go beyond the simple full-sky power spectrum estimation: if the post-cleaning residuals are strongly $E$-dominated, a mask based on the total amplitude $P$ may preferentially remove bright $E$-family residuals that are not the dominant contaminants of the $BB$ spectrum. In that case, a $B$-family tracer such as $P_B$ can target the residual structures that are more directly relevant for $B$-mode power. The next section tests this expectation explicitly.

\subsection{Application 3: masking with $P_B$ on $\underline{P}_B$}
\label{sec:masking_PEPB}

In this final application, we consider foreground masking. The motivation follows directly from the toy model of Fig.~\ref{fig:toy_model_maps} and from the post-ILC residual maps of Fig.~\ref{fig:map-space-residuals-postILC}. If the contamination relevant for $B$-mode power spectra is more compact, or more cleanly isolated, in the $\textit{B}$-family fields than in the full polarization amplitude $P$, then a mask built from a $P_B$ tracer and applied to $(Q_B,U_B)$ may remove foreground power more efficiently than a standard mask built from $P$ and applied to $(Q,U)$. We also compare this spin-preserving strategy to a scalar strategy in which the scalar $B$ map is masked with a $|B|$ tracer.

For a tracer field $T$ and a retained sky fraction $f_{\rm sky}$, we build a mask by removing the brightest pixels of the tracer $T$ and keeping the remaining fraction $f_{\rm sky}$. We compare three target-field/tracer choices:
\begin{itemize}
    \item the standard full-polarization case: masking $(Q,U)$ with a $P$ tracer;
    \item the spin-preserving $\textit{B}$-family case: masking $(Q_B,U_B)$ with a $P_B$ tracer;
    \item the scalar case: masking scalar $B$ with a $|B|$ tracer.
\end{itemize}
The first case is used as the reference. We define the relative improvement as the ratio between the variance left after the reference mask and the variance left after an alternative strategy. This variance is only a simple proxy for the subsequent contamination in $C_\ell^{BB}$ and in the tensor-to-scalar ratio, but it provides a very intuitive and useful first diagnostic. Values below unity therefore indicate an improvement over the standard $P$-based mask.

\begin{figure}
    \centering
    \includegraphics[width=\linewidth]{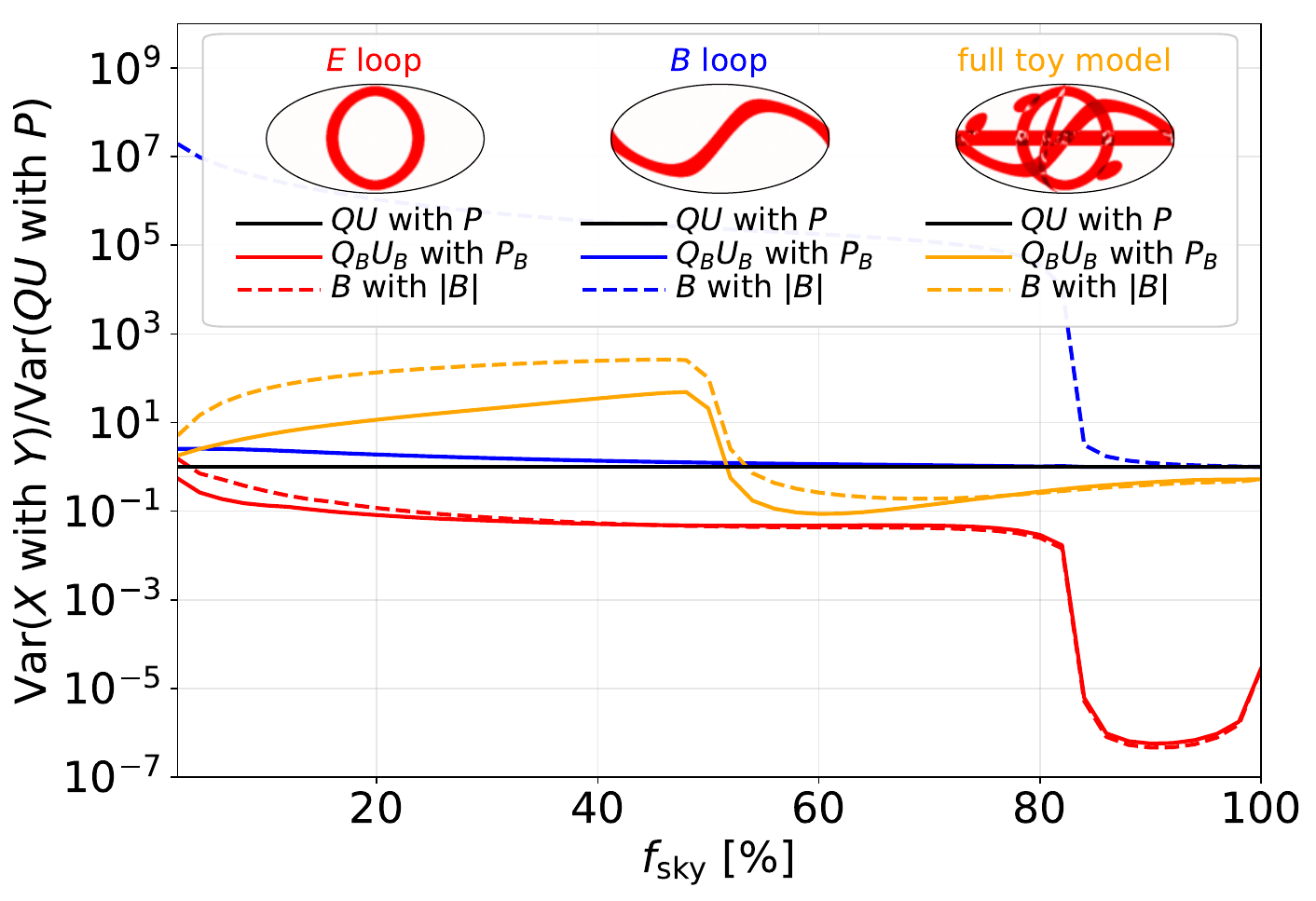}
    \caption{Relative masking improvement for the variance of the masked field in three toy model skies, as a function of retained sky fraction $f_{\rm sky}$. The reference is the variance left after masking $(Q,U)$ with the full polarization amplitude $P$ tracer. Solid curves show the improvement obtained by masking $(Q_B,U_B)$ with $P_B$ tracer; dashed curves show the improvement obtained by masking scalar $B$ with $|B|$ tracer. Red, blue, and orange curves correspond respectively to an $E$-loop-only toy sky, a $B$-loop-only toy sky, and the full toy model. Values above unity indicate smaller residual variance than the reference $P$-based mask. The toy models show that $B$-family masks are useful when the relevant contamination is better localized in $B$-like structures and when the total polarization amplitude is also contaminated by $E$-like structures.}
    \label{fig:full_toy_model_masking_strategy}
\end{figure}

Fig.~\ref{fig:full_toy_model_masking_strategy} first illustrates this idea on the controlled toy skies of Sec.~\ref{sec:simple_toy_model}, where the conclusions can already be anticipated from the morphology shown in Fig.~\ref{fig:toy_model_maps}. In the $E$-loop-only case (red), the foreground is absent from the $B$-family fields, so analysing or masking in $B$-separated fields strongly reduces the relevant variance, in principle removing it entirely. Conversely, in the $B$-loop-only case (blue), masking $B$-separated fields is suboptimal: the separation does not further isolate the contaminating structure compared to the full $P$ field, and the $B$ variance is redistributed non-locally over the sky, especially for the scalar $B$ map. The full toy model (orange), which contains both $E$- and $B$-dominated structures, is less trivial and more informative: at high retained sky fraction, when localized $\textit{B}$-like structures remain to be masked, targeting $(Q_B,U_B)$ or scalar $B$ can outperform targeting $(Q,U)$. Moreover, in this regime, the spin-preserving $(Q_B,U_B)$ strategy is slightly more efficient than the scalar-$B$ strategy.

These toy examples clarify the mechanism. If the sky contains structures that are bright in $P$ but mostly $\textit{E}$-like, a $P$ mask removes regions that are not necessarily the dominant contaminants for $BB$. A $\textit{B}$-family mask can instead focus on the structures that actually contaminate $BB$. The scalar $B$ map and the spin-preserving $(Q_B,U_B)$ fields do not behave identically, however: in this test, masked scalar $B$ retains more contaminating variance than masked $(Q_B,U_B)$ (as can be seen in Fig.~\ref{fig:full_toy_model_masking_strategy}, the dashed curves are consistently higher than the  solid curves). This indicates that $(Q_B,U_B)$ preserves the location of the relevant bright structures more faithfully than scalar $B$, which is precisely the property needed for efficient masking.

Finally, after this proof of concept, we repeat the masking exercise on more realistic foreground residuals from the ILC examples of Sec.~\ref{sec:EB_separated_ILC}. This is closer to the practical situation in which one masks residual contamination after component separation, rather than the input foreground sky. We use the residual map from $\mathbb{C}\underline{P}_E\underline{P}_B$ILC applied to the modified \texttt{s5}-like model, corresponding to the solid orange curve of Fig.~\ref{fig:PEPBILC} and to the fifth row of Fig.~\ref{fig:map-space-residuals-postILC}. In this case the $\textit{B}$-family residuals are more localized than the full-polarization residuals, making it a useful test case for comparing mask tracers and target fields.}

\begin{figure}
    \centering
    \includegraphics[width=\linewidth]{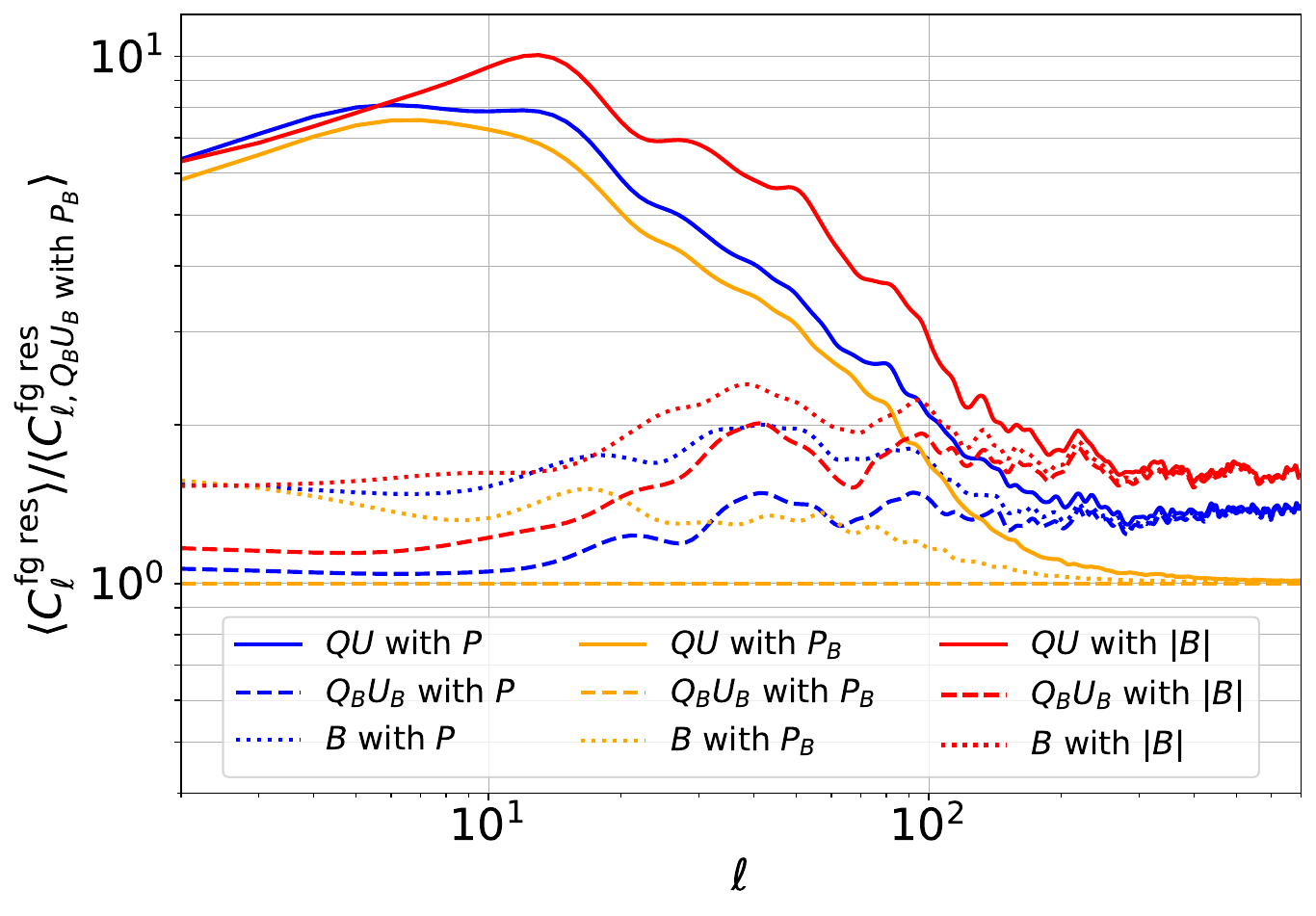}
    \caption{Foreground residual $BB$ pseudo-power spectra after 50\% masking, averaged over simulations and normalized to the reference case in which $(Q_B,U_B)$ are masked with $P_B$ before estimating $C_\ell^{BB}$. The input maps are the residuals obtained after $\mathbb{C}\underline{P}_E\underline{P}_B$ILC cleaning of CMB+\texttt{d0}+modified \texttt{s5}, as described in Sec.~\ref{sec:EB_separated_ILC}. The horizontal black line at unity marks the reference. Blue curves use the full-polarization tracer $P$, orange curves use the $B$-family tracer $P_B$, and red curves use the scalar tracer $|B|$. Solid, dashed, and dotted lines correspond respectively to estimating the residual spectrum from $(Q,U)$, from $(Q_B,U_B)$, and from scalar $B$. Values above unity indicate larger residuals than the reference strategy. For visualization purposes, the curves have been smoothed with a Gaussian taper in $\ell$ space of $\sigma_\ell=2$. For this residual morphology, masking $(Q_B,U_B)$ with $P_B$ gives the lowest residual $BB$ spectrum among the tested strategies.}
    \label{fig:PEPBILC_residuals_masking_strategy}
\end{figure}

{Fig.~\ref{fig:PEPBILC_residuals_masking_strategy} shows the ratio of residual pseudo-spectra, $C_{\ell}^{BB,{\rm fg\,res}}/C_{\ell,{\rm ref}}^{BB,{\rm fg\,res}}$, for the different masking choices, all constructed at a common sky fraction of $f_{\rm sky}=50\%$ and evaluated with \texttt{Xpol}\footnote{\url{https://gitlab.in2p3.fr/tristram/Xpol}}. The orange dashed curve, which corresponds to masking $(Q_B, U_B)$ with a $P_B$ tracer, is the reference and is therefore equal to unity. All other combinations of tracer and target field lie above this reference over the plotted multipole range. For example, around $\ell\simeq 30$, any other strategy increases the masked residual $BB$ power by at least about $40\%$. For this particular residual morphology, masking $(Q_B,U_B)$ with a $P_B$-based mask is therefore the most efficient of the tested strategies for reducing the residual foreground bias in $C_\ell^{BB}$, and hence in the inferred tensor-to-scalar ratio.

The conclusion is not that $P_B$ masks are universally superior. Rather, the optimal mask should be constructed in the field in which the relevant foreground residuals are most localized and most directly connected to the target $BB$ contamination. In particular, when the residuals after component separation are strongly concentrated in the $E$ family, masking $(Q_B,U_B)$ with a $P_B$ tracer can reduce the residual $BB$ bias more efficiently than masking $(Q,U)$ with $P$ or masking scalar $B$ with $|B|$. Perhaps unsurprisingly, our tests confirm that component-separation methods yielding $BB$ residuals that are orders of magnitude smaller than the corresponding $EE$ residuals naturally benefit from masking strategies that explicitly distinguish between the two parity families.

Together, the three applications in this section lead to a coherent practical picture. The $E/B$-separated fields are not automatically preferable in every analysis step. They become useful when the foreground complexity, cleaning residuals, or masking-relevant structures are better organized in the $E$- and $B$-family maps than in the total polarization field. In particular, synchrotron emission is known to contain large coherent loop- and filament-like structures that are predominantly $E$-dominated, a morphology that naturally motivates distinguishing between the two parity families when modelling, cleaning, or masking the sky. This motivates applying the diagnostics developed here to real multi-frequency data before choosing a modelling, cleaning, or masking strategy. A natural next step, beyond the simple pseudo-$C_\ell$ estimation considered here, would be to propagate these ideas into real-space quadratic power-spectrum estimators, such as QML methods \citep{Tegmark:1996qt}, which we leave for future work.}

{\subsection{Scope and limitations: leakage in partial sky}
\label{sec:limitations_partial_sky}

As mentioned in Sec.~\ref{sec:introduction}, all applications presented in this section use full-sky harmonic $E/B$ projectors. This is an intentional simplification. On a cut sky, the separation into $E$- and $B$-family fields is no longer exact, modes become ambiguous and can be misinterpreted. In such a setting, the spin-preserving fields $(Q_E,U_E)$ and $(Q_B,U_B)$, just like the scalar maps $E$ and $B$, should be constructed only after adopting a specific leakage-mitigation prescription, such as purification, inpainting, or forward-modelling of the survey mask. The diagnostics proposed here remain conceptually applicable, but their quantitative performance will depend on that choice.

The results of Secs.~\ref{sec:distinguishing_modelling_pysm}--\ref{sec:masking_PEPB} should therefore be interpreted as full-sky proofs of principle. They show that, when foreground morphology and SEDs are better organized in $\textit{E}$- and $\textit{B}$-family fields than in the total polarization field, modelling, cleaning, and masking can benefit from these representations. Demonstrating the same gains under realistic sky cuts, anisotropic noise, beams, and leakage control is left for future work.}

\section{Summary and conclusions}
\label{sec:conclusions}

{In this work we have developed a unified map--space framework to describe the frequency dependence of polarized synchrotron emission, with particular emphasis on field representations that preserve the spin--2 nature of the signal. Our motivation is practical. The spin-2 $E/B$ decomposition organizes the polarization morphology into two parity families, isolating the $B$-family contribution that is directly relevant for primordial tensor searches from the $E$-family contribution that often traces coherent Galactic structures. This naturally suggests studying spectral behaviour not only in the total complex polarization field $\underline{P}=Q+iU$, but also in its spin-preserving projections $\underline{P}_E$ and $\underline{P}_B$. The central difficulty is that the spin-2 and spin-0 $E/B$ transforms are non-fully-local on the sphere, and can therefore induce apparent spectral deformations in projected fields even when the underlying sky is spectrally simple in $\underline{P}$. This point is important both for Galactic science and for CMB analyses: additional effective spectral complexity increases the number of parameters required for foreground modelling, can destabilize extrapolation or component separation when only a limited number of frequency channels are available, and may change the apparent morphology of foreground structures with frequency.}

\smallskip

{We introduced complex log--Taylor and complex moment expansions for generic complex polarization fields. These parametrizations treat amplitude and polarization angle on the same footing. The complex log--Taylor expansion provides a direct generalization of the familiar amplitude--tilt--curvature description, while the moment expansion is linear in the field and therefore transforms simply under any linear operator. In this language, the closure relation between $\underline{P}$, $\underline{P}_E$ and $\underline{P}_B$ extends to all spectral orders: the moments of the full field are the sum of the corresponding $E$- and $B$-family moments. This property is specific to the spin-preserving decomposition and has no direct analogue for the scalar spin-0 maps $E$ and $B$, or for the complex scalar $\underline{S}=E+iB$.}

\smallskip

{Building on this formalism, we derived analytic predictions for the spectral moments generated by several synchrotron mechanisms: spatial variations of the spectral index, line-of-sight superposition of components with different indices and angles, intrinsic curvature, synchrotron ageing, and internal and external Faraday rotation and depolarization. These mechanisms leave different signatures in the complex spectral parameters. Ageing generates real negative curvature without frequency-dependent angle evolution; Faraday effects can generate negative curvature and rotation of the polarization angle; and line-of-sight superposition can induce both amplitude curvature and frequency-dependent angle changes. These amplitude and angle effects imply frequency-dependent changes in the local polarization pattern and therefore produce effective $\textit{E}\leftrightarrow\textit{B}$ mixing. In addition, distinct emitters may coexist in the same sky region while having both different spectral properties and different parity content. This coexistence provides a simple physical route by which the total polarization field $\underline{P}$ can become spectrally more complex than its $\textit{E}$- and $\textit{B}$-family components separately. The spin-2 $\textit{E}/\textit{B}$ decomposition does not by itself identify the physical origin of a spectral deformation, nor does it create new information beyond $(Q,U)$. Its role is complementary: it projects these spectral signatures into $\textit{E}$- and $\textit{B}$-family morphology, thereby showing whether a given effect is mainly associated with gradient-like structures, curl-like structures, or both.}

\smallskip

{We validated the formalism on a controlled toy model, showing that the predicted complex moments and log--Taylor parameters accurately reproduce those fitted from multi-frequency maps. The same toy model also illustrates how the spin-2 $\textit{E}/\textit{B}$ transform redistributes spectral behaviour between parity families. In much of this particular toy sky, the total field $\underline{P}$ remains relatively simple while $\underline{P}_E$ and $\underline{P}_B$ display richer effective spectral structure, as expected from the non-fully local nature of the projection. However, the opposite situation occurs where $\textit{E}$- and $\textit{B}$-dominated structures overlap with different spectral indices. There, the total field $\underline{P}$ mixes components with different SEDs, whereas $\underline{P}_E$ and $\underline{P}_B$ partially disentangle them. In these regions, the separated fields are visibly simpler than $\underline{P}$ in terms of curvature $\gamma$, angle rotation $\psi'$, and angle curvature $\psi''$. This confirms, in a controlled setting, the coexistence mechanism described above: $\textit{E}/\textit{B}$-separated spin-2 fields can be the simpler spectral variables when different physical emitters have different parity distributions. We further showed that scalar quantities, in particular $|E|$ and $|B|$, display the largest induced spatial variability of spectral parameters. The complex scalar $\underline{S}=E+iB$ is less affected than $|E|$ and $|B|$, but lacks a directly interpretable polarization amplitude and angle. By contrast, the spin-2 fields $\underline{P}_E$ and $\underline{P}_B$ exhibit moderate, though non-negligible, induced spectral deformations while retaining a clear map-space interpretation in terms of amplitudes and angles.}

\smallskip

A more realistic \texttt{PySM} synchrotron sky, spectrally simpler but morphologically richer than the toy model, confirmed this conclusion about the distinction between scalar and spin-2 decompositions. In the standard \texttt{s5} model, where the input sky is a rigid-angle power law in $P$, the spin-2 and spin-0 $E/B$ transforms induce non-zero effective curvature and angle evolution in the projected fields. These spectral deformations are not removed by changing the multipole cut of the transform, indicating that they are not primarily numerical artefacts of the truncation but consequences of the geometry of the decomposition. When only three frequency channels are available, low-order log--Taylor and moment truncations both provide useful descriptions, but their relative performance depends on the field considered. This reinforces the practical message that the optimal low-order parametrization should be tested on the data, rather than assumed to be the same for $P$, $P_E$ and $P_B$.

\smallskip

{We then explored three simple CMB-oriented applications. First, by contrasting a standard \texttt{PySM} model, in which synchrotron is a power law in $\underline{P}$, with a modified model in which simple power laws are imposed in $(\underline{P}_E,\underline{P}_B)$, we showed that three noiseless frequency channels are already enough to discriminate between these two limiting descriptions. This test should be understood as a proof of principle for field-level model comparison. In real data, the preferred representation may even vary across the sky. Applying such diagnostics would be valuable for building more spectrally realistic foreground models, choosing how to extrapolate foreground templates, and selecting the parametrization used in parametric component-separation analyses.}

\smallskip

{Second, we illustrated how internal linear combinations can be applied to different map-space representations: the full polarization field, the spin-2 $E$- and $B$-family fields, the scalar spin-0 maps, or the complex scalar $\underline{S}$. In our idealized examples, ILCs perform best when the weights act on fields in which the foreground SEDs are comparatively simple. If the sky is a rigid-angle power law in $P$, separating into $P_E$ and $P_B$ does not help and can even degrade the residuals. Conversely, when the simple spectral behaviour is imposed in $(P_E,P_B)$, the separated-field ILCs can yield substantially smaller residual $BB$ foreground power than a $P$-based ILC. We also showed that different ILC choices can leave similar full-sky residual $BB$ power while producing residual maps with different parity content and different spatial localization. Thus, the choice of ILC field affects not only the total residual level, but also where the residual foregrounds live on the sky and whether they are predominantly $E$- or $B$-family residuals.}

\smallskip

{Third, we showed that this residual morphology matters for masking. In controlled toy examples, masks based on $B$-family tracers are useful when the contamination relevant for $BB$ is better localized in $B$-like structures than in the total polarization amplitude $P$. In post-ILC residual maps, we found cases in which masking $(Q_B,U_B)$ with a $P_B$ tracer reduces the residual $BB$ spectrum more efficiently than masking $(Q,U)$ with $P$ or masking scalar $B$ with $|B|$. This result should not be interpreted as a universal preference for $P_B$ masks. Rather, it demonstrates the general principle that the mask should be built from the field in which the relevant foreground residuals are most localized and most directly connected to the target $BB$ contamination.}

\smallskip

{Taken together, these applications lead to a coherent practical recommendation. The spin-2 $E/B$-separated fields are not automatically preferable in every analysis step. They become useful when the foreground SEDs, cleaning residuals, or masking-relevant structures are better organized in the $E$- and $B$-family fields than in the total polarization field. This situation is naturally motivated by synchrotron morphologies containing coherent, often $E$-dominated loops or filaments together with more fragmented $B$-family structures, especially when these components have their own independent spectral behaviour. Therefore, before fixing a map-space analysis strategy, one should compare the spectral properties of $\underline{P}$, $\underline{P}_E$, and $\underline{P}_B$ on the data themselves, ideally locally on the sky and at fixed modelling complexity. The relevant question is not whether $\underline{P}_E$ and $\underline{P}_B$ are intrinsically better than $\underline{P}$, but where, and for which foreground conditions, one representation provides a simpler and more interpretable description than another. This is true for Galactic science, where such comparisons can help localize the signatures of physical mechanisms in parity space, and for CMB science, where they can guide foreground extrapolation, parametric component separation, ILC design, masking, and ultimately the control of biases on $C_\ell^{BB}$ and on the tensor-to-scalar ratio.}

\smallskip

{The tests presented in Sec.~\ref{sec:applications_cmb} were deliberately kept simple. They use noiseless mock data, full-sky harmonic projectors, and a limited number of idealized choices. They are therefore not intended to constitute a complete CMB foreground pipeline. Their purpose is to demonstrate that the field representation chosen for modelling, cleaning, and masking can matter in practice, and that this choice can be diagnosed from multi-frequency data. In realistic applications, the same ideas should be embedded in proper statistical model comparison, including noise, beams, bandpasses, sky cuts, anisotropic coverage, and $E/B$ leakage mitigation.} {In particular, the spin-2 fields $(Q_E,U_E)$ and $(Q_B,U_B)$ should be constructed on a cut sky only after adopting an appropriate purification, inpainting, forward-modelling strategy, or large-scale mode filtering. Quantifying the resulting gains for specific surveys is left to future work.}

\smallskip

{In forthcoming work, we will apply this framework to observational synchrotron polarization data, including new \textit{C-BASS} data \citep{C-BASS:2018dwc, C-BASS:2026}, complemented by \textit{S-PASS} \citep{Krachmalnicoff:2018imw, Carretti:2019yzm}, \textit{WMAP} \citep{WMAP:2012fli}, and \textit{Planck} observations \citep{Planck:2018nkj}. The same diagnostics will also be relevant for upcoming CMB analyses using, for example, the \textit{Simons Observatory} \citep{SimonsObservatory:2019qwx} and \textit{LiteBIRD} \citep{LiteBIRD:2022cnt}. In that context, the central goal will be to determine, directly from the data, which field representation offers the simplest foreground description in each sky region and at each stage of the analysis, so as to maximize Galactic interpretability while minimizing residual foreground bias in cosmological constraints.}

\section*{Acknowledgements}

We thank the anonymous referee for their constructive comments and suggestions, which have greatly improved this manuscript.

We thank Alessandro Carones, Ishaque Khan and the \emph{RadioForegroundsPlus} collaboration, as well as members of the BIPAC and CMB-OX in Oxford, for useful comments and discussions. 

We acknowledge support from the Horizon Europe Project RadioForegroundsPlus HORIZON-CL4-2023-SPACE-01, GA 101135036, which is supported in the UK by UKRI grant number 10101603.  We acknowledge partial support by the Italian Space Agency LiteBIRD Project (ASI Grants No. 2020-9-HH.0 and 2016-24-H.1-2018), as well as the InDark and LiteBIRD Initiative of the National Institute for Nuclear Physics, and Project SPACE-IT-UP  by the Italian Space Agency and Ministry of University and Research, Contract Number  2024-5-E.0 and the CMB-Inflate project funded by the European Union’s Horizon 2020 Research and Innovation Staff Exchange under the Marie Skłodowska-Curie grant agreement No 101007633.

We made extensive use of the \texttt{numpy} \citep{vanderWalt:2011bqk}, \texttt{astropy} \citep{Astropy:2018wqo}, \texttt{healpy} \citep{Zonca:2019vzt} and \texttt{matplotlib} \citep{Hunter:2007ouj} Python packages, as well as the \texttt{HEALPix} package \citep{Gorski:2004by}. The AI tool ChatGPT 5.2, accessed through the \textit{Oxford University ChatGPT Edu Workspace}, was used solely for initial language editing and stylistic refinement. We, the human authors, take full responsibility for the content of the manuscript. 

\section*{Data availability}

The (mock) data that support the figures and plots of this paper are available from the corresponding author upon request.

\bibliographystyle{mnras}
\bibliography{references}

\appendix
\section{Why \PabEtitle{} and \PabBtitle{} are spin-2 tensors}

\renewcommand{\thefigure}{\arabic{figure}}

\label{app:details_spin2_prop_PEPB}

We show below that the $E$/$B$ parts of the linear polarization inherit the same spin-2 transformation law as the full polarization tensor $\mathcal{P}_{ab}$.

The linear polarization at a point on the sphere can be represented by the symmetric, trace-free tensor, written in the local orthonormal tangent basis $(\hat e_\theta,\hat e_\phi)$ as Eq.~\ref{eq:polar}. A local rotation of this tangent basis by angle $\alpha$, $(\hat e_\theta,\hat e_\phi)\mapsto(\hat e'_\theta,\hat e'_\phi)=R(\alpha)\,(\hat e_\theta,\hat e_\phi)$ changes the Stokes parameters according to
\begin{equation}
\label{eq:spin2_def}
(Q\pm iU)'(\mathbf n)=e^{\mp 2i\alpha}\,(Q\pm iU)(\mathbf n).
\end{equation}
This is the defining property of a spin-$\mp2$ field on the sphere. Equivalently, $\mathcal P_{ab}$ transforms under the action of the rotation on its tensor indices as a rank-2 symmetric traceless object.

Because $Q\pm iU$ are spin-$\mp2$ fields they admit the spin-weighted harmonic expansion
\begin{equation}
(Q\pm iU)(\mathbf n)=\sum_{\ell m} a_{\ell m}^{\pm 2}\;{}_{\pm 2}Y_{\ell m}(\mathbf n),
\end{equation}
where ${}_{2}Y_{\ell m}$ denotes the spin-$2$ spherical harmonics. The standard $E$ and $B$ harmonic coefficients are defined by the linear combinations,
\begin{align}
a_{\ell m}^{E} &= -\tfrac{1}{2}\bigl(a_{\ell m}^{2}+a_{\ell m}^{-2}\bigr)\quad\mathrm{and}\quad a_{\ell m}^{B} = \tfrac{i}{2}\bigl(a_{\ell m}^{2}-a_{\ell m}^{-2}\bigr).
\end{align}
The inverse transform reconstructs the $E$ and $B$ contributions to the spin-2 fields:
\begin{align}
&(Q_E\pm iU_E)(\mathbf n) = \sum_{\ell m} a_{\ell m}^{E}\;{}_{\pm2}Y_{\ell m}(\mathbf n)\quad \mathrm{and}\\&
(Q_B\pm iU_B)(\mathbf n) = \sum_{\ell m} (\mp i)\,a_{\ell m}^{B}\;{}_{\pm2}Y_{\ell m}(\mathbf n).
\end{align}
From there, we easily verify that $Q\pm iU = (Q_E\pm iU_E)+(Q_B\pm iU_B)$. Furthermore, under a local rotation by $\alpha$ the spin-2 weighted harmonics transform as
\begin{equation}
{}_{2}Y_{\ell m}'(\mathbf n)=e^{-2i\alpha}\;{}_{2}Y_{\ell m}(\mathbf n).
\end{equation}
Therefore each harmonic term in the sums for $Q_E\pm iU_E$ and $Q_B\pm iU_B$ acquires the same phase factor $e^{\mp 2i\alpha}$ as the corresponding term in the full field. Concretely, using the inverse transforms above,
\begin{align}
(Q_E\pm iU_E)'(\mathbf n)
&= \sum_{\ell m} a_{\ell m}^{E}\;{}_{\pm2}Y_{\ell m}'(\mathbf n)
\\&= e^{\mp 2i\alpha}\sum_{\ell m} a_{\ell m}^{E}\;{}_{\pm2}Y_{\ell m}(\mathbf n)
\\&= e^{\mp 2i\alpha}\,(Q_E\pm iU_E)(\mathbf n),
\end{align}
and similarly $Q_B\pm iU_B$.

We then define the $E$ and $B$ polarization tensors as in Eq.~\ref{eq:polar} but replacing $(Q,U)$ by $(Q_E,U_E)$ and $(Q_B,U_B)$ respectively. Since $Q_E\pm iU_E$ and $Q_B\pm iU_B$ each satisfy the spin-2 transformation law $(Q\pm iU)'\!=\!e^{\mp2i\alpha}(Q\pm iU)$, the tensors $\mathcal{P}_{ab}^{(E)}$ and $\mathcal{P}_{ab}^{(B)}$ transform in the same way under local rotations of the tangent frame as the full $\mathcal{P}_{ab}$. Therefore each is a spin-2 (symmetric, traceless) tensor field on the sphere.

\section{Convolution kernels}
\label{app:convolution_kernel}

The real-space kernels that relate $(Q,U)$ to the $E$- and $B$-family fields $(Q_E,U_E)$ and $(Q_B,U_B)$ are derived in \citet{Rotti:2018pzi}. We denote by $\mathcal{K}_{\mathcal I}$ and $\mathcal{K}_{\mathcal D}$ the two radial kernels introduced in their Eq.~(3.23) (under the notation $\,_{\mathcal I}f$ and $\,_{\mathcal D}f$).

\setcounter{figure}{17}
\begin{figure}
    \centering
    \includegraphics[width=\linewidth]{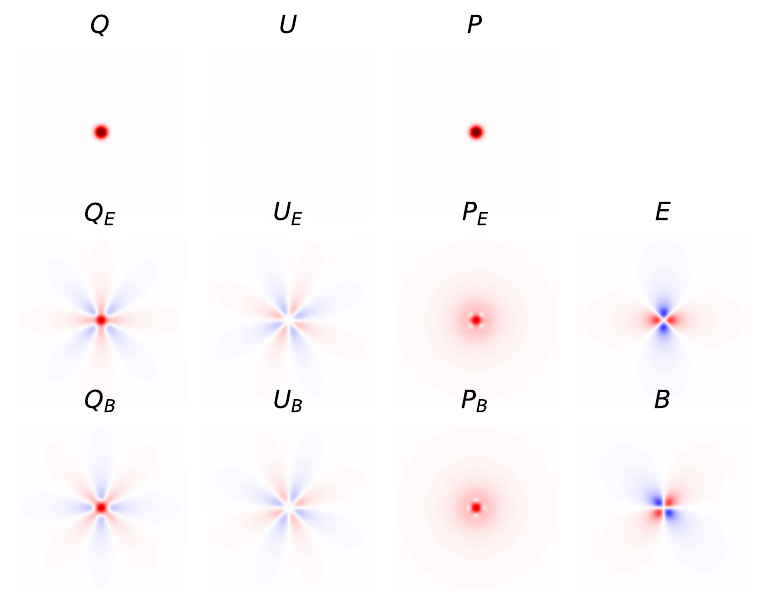}
    \caption{Transformation of a two-dimensional positive Gaussian feature in $Q$ into the various $E$- and $B$-mode fields. The panels show, from left to right, the input Stokes fields $(Q,U,P,\psi)$ followed by the derived $E$-family fields $(Q_E,U_E,P_E,E)$ and the $B$-family fields $(Q_B,U_B,P_B,B)$. The colour map is the same for all panels (blue = negative, red = positive).}
    \label{fig:full_convolution_kernel}
\end{figure}

\begin{figure*}
    \centering
    \includegraphics[width=\linewidth]{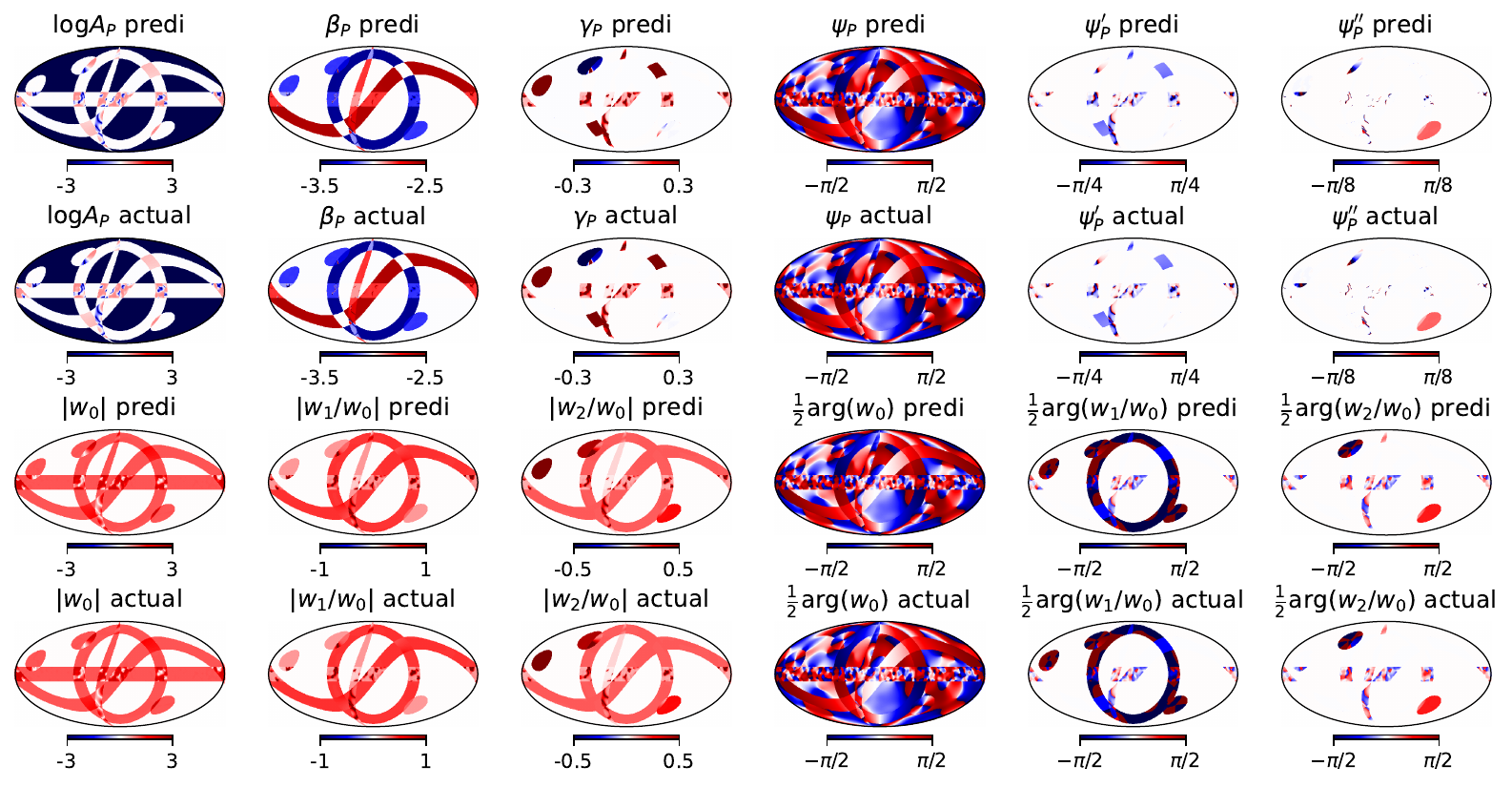}
    \caption{Full-sky Mollweide maps of spectral quantities for the toy-model total polarization field $P$. From left to right, columns show $\log A_P$, $\beta_P$, $\gamma_P$, $\psi_P$, $\psi'_P$ and $\psi''_P$ (first and second columns of panels) and, in the lower part, $|w_0|$, $|w_1/w_0|$, $|w_2/w_0|$, $\tfrac{1}{2}\arg(w_0)$, $\tfrac{1}{2}\arg(w_1/w_0)$ and $\tfrac{1}{2}\arg(w_2/w_0)$. The first and third rows display the quantities predicted from the analytic relations of Sec.~\ref{sec:predictions_from_physics}, while the second and fourth rows display the corresponding quantities obtained by fitting a complex log–Taylor expansion to the three toy-model frequency maps. Each panel includes its own colour bar, with units matching the corresponding parameter.}
    \label{fig:predi_vs_actual_P}
\end{figure*}

On a discrete pixelization, let indices $p$ and $q$ correspond to sky directions $\mathbf{n}$ and $\mathbf{n}'$, respectively, and let $\alpha_{qp},\beta_{qp},\gamma_{qp}$ be the Euler angles of the rotation mapping $\mathbf{n}'\to\mathbf{n}$ in the $\mathrm{SO}(3)$ convention of \citet{Rotti:2018pzi}.  Then the complex $E$/$B$ family fields in pixel $p$ can be written as
\begin{align}
\underline{P}_{E,p} = \sum_q  \Delta\Omega\Big[
    &\mathcal{K}_{\mathcal I}(\beta_{qp})\,e^{-2i(\alpha_{qp}+\gamma_{qp})}\,\underline{P}_{q}
  \ +\nonumber\\& \mathcal{K}_{\mathcal D}(\beta_{qp})\,e^{2i(\alpha_{qp}-\gamma_{qp})}\,\underline{P}_{q}^{*}
\label{eq:kernel_QU_app_E}
\Big],
\end{align}
\begin{align}    
\underline{P}_{B,p} = \sum_q \Delta\Omega\Big[
    &\mathcal{K}_{\mathcal I}(\beta_{qp})\,e^{-2i(\alpha_{qp}+\gamma_{qp})}\,\underline{P}_{q}
  - \nonumber\\&\mathcal{K}_{\mathcal D}(\beta_{qp})\,e^{2i(\alpha_{qp}-\gamma_{qp})}\,\underline{P}_{q}^{*}
\Big],
\label{eq:kernel_QU_app_B}
\end{align}
where $\Delta\Omega$ is the pixel area and $\mathcal{K}_{\mathcal I}$ and $\mathcal{K}_{\mathcal D}$ are the two radial kernels introduced in Eq.~(3.23) of \cite{Rotti:2018pzi} (under the notation $\,_{\mathcal I}f$ and $\,_{\mathcal D}f$).

It can be convenient to extract the purely geometric kernels that multiply
$\underline{P}_q$ and $\underline{P}_q^*$:
\begin{align}
&\underline{\mathcal{K}}^{(+)}_{E,pq}
\equiv \Delta\Omega\,\mathcal{K}_{\mathcal I}(\beta_{qp})\,
        e^{-2i(\alpha_{qp}+\gamma_{qp})},
\\
&\underline{\mathcal{K}}^{(-)}_{E,pq}
\equiv \Delta\Omega\,\mathcal{K}_{\mathcal D}(\beta_{qp})\,
        e^{2i(\alpha_{qp}-\gamma_{qp})},
\\
&\underline{\mathcal{K}}^{(+)}_{B,pq}
\equiv \Delta\Omega\,\mathcal{K}_{\mathcal I}(\beta_{qp})\,
        e^{-2i(\alpha_{qp}+\gamma_{qp})},
\\
&\underline{\mathcal{K}}^{(-)}_{B,pq}
\equiv -\,\Delta\Omega\,\mathcal{K}_{\mathcal D}(\beta_{qp})\,
        e^{2i(\alpha_{qp}-\gamma_{qp})}.
\end{align}
Equations~\ref{eq:kernel_QU_app_E} and \ref{eq:kernel_QU_app_B} then become
\begin{align}
\underline{P}_{E,p}
&=
\sum_q
\Big[
    \underline{\mathcal{K}}^{(+)}_{E,pq}\,\underline{P}_q
  + \underline{\mathcal{K}}^{(-)}_{E,pq}\,\underline{P}_q^*
\Big],
\label{eq:PE_kernel_geom}
\\
\underline{P}_{B,p}
&=
\sum_q
\Big[
    \underline{\mathcal{K}}^{(+)}_{B,pq}\,\underline{P}_q
  + \underline{\mathcal{K}}^{(-)}_{B,pq}\,\underline{P}_q^*
\Big].
\label{eq:PB_kernel_geom}
\end{align}
These expressions make explicit that the \textit{E}/\textit{B} projectors are linear in
$(Q,U)$ (equivalently in $\underline{P}$ and $\underline{P}^*$) with
geometry-only kernels $\underline{\mathcal{K}}^{(\pm)}_{E/B,pq}$.

Fig.~\ref{fig:full_convolution_kernel} illustrates the action of the real-space \textit{E}/\textit{B} convolution kernels on a simple test case consisting of a two-dimensional positive Gaussian feature in $Q$. The first set of panels shows the input polarization fields $(Q,U,P)$, where the Gaussian structure in $Q$ produces a localized polarization pattern with a well-defined orientation. Applying the convolution kernels yields the $E$-family maps $(Q_E,U_E,P_E, E)$ and the $B$-family maps $(Q_B,U_B,P_B,B)$ shown on the right (here, in practice, we performed the transformation through harmonic space).

Although the Gaussian is localized, the resulting $E$- and $B$-family fields extend over a larger area due to the non-fully-local nature of the convolution. The figure illustrates that $P_E$ and $P_B$ are approximately rotation- and translation-invariant representations of the polarization morphology, highlighting how even a simple structure is redistributed across neighbouring pixels.

\begin{figure*}
    \centering
    \includegraphics[width=\linewidth]{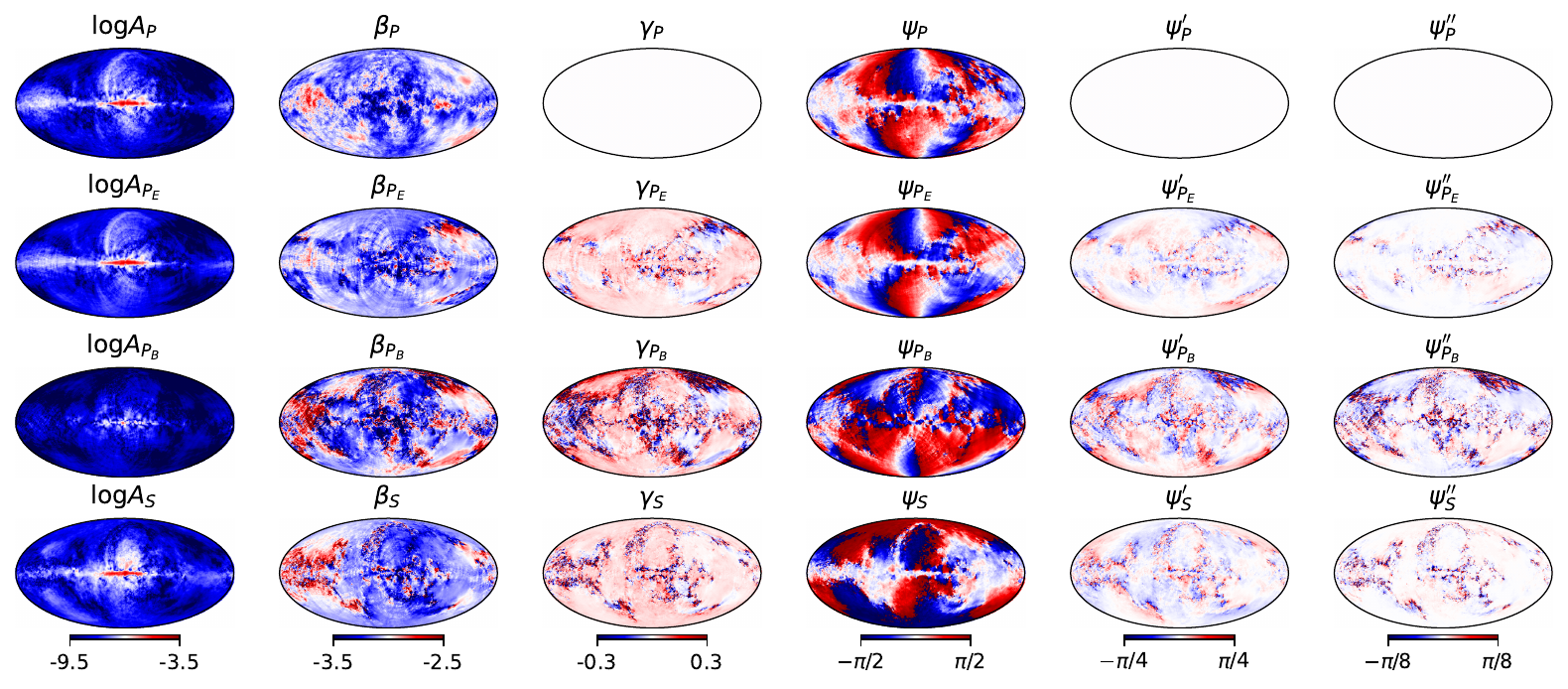}
    \includegraphics[width=\linewidth]{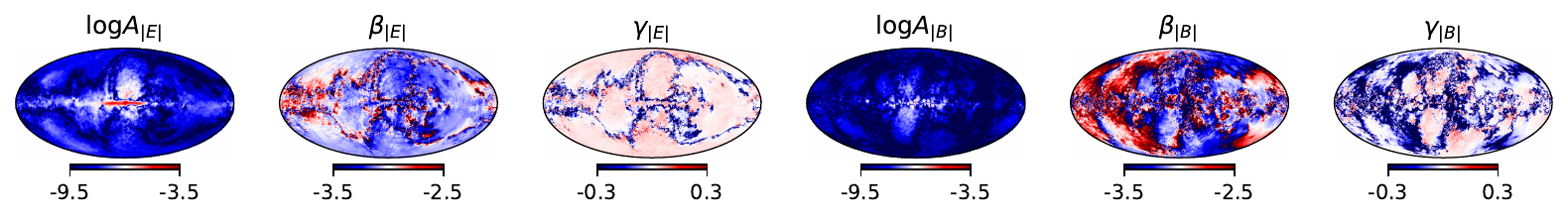}
    \caption{Similar to Fig.~\ref{fig:all_fields_logtaylor_TM} but for \texttt{PySM s5} around 10 GHz. On the lower row, we also show the log-Taylor spectral parameters one would obtain by fitting the absolute value of $E$ or $B$ scalar fields.}
    \label{fig:all_fields_logtaylor_pysm}
\end{figure*}

\section{Verification of relations for the toy model}
\label{sec:app_verif_rel}

In this Appendix, we explicitly verify the relations derived in Sec.~\ref{sec:predictions_from_physics} and used throughout Sec.~\ref{sec:comparaison_toy_model}. For the eight–component toy model described in Sec.~\ref{sec:comparaison_toy_model}, we first compute the complex spectral moments $\underline{w}_0$, $\underline{w}_1$ and $\underline{w}_2$ of the total polarization field $\underline{P}$ using the analytic combination rules of Eqs.~\ref{eq:w0_combination}–\ref{eq:w2_combination}, and convert them to complex log–Taylor parameters $(\underline{A},\underline{\beta},\underline{\gamma})$ with Eq.~\ref{eq:moments_2_taylor}. This yields the \emph{predicted} maps of amplitude, spectral index, curvature and angle derivatives. Independently, we generate three closely spaced frequency maps at $\nu_a=9.999~\mathrm{GHz}$, $\nu_b=\nu_0=10~\mathrm{GHz}$ and $\nu_c=10.001~\mathrm{GHz}$, and fit in each pixel a second–order complex log–Taylor expansion of $\underline{P}_\nu$ in $s$ (\textit{cf.} Eq.~\ref{eq:log_freq_s}). From these fits we obtain an \emph{actual} set of complex log–Taylor parameters and, by inverting Eq.~\ref{eq:moments_2_taylor}, an actual set of moments.

Fig.~\ref{fig:predi_vs_actual_P} compares these two constructions. The first two rows show, for the total polarization field $P$, the predicted and actual maps of $\log A_P$, $\beta_P$, $\gamma_P$, the polarization angle $\psi_P$ and its first two derivatives with respect to $s$. The last two rows show the corresponding moduli and phases of the moments, displayed as $|w_0|$, $|w_1/w_0|$, $|w_2/w_0|$ and the half–phases $\tfrac{1}{2}\arg(w_0)$, $\tfrac{1}{2}\arg(w_1/w_0)$, $\tfrac{1}{2}\arg(w_2/w_0)$. The near–identity of the predicted and actual maps for all these quantities confirms that the line–of–sight combination rules and the log–Taylor/moment relations provide an accurate description of the toy model’s spectral behaviour at the pivot frequency.

\section{Maps of the \texttt{s5} log-Taylor parameters}
\label{sec:log_taylor_pysm}

Fig.~\ref{fig:all_fields_logtaylor_pysm} presents the full-sky maps of the complex log--Taylor spectral parameters for the \texttt{PySM s5} synchrotron model around a reference frequency of 10\,GHz, in direct analogy with Fig.~\ref{fig:all_fields_logtaylor_TM} for the toy model. For each field, we show the logarithmic amplitude $\log A$, the spectral index $\beta$, the curvature parameter $\gamma$, and the successive orders of the polarization-angle expansion $\psi$, $\psi'$, and $\psi''$. 

While, as expected from the rigid-angle power-law in total polarization, there is no curvature or rotation of the polarization angle with frequency in the model, spatial variations of the spectral index in $P$ are clearly visible, and give rise to curvature, $\psi'$, and $\psi''$ in $E$/$B$-separated fields, because of the mechanism explained in Sec.~\ref{sec:spatial_beta}. This spectral complexity arises solely from this mechanism, as there is no intrinsic curvature or realistic physical effects such as Faraday rotation or synchrotron ageing in the model. 

The scalar field $\underline{S}$ displays a comparable level of spectral complexity as $\underline{P}_E$ and $\underline{P}_B$, with large-amplitude fluctuations in both $\beta_S$ and $\gamma_S$ and strong angular evolution. The spin-2 \textit{E}/\textit{B} separation redistributes spectral complexity between the two parity families: $\underline{P}_E$ typically shows more coherent large-scale spectral patterns, while $\underline{P}_B$ exhibits stronger small-scale fluctuations in the curvature and angle-derivative maps. Finally, the lower row illustrates that the absolute values of the scalar fields $E$ or $B$ individually exhibit the highest level of spectral complexity, with large-amplitude fluctuations in both $\beta$ and $\gamma$.

A more quantitative comparison of the different fields, based on the statistical distributions of these spectral parameters, is presented in Sec.~\ref{sec:spec_complexity_in_pysm_hist}.

\bsp	
\label{lastpage}
\end{document}